\documentclass[%
 preprintnumbers,
 amsmath,amssymb,
 prb, 
 twocolumn]{revtex4-2}

\usepackage{SIunits}
\usepackage{graphicx}
\usepackage{xspace}
\usepackage{braket}
\usepackage{bm}
\usepackage{placeins}
\usepackage{leftidx}

\usepackage[utf8]{inputenc}

\usepackage[usenames,dvipsnames]{color}

\usepackage[pdftex, colorlinks=true, urlcolor=blue, linkcolor=blue, citecolor=blue, pdfstartview=FitH, plainpages=false]{hyperref}

\newcommand{\kb}{\mathbf{k}}
\newcommand{\qb}{\mathbf{q}}
\newcommand{\qq}{\mathbf{q}}
\newcommand{\kk}{\mathbf{k}}
\newcommand{\rb}{\mathbf{r}}
\newcommand{\Gb}{\mathbf{G}}
\newcommand{\im}{i}
\newcommand{\enk}{\varepsilon_{n\kb}}
\newcommand{\emkq}{\varepsilon_{m\kb+\qb}}
\newcommand{\ef}{\varepsilon_F}
\newcommand{\wnuq}{\omega_{\qb\nu}}
\newcommand{\qnu}{{{\qb\nu}}}
\newcommand{\wqnu}{{\omega_{\qnu}}}

\newcommand{\vnk}{\mathbf{v}_{n\kb}}

\newcommand{\vnka}{\mathrm{v}_{n\kb,\alpha}}
\newcommand{\vnkb}{\mathrm{v}_{n\kb,\beta}}

\newcommand{\Lc}{\mathcal{L}}

\newcommand{\abinit}{\textsc{Abinit}\xspace}

\newcommand{\epw}{\textsc{EPW}\xspace}

\newcommand{\yambo}{Yambo\xspace}

\newcommand{\gkkp}{g_{mn\nu}(\kb,\qb)}

\newcommand{\abinitio}{\textit{ab initio}\xspace}
\newcommand{\grid}[1]{${#1}\times{#1}\times{#1}$}
\newcommand{\RR}{{\mathbf R}}
\newcommand{\rr}{{\mathbf r}}
\newcommand{\ie}{{\emph{i.e.}}\;}

\newcommand{\PDER}[2]{\dfrac{\partial #1}{\partial #2}}
\newcommand{\GG}{{\bf G}}

\newcommand{\qG}{{\bf q+G}}

\newcommand{\FM}{{\text{FM}}}

\newcommand{\KS}{{\text{KS}}}
\newcommand{\VH}{{V^{\text{H}}}}
\newcommand{\vH}{{v^{\text{H}}}}
\newcommand{\Vxc}{{V^{\text{xc}}}}
\newcommand{\vxc}{{v^{\text{xc}}}}
\newcommand{\Vscf}{{V^{\text{scf}}}}
\newcommand{\vscf}{{v^{\text{scf}}}}
\newcommand{\Vloc}{{V^{\text{loc}}}}
\newcommand{\vloc}{{v^{\text{loc}}}}
\newcommand{\Vnl}{{V^{\text{nl}}}}
\newcommand{\vnl}{{v^{\text{nl}}}}

\newcommand{\Zstar}{{\bf Z}^*}
\newcommand{\BZ}{{\text{BZ}}}
\newcommand{\epsinf}{{\bm{\epsilon}}^\infty}
\newcommand{\zero}{\mathbf{0}}

\newcommand{\mcS}{{\mathcal{S}}}

\newcommand{\omcS}{{\hat\mcS}}
\newcommand{\HH}{{\hat H}}
\newcommand{\ee}{\varepsilon}

\newcommand{\Si}{\mcS^{-1}}

\newcommand{\LL}{{\bf L}}

\newcommand{\btau}{{\bf \bm \tau}}
\newcommand{\bvec}{{\bf f}}

\newcommand{\nk}{{n\kk}}
\newcommand{\dd}{{\,\text{d}}}

\newcommand{\oSv}{{\hat\mcS_\bvec}}
\newcommand{\etal}{\textit{et al.}\xspace}
\newcommand{\dec}{{\Delta\varepsilon_c}}
\newcommand{\dev}{{\Delta\varepsilon_v}}

\begin{document}

\title{Phonon-limited electron mobility in Si, GaAs and GaP 
with exact treatment of dynamical quadrupoles}
\date{\today}
\author{Guillaume Brunin$^1$}
\author{Henrique Pereira Coutada Miranda$^1$}
\author{Matteo Giantomassi$^1$}
\author{Miquel Royo$^2$}
\author{Massimiliano Stengel$^{2,3}$}
\author{Matthieu J. Verstraete$^{4,5}$}
\author{Xavier Gonze$^{1,6}$}
\author{Gian-Marco Rignanese$^1$}
\author{Geoffroy Hautier$^1$}

\affiliation{$^1$UCLouvain, Institute of Condensed Matter and Nanosciences (IMCN), Chemin des \'Etoiles~8, B-1348 Louvain-la-Neuve, Belgium}
\affiliation{$^2$Institut de Ci\`encia de Materials de Barcelona (ICMAB-CSIC), Campus UAB, 08193 Bellaterra, Spain}
\affiliation{$^3$ICREA-Instituci\'o Catalana de Recerca i Estudis Avan\c{c}ats, 08010 Barcelona, Spain}
\affiliation{$^4$NanoMat/Q-Mat/CESAM, Universit{\'e} de Li{\`e}ge (B5), B-4000 Li{\`e}ge, Belgium}
\affiliation{$^5$Catalan Institute of Nanoscience and Nanotechnology (ICN2), Campus UAB, 08193 Bellaterra, Spain}
\affiliation{$^6$Skolkovo Institute of Science and Technology, Moscow, Russia}


\begin{abstract}

We describe a new approach to compute the electron-phonon self-energy and carrier mobilities in semiconductors.
Our implementation does not require a localized basis set to interpolate the electron-phonon matrix
elements, with the advantage that computations can be easily automated.
Scattering potentials are interpolated on dense $\qq$ meshes using Fourier transforms and \abinitio models to describe the long-range potentials generated by dipoles and quadrupoles.
To reduce significantly the computational cost, we take advantage of crystal symmetries and employ the
linear tetrahedron method and double-grid integration schemes, in conjunction with filtering techniques in the Brillouin zone.
We report results for the electron mobility in
Si, GaAs, and GaP obtained with this new methodology.
\end{abstract}

\maketitle

\section{Introduction}

Electron-phonon (e-ph) interactions play an important role in various physical phenomena~\cite{Giustino2017} such as
conventional phonon-mediated superconductivity~\cite{Mcmillan1968,Choi2002,Luders2005,Marques2005,Margine2014,Margine2016},
phonon-assisted light absorption~\cite{Kioupakis2010,Noffsinger2012,Peelaers2012},
temperature-dependent band structures, zero-point renormalization of the 
band gap in semiconductors~\cite{Allen1976,Allen1983,Ponce2014,Antonius2015,Ponce2015}, 
and thermal~\cite{Broido2007,Liao2015,Kim2016}
and electrical conductivities~\cite{Xu2014,Bernardi2014,Krishnaswamy2017,Ma2018,Li2015,Mustafa2016,Liao2015b,Ponce2018,Zeng2018,Zhou2016}.
Over the past years, several different open-source codes have been developed to compute e-ph related physical properties from first principles.
For example, the e-ph self-energy (SE) and the renormalization of band energies can be
obtained either with \abinit~\cite{Gonze2011,Gonze2016,Gonze2019}
or \yambo~\cite{Marini2008,Marini2009}.
The latter relies on the e-ph matrix elements
computed with Quantum Espresso~\cite{Giannozzi2009}
and post-processes the data to compute the SE and quasi-particle corrections.
With the \epw software~\cite{Ponce2016,Ponce2018}, one can study several e-ph quantities
using an interpolation scheme that exploits the localization in real space 
of the Wannier representation~\cite{Marzari2012} to obtain the e-ph matrix elements in the full Brillouin zone (BZ) at a relatively 
small computational cost.
Unfortunately the generation of maximally-localized Wannier functions (MLWFs) is not always trivial
and non-negligible effort may therefore be needed to obtain an appropriate set of 
MLWFs spanning the energy region of interest~\footnote{
Considerable efforts have been made in recent years to develop
more robust algorithms to localize Wannier orbitals using few tunable parameters.
See, for example, the SCDM method proposed in Ref.~\cite{Damle2018}.
}.
This is especially true for 
systems 
whose band structure
cannot be easily interpreted in terms of standard chemistry concepts, or when
high-energy states must be included 
to compute the real part of the SE whose 
convergence with the number of empty states is notoriously slow~\cite{Vansetten2018}.
It is therefore not surprising that recent works
proposed to replace Wannier functions with atomic orbitals~\cite{Agapito2018}.

In this work, we present an alternative fully \abinitio method to compute the e-ph SE using plane waves and Bloch wave functions as implemented in \abinit~\cite{Gonze2009,Gonze2016,Gonze2019} thus
bypassing the transformation to localized orbitals.
Particular emphasis is given to the calculation of the imaginary part of the SE and phonon-limited mobilities although a similar methodology can be employed for the real part of the SE and temperature-dependent band structures~\cite{Allen1976,Giustino2017}.
While our approach requires, in principle, more floating-point operations and more memory than
a Wannier-based approach, it has the advantage that the computation can be
automated with minimal input from the user and that systematic convergence studies can be easily performed.
To make our approach competitive with Wannier-based implementations,
we take full advantage of crystal symmetries and employ the linear tetrahedron method in conjunction with 
double-grid and filtering techniques both in $\kk$  and $\qq$ space.
As a result, the number of
e-ph matrix elements needed to reach convergence is significantly reduced.
The scattering potentials are interpolated on arbitrary $\qq$ grids using a Fourier-based interpolation scheme~\cite{Eiguren2008} in which the long-range dipolar and quadrupolar fields are treated using nonanalytic expressions depending on the high-frequency dielectric tensor, Born effective charges, dynamical quadrupoles tensors and the response to a homogeneous static electric field. The importance of the dynamical quadrupoles for obtaining a reliable interpolation at small $\qq$ is discussed in our accompanying letter~\cite{Brunin2019prl}.
Parallel scalability with memory distribution is achieved thanks to five different levels of MPI parallelism.

The paper is organized as follows.
In Sec.~\ref{sec:eph_interaction}, we introduce the theory of e-ph interactions and its connection with state-of-the-art first-principles methods. 
We also provide more details on the Fourier interpolation of the scattering potentials and derive the expression for the long-range components including quadrupolar fields.
In Sec.~\ref{sec:transport_properties}, we lay out the formalism to compute phonon-limited transport properties, focusing on carrier mobilities within the linearized Boltzmann equation.
The different strategies implemented to reduce the computational cost are discussed in Sec.~\ref{sec:implementation}. 
Finally, in Sec.~\ref{sec:results}, we illustrate all these techniques by computing the phonon-limited mobility of electrons in Si, GaAs and GaP within the self-energy relaxation time approximation (SERTA).
Atomic (Hartree) units are used throughout the paper.

\section{Electron-phonon interaction}
\label{sec:eph_interaction}
In periodic solids, 
electron-phonon coupling effects 
are usually discussed in terms of the Hamiltonian~\cite{Mahan2014,Giustino2017}
\begin{equation}
\label{eq:h_tot}
\hat{H} = \hat{H}_\text{e} + \hat{H}_\text{ph} + \hat{H}_\text{e-ph}
\end{equation}
where 
\begin{equation}
\label{eq:h_e}
\hat{H}_\text{e} = \sum_{n\kb} 
    \ee_{n\kb}\,\hat{c}^\dagger_{n\kb} \hat{c}_{n\kb}
\end{equation}
describes noninteracting quasiparticles with crystalline momentum $\kk$, band index $n$ and dispersion $\ee_\nk$, while
\begin{equation}
\label{eq:h_ph}
\hat{H}_\text{ph} = \sum_{\qnu} 
    \omega_{\qnu}(\hat{a}^\dagger_\qnu\hat{a}_{\qnu} + \frac{1}{2})
\end{equation}
is the Hamiltonian associated to noninteracting phonons with wave vector $\qq$, mode index $\nu$ and frequency $\wqnu$.
$\hat{c}^\dagger_{n\kb}$ and $\hat{c}_{n\kb}$ ($\hat{a}^\dagger_{\qnu}$ and $\hat{a}_\qnu$) are fermionic (bosonic) creation and destruction operators.
Finally,
\begin{equation}
\label{eq:h_eph}
    \hat{H}_\text{e-ph} = \frac{1}{\sqrt{N_p}} \sum_{\substack{\kb,\qb \\ mn\nu}} 
                   \gkkp\, \hat{c}^\dagger_{m\kb+\qb} \hat{c}_{n\kb}
                   (\hat{a}_\qnu + \hat{a}^\dagger_{-\qnu})
\end{equation}
describes the coupling to first order in the atomic displacements, with $N_p$ the number of unit cells in the Born-von K\'arm\'an supercell
and $\gkkp$ the e-ph coupling matrix elements.
In principle,~Eq.~\eqref{eq:h_tot} should include the so-called Debye-Waller (DW) Hamiltonian that gives the coupling to second-order in the atomic displacement~\cite{Allen1976,Giustino2017}.
This term contributes importantly to the real part of the SE and must therefore be included when computing quasiparticle corrections and temperature dependent band structures.
In this work, however, we will be mainly focusing on the imaginary part of the e-ph SE hence the DW term will be ignored~\cite{Giustino2017}.

\subsection{Connection with DFT and DFPT}

The connection between the many-body Hamiltonian in Eq.~\eqref{eq:h_tot} and density-functional theory (DFT) is established by using the Kohn-Sham (KS) band structure in Eq.~\eqref{eq:h_e} and phonons from density-functional perturbation theory (DFPT) in Eq.~\eqref{eq:h_ph}~\cite{Marini2015,Giustino2017}.
Finally, the e-ph matrix element, $\gkkp$, is computed using
\begin{equation}
\gkkp =
    \braket{\psi_{m\kb+\qb}|\Delta_\qnu V^\KS|\psi_{n\kb}}
\label{eq:elphon_mel}
\end{equation}
where $\psi_{n\kb}$ is the KS Bloch state and $\Delta_\qnu V^\KS$ is the first-order variation of the
self-consistent KS potential,
\begin{equation}
    \Delta_{\qb\nu} V^\KS = 
    \dfrac{1}{\sqrt{2\wnuq}} \sum_{p\kappa\alpha} 
    \PDER{V^\KS}{{\tau}_{\kappa\alpha}} \dfrac{{e}_{\kappa\alpha,\nu}(\qb)}{\sqrt{M_\kappa}}\,e^{i\qb\cdot\RR_p} \label{eq:delta_vks_a}
\end{equation}
where ${e}_{\kappa\alpha,\nu}(\qb)$ is the $\alpha$-th Cartesian component of the phonon eigenvector 
for the atom $\kappa$ in the unit cell,
$M_\kappa$ its atomic mass, $\RR_p$ are lattice vectors identifying the unit cell $p$, and $\bm{\tau}_{\kappa}$ the ion position in the unit cell.
Following the convention of Ref.~\cite{Giustino2017}, we can write Eq.~\eqref{eq:delta_vks_a} in terms of a lattice-periodic function $\Delta_{\qb\nu} v^\KS$ defined as:
\begin{equation}
    \Delta_{\qb\nu} V^\KS = e^{i\qb\cdot\rr} \Delta_{\qb\nu} v^\KS.
\label{eq:delta_vks}
\end{equation}
The latter can be obtained as
\begin{equation}
    \Delta_{\qb\nu} v^\KS = 
    \dfrac{1}{\sqrt{2\wnuq}} \sum_{\kappa\alpha} 
    \dfrac{{e}_{\kappa\alpha,\nu}(\qb)}{\sqrt{M_\kappa}}\,\partial_{\kappa\alpha,\qq} v^\KS,
\end{equation}
where
\begin{equation}
    \partial_{\kappa\alpha,\qq} v^\KS = \sum_p e^{-i\qq\cdot (\rr - \RR_p)}\,
      \PDER{V^\KS}{\tau_{\kappa\alpha}} \Bigr|_{(\rr - \RR_p)}
\label{eq:DFPTpot}
\end{equation}
is the first-order derivative of the KS potential that can be obtained with DFPT
by solving self-consistently a system of Sternheimer equations for a given $(\kappa\alpha, \qq)$ perturbation~\cite{Gonze1997,Baroni2001}.

\subsection{Interpolation of the e-ph matrix elements}

Accurate calculations of e-ph properties require the coupling matrix elements on very dense $\kb$- and $\qb$-point grids.
The explicit DFPT computation of $\partial_{\kappa\alpha,\qq} v^\KS$ for many $\qq$ points thus represents the main bottleneck of the entire process.
To reduce the computational cost, 
following Ref.~\cite{Eiguren2008},
we interpolate the scattering potentials in $\qq$ space using Fourier transforms while the KS wave functions are obtained by performing a non-self-consistent (NSCF) calculation on arbitrary $\kk$ meshes.
In a pseudopotential-based implementation~\footnote{We assume norm-conserving pseudopotentials.}, 
the KS potential is given by
\begin{equation}
    \begin{split}
    V^\KS(\rr,\rr') = \underbrace{\Bigl[ \VH[n](\rr)  + \Vxc[n](\rr) + \Vloc(\rr) \Bigr] }_{\Vscf(\rr)} & \\
    \times \delta(\rr-\rr') + \Vnl(\rr,\rr')
    \end{split}
\end{equation}
and consists of contributions from the Hartree part ($\VH$), the exchange-correlation (XC) potential ($\Vxc$), 
and the bare pseudopotential term that, in turn, consists of the local ($\Vloc$) and nonlocal ($\Vnl$) parts~\cite{Payne1992}.
Following the internal \abinit convention, we group the Hartree, XC 
and local terms in a single potential, $\Vscf$, although only the first two terms are computed self-consistently.
The lattice-periodic part of the first-order derivative of the KS potential thus reads
\begin{equation}
    \begin{split}
    \partial_{\kappa\alpha,\qb} v^\KS = \underbrace{ \Bigl [ 
        \partial_{\kappa\alpha,\qb} \vH + 
        \partial_{\kappa\alpha,\qb} \vxc +
        \partial_{\kappa\alpha,\qb} \vloc \Bigr ]}_{\partial_{\kappa\alpha,\qb} \vscf} & \\
    +\, \partial_{\kappa\alpha,\qb} \vnl.
    \end{split}
    \label{eq:dvKS}
\end{equation}
The real space representation of $\partial_{\kappa\alpha,\qb} \vscf$ is obtained through the following Fourier transform:
\begin{equation}
	\label{eq:dfpt_pot_realspace}
    W_{\kappa\alpha}(\rr - \RR_p) = \dfrac{1}{N_\qb} \sum_\qb e^{-i\qb\cdot(\RR_p -\rr)}\,\partial_{\kappa\alpha,\qb}\vscf(\rr),
\end{equation}
where the BZ sum is over the $N_\qb$ $\qb$ points of the initial grid employed in the DFPT calculation.
$W_{\kappa\alpha}(\rr - \RR_p)$ represents the variation of the SCF potential associated 
with the displacement of the atom $\kappa$ in the unit cell identified by $\RR_p$ along the Cartesian direction $\alpha$.
Once $W_{\kappa\alpha}(\rr - \RR_p)$ is known, one can interpolate the potential at an arbitrary point $\tilde{\qb}$ 
using the inverse Fourier transform:
\begin{equation}
	\label{eq:dfpt_pot_interpolation}
    \partial_{\kappa\alpha,\tilde\qb}\vscf(\rr)
    \approx \sum_{\RR_p} 
    e^{i\tilde{\qb}\cdot(\RR_p - \rr)} W_{\kappa\alpha}(\rr - \RR_p),
\end{equation}
where the sum is over the real-space lattice vectors in the supercell associated to the $\qq$ mesh used in Eq.~\eqref{eq:dfpt_pot_realspace}.
Note that only the SCF part of the DFPT potential needs to be interpolated as the derivative on the nonlocal part $\Vnl$ can be computed analytically~\cite{Gonze1997}.

Our approach therefore differs from Wannier-based methods in which electrons and phonons are treated on the same footing and the e-ph vertex is expressed in real space as a two-point function~\cite{Giustino2007}.
The advantage of our method is that the Wannierization step is completely avoided since the electronic wave functions are treated exactly and only phonon-related quantities (scattering potentials, phonon eigenvectors and frequencies) need to be interpolated.
The price to pay is that one must perform an explicit NSCF calculation of the Bloch states 
on the dense $\kk$ mesh and the computation of the e-ph matrix elements (Eq.~\eqref{eq:elphon_mel}) now requires the application of the first order KS Hamiltonian.
It should be noted, however, that the NSCF part represents a small fraction of the overall computing time when compared to the DFPT calculations that must be performed for each $\qq$ point in the IBZ and all irreducible atomic perturbations.
The application of the first order Hamiltonian is floating-point intensive but one can benefit from highly efficient fast Fourier transforms to apply 
$\partial_{\kappa\alpha,\qb} \vscf$ in $\rr$ space while 
the derivative of the nonlocal part is computed directly in $\GG$ space
using pseudopotentials in the fully-separable Kleinman-Bylander form~\cite{Kleinman1982}.
In Sec.~\ref{sect:BZ_filtering}, we also explain
how to take advantage of homogeneous BZ meshes and filtering techniques to reduce considerably the number of matrix elements that must be computed.

\subsection{Long-wavelength limit in semiconductors}
\label{sec:lr_limit}

The accuracy of the interpolation in Eq.~\eqref{eq:dfpt_pot_interpolation} depends on the density of the initial \abinitio $\qq$ mesh that in turn defines the size of the Born-von K\'arm\'an supercell.
This technique was initially proposed to study superconducting properties in metals and initial $\qq$ meshes of the order of \grid{8} were found sufficient to obtain accurate results~\cite{Eiguren2008}.
In metals, indeed, the screened potential is usually short ranged provided one excludes pathological cases associated to Kohn anomalies~\cite{Kohn1959}.
In semiconductors, on the contrary, 
the long-range (LR) behavior of the screened potential is associated to nonanalyticities for $\qq \rightarrow \zero$ that must be avoided in the Fourier interpolation, and thus be handled separately. 

We now generalize the standard technique used in polar semiconductors to cope with these LR
nonanalytic contributions to the scattering potentials,
and analyze in detail the contribution of quadrupolar terms.
We recover 
the formal study 
done by Vogl in 1976 \cite{Vogl1976} in the context of many-body perturbation theory (MBPT), and update it to the DFPT treatment of atomic displacement and electric field response~\cite{Gonze1997}, as currently implemented in \abinit and other first-principles packages. 
We also establish the connection with the treatment of Born effective charges and dynamical quadrupoles by Stengel~\cite{Stengel2013}. 
Finally, we show how the nonanalytic terms are treated
in the relevant interpolations, as already done for the lowest-order dipole-dipole nonanalytic contribution
to the interatomic force constants~\cite{Gonze1997} and for the Fr\"ohlich-type
dipole-induced scattering potential~\cite{Verdi2015}, both relying
on the Born effective charges. 
We extend the latter work to the next order by including dynamical quadrupoles and local-field potentials.

For this purpose, the nonanalytic long-wavelength components ($\qb \rightarrow \zero$) of the first-order derivative of the SCF potential $\partial_{\kappa\alpha,\qb} v^\text{scf}(\rr)$ must be identified. 
Our aim is to retain all contributions that are 
${\cal{O}}(q_{\alpha}/q^2)$ with $q$ the norm of $\qq$ (the strongest divergence is
dipole-like and has already been
treated, e.g., in Ref.~\cite{Verdi2015}), as well as
${\cal{O}}(q_{\alpha}q_{\beta}/q^2)$, linked to quadrupoles and local field potentials with nonanalytic directional behavior, as shown in the following.
On the other hand, we will neglect all ${\cal{O}}(q_{\alpha}q_{\beta}q_{\gamma}/q^2)$ contributions (associated with octupoles) and analytic ${\cal{O}}(1)$
contributions that can already be Fourier interpolated. 
We will obtain the above-mentioned behaviors
after alignment and rescaling of the wave vector components. 
In the following, for a real-space function $f_{\qb}(\rb)$, we denote generically with $f_{\qb}(\Gb)$ the components of its Fourier decomposition:
\begin{equation}
f_{\qb}(\rb)=\sum_{\Gb}f_{\qb}(\Gb) e^{i (\qG) \cdot \rr}.
\end{equation}

Let us first focus on the $\qb \rightarrow \zero$ behavior of the first-order derivatives of the wave functions with respect to collective atomic displacements, $|u^{\tau_{\kappa\alpha}}_{n\kb,\qb \rightarrow \zero}\rangle$~\cite{Stengel2013,Ponce2015}.
The computation of these quantities implies auxiliary DFPT calculations, 
delivering 
$|u^{\tau_{\kappa\alpha}^\prime}_{n\kb,\qb=\zero}\rangle$ 
and
$|u^{\cal{E}_{\lambda}^\prime}_{n\kb}\rangle$.
These are obtained by considering 
perturbations of the collective atomic displacement
and of the electric field types, at 
$\qb = \zero$,
in which the divergent
$\Gb=\zero$ component of the first-order derivative of the
Hartree potential $\frac{4 \pi}{q^2}$ 
has been removed, see Eq.~\eqref{eq:Hartree_change}.
From Ref.~\cite{Ponce2015}, we can express the derivatives of the wave functions {\it with} the ${\Gb=\zero}$ contribution
to the Hartree potential as
\begin{equation}
\begin{split}
|u^{\tau_{\kappa\alpha}}_{n\kb,\qb \rightarrow \zero}\rangle  = &
\,|u^{\tau_{\kappa\alpha}^\prime}_{n\kb,\qb=\zero}\rangle \\
 & +
\frac{4 \pi}{\Omega}
\frac{(-{\cal{Q}})
     (q_{\gamma}Z^*_{\kappa\alpha,\gamma})
     q_{\lambda} |u^{\cal{E}_{\lambda}^\prime}_{n\kb}\rangle}
{q^2\epsilon_M(\qq)}e^{-iq_\eta\tau_{\kappa\eta}},
\end{split}
\label{eq:wfNA_q_rightarrow_0}
\end{equation}
where $\epsilon_M(\qq)$ is the $\qq$-dependent macroscopic dielectric function,
$\Zstar_{\kappa}$ is the Born effective charge tensor,
$\eta$, $\gamma$ and $\lambda$, are Cartesian coordinates,
and ${\cal{Q}}$ is the electronic charge in atomic units, that is, $-1$. The sum over repeated indices is implied in this equation and in the remaining of this section. See Appendix~\ref{app:lr_limit} for more details about the derivation of Eq.~\eqref{eq:wfNA_q_rightarrow_0}.

We now generalize Eq.~\eqref{eq:wfNA_q_rightarrow_0} to arbitrary $\qb$ wave vectors.
We consider a first auxiliary DFPT calculation
with respect to collective atomic displacements, 
now at nonzero $\qb$, where the $\Gb=\zero$ component of the first-order derivative of the
Hartree potential has been removed, delivering
$|u^{\tau_{\kappa\alpha}'}_{n\kb,\qb}\rangle$.
The second auxiliary DFPT calculation is with respect to a 
monochromatic scalar potential $\phi$, with
nonzero $\qb$, as defined in Ref.~\cite{Royo2019},
again without the $\Gb=\zero$ contribution
to the first-order derivative of the Hartree potential. The corresponding first-order
wave function is $|u^{\phi'}_{n\kb,\qb}\rangle$.
The first-order DFPT response of the wave functions
to collective atomic displacements, 
at nonzero $\qb$ {\it with} the $\Gb=\zero$ contribution
to the Hartree potential is then obtained as
\begin{equation}
|u^{\tau_{\kappa\alpha}}_{n\kb,\qb}\rangle=
|u^{\tau_{\kappa\alpha}'}_{n\kb,\qb}\rangle
+
\frac{4 \pi}{\Omega}
\frac{{\cal{Q}}Q^{\qb}_{\kappa\alpha}}
{q^2\epsilon_M({\qb})
}|u^{\phi'}_{n\kb,\qb}\rangle e^{-iq_\eta\tau_{\kappa\eta}},
\label{eq:wfNA_q}
\end{equation}
see Appendix~\ref{app:lr_limit} for more details.
In the long-wavelength limit, the cell-integrated charge response to a monochromatic atomic displacement $Q^{\qb}_{\kappa\alpha}$
is~\cite{Royo2019} 
\begin{equation}
Q^{\qb}_{\kappa\alpha}=
-i q_{\gamma} Z^*_{\kappa\alpha,\gamma}
-\frac{q_{\beta}q_{\gamma}}{2}
Q^{\beta\gamma}_{\kappa\alpha}
+...,
\label{eq:Qqka_expansion}
\end{equation}
where $Q^{\beta\gamma}_{\kappa\alpha}$ is the dynamical quadrupole.

The self-consistency loop is linear in the first-order derivatives of the different quantities that 
are connected through it. 
So, the first-order change of the density, the first-order derivatives of the SCF
and Kohn-Sham potentials
of the $\tau_{\kappa \alpha}$ perturbation
bear the same relation to those of the 
$\tau_{\kappa \alpha}'$ and $\phi'$ perturbation,
than the one obtained for the wave functions,
Eq.~\eqref{eq:wfNA_q}.
For the phase-factorized periodic part of the KS potential (see Eqs.~\eqref{eq:delta_vks}--\eqref{eq:DFPTpot}), 
we obtain
\begin{equation}
\begin{split}
\partial_{\kappa\alpha,\qb}v^{\text{KS}} = & 
\,\partial_{\kappa\alpha,\qb}'v^{\text{KS}} \\
 & +
\frac{4 \pi}{\Omega}
\frac{Q^{\qb}_{\kappa\alpha}}
{q^2\epsilon_M({\qb})
}
(1+{\cal{Q}}\partial_{\phi}'v_\qb^{\text{Hxc}}) e^{-iq_\eta\tau_{\kappa\eta}},
\end{split}
\label{eq:vKS}
\end{equation}
where the prime refers again to the auxiliary DFPT calculations and we have taken into account that the $\phi$ 
perturbation has an external scalar ${\cal{Q}}$ contribution
and that the nonlocal term
is not present in the $\phi$ perturbation or in the
corresponding Hartree or XC potentials.
This formula allows one to compute the e-ph matrix elements in the vicinity of
$\qb={\zero}$, to all orders in $q$.

We can now derive the dipole and quadrupole-type nonanalyticities through
the analysis of the different contributing terms.
The first term in Eq.~\eqref{eq:vKS} is analytic.
The denominator of the second term, $q^2\epsilon_M({\qb})$,
expands as~\cite{Ponce2015}
\begin{equation}
q^2\epsilon_M({\qb})=
q_{\delta}\epsilon^{\infty}_{\delta\delta'}q_{\delta'} +{\cal{O}}(q^4),
\label{eq:q2epsilon}
\end{equation}
where $\epsinf$ is the high-frequency dielectric tensor.
Appearing in the denominator, it gives the nonanalytic behavior of the second term.
There is no cubic contribution in $q$ in this expansion, as emphasized
by Vogl~\cite{Vogl1976}.
The $\epsinf$ tensor might be
used as a metric, to highlight the ${\cal{O}}(q^2)$
character of this denominator~\cite{Gonze1997}.

Equation~\eqref{eq:Qqka_expansion} shows that the expansion of $Q^{\qb}_{\kappa\alpha}$
is analytic, and likewise
for $\partial_{\phi}'v^{\text{Hxc}}$, whose expansion
starts with the term proportional to $q$:
\begin{equation}
\partial_{\phi}'v_\qb^{\text{Hxc}}=
iq_{\alpha}v^{\text{Hxc},\cal{E}_\alpha}+{\cal{O}}(q^2).
\label{eq:dphi'vHxc}
\end{equation}
Thus, the dominant nonanalyticity is a real-space constant shift in the potential, of dipolar character (i.e. of order
$\frac{q_\alpha}{q^2}$ considering the appropriate metric in reciprocal space):
\begin{equation}
\frac{4 \pi}{\Omega}
\frac{i q_{\gamma} Z^*_{\kappa\alpha,\gamma}}
{{q_{\delta}\epsilon^{\infty}_{\delta\delta'}q_{\delta'}}} e^{-iq_\eta\tau_{\kappa\eta}}.
\label{eq:NAdipole}
\end{equation}
At the quadrupole level (i.e. of order
$\frac{q_\beta q_\gamma}{q^2}$ in the proper metric), there are two contributions:
a real-space constant shift,
\begin{equation}
\frac{4 \pi}{\Omega}
\frac{(\frac{q_{\beta}q_{\gamma}}{2})
Q^{\beta\gamma}_{\kappa\alpha}}
{{q_{\delta}\epsilon^{\infty}_{\delta\delta'}q_{\delta'}}
} e^{-iq_\eta\tau_{\kappa\eta}},
\label{eq:NAquadrupole1}
\end{equation}
and a nonconstant real-space potential,
\begin{equation}
\frac{4 \pi}{\Omega}
\frac{(-q_{\beta}q_{\gamma}) Z^*_{\kappa\alpha,\beta}}
{{q_{\delta}\epsilon^{\infty}_{\delta\delta'}q_{\delta'}}}
{\cal{Q}}v^{\text{Hxc},\cal{E}_\gamma}(\rr) e^{-iq_\eta\tau_{\kappa\eta}}.
\label{eq:NAquadrupole2}
\end{equation}
The latter is obtained from \abinit, considering the electric field (${\cal{QE}}_\alpha$) perturbation.

These terms parallel those found in the MBPT context by Vogl~\cite{Vogl1976}. 
Note, however, that Vogl was focusing on the e-ph scattering matrix elements
for some well-defined normal mode of vibration
while the present work focuses on the collective displacement of a
given single sublattice $\kappa$ in a specific direction $\alpha$. 
In this context, Vogl obtained a third quadrupole contribution that is not present here.
The generalization of Eq.~\eqref{eq:NAdipole} to finite $\Gb$ has already been done in Ref.~\cite{Verdi2015}. The generalization of Eqs.~\eqref{eq:NAquadrupole1} and \eqref{eq:NAquadrupole2} follows the same derivation.
The expression for the LR model finally reads:
\begin{widetext}
    \begin{equation}
        V^{\Lc}_{\kappa\alpha,\qb}(\rb) = 
        \frac{4\pi}{\Omega} \sum_{\Gb\neq\mathbf{-q}}
        \frac{ i(q_{\beta}+G_{\beta}) 
        Z^*_{\kappa\alpha,\beta} 
        -
        (q_{\beta}+G_{\beta})(q_{\gamma}+G_{\gamma})
        ({Z^*_{\kappa\alpha,\beta}\cal{Q}}v^{\text{Hxc},\cal{E}_{\gamma}}(\rr)- \frac{1}{2}Q_{\kappa\alpha}^{\beta\gamma})
        }
        {
        {(q_{\delta}+G_{\delta})\epsilon^{\infty}_{\delta\delta'}(q_{\delta'}+G_{\delta'})}}
        e^{i (q_\eta + G_\eta) (r_\eta - \tau_{\kappa\eta})}.
    \label{eq:LRpot}
    \end{equation}
\end{widetext}
In the actual implementation, following previous approaches~\cite{Sjakste2015,Giustino2017},
each component is multiplied 
by the Gaussian filter $e^{-\frac{|\qG|^2}{4\alpha}}$~\footnote{
The $\alpha$ parameter determines the separation between the long-range and the short-range parts of the interaction and is used 
to express
Ewald sums~\cite{Pickett1989,Gonze1997} 
in terms of a sum in $\GG$ space (long-range part) and a sum in real space that, being short ranged, is not relevant for the definition of the LR model.
The optimal value of $\alpha$ is material-dependent
and should therefore be subject to convergence studies.
In our systems, 
we observed small changes in the physical observables ($\sim$1\%) with $\alpha$.
A value of $0.1$~Bohr$^{-2}$ is used in all calculations.
}.
%
%

The impact of the electric-field term in Eq.~\eqref{eq:LRpot} has been analyzed in details in Ref.~\cite{Brunin2019prl}. In Si, GaP, and GaAs, the effect of the $\cal{E}$ term 
on the Fourier interpolation is negligible compared to the role played by $Q^{\beta\gamma}_{\kappa\alpha}$.
Therefore we do not include the
$\cal{E}$ term in the computations reported in this work.
We do treat, however,
the dipole-quadrupole and quadrupole-quadrupole interactions 
when interpolating the phonon frequencies following the formalism
detailed in Ref.~\cite{Stengel2013}.

\section{Phonon-limited transport properties} 
\label{sec:transport_properties}

\subsection{Electron lifetimes}
\label{sec:e_lifetimes}

The electron lifetime due to the e-ph scattering is related to the inverse of the 
imaginary part of the e-ph Fan-Migdal SE~\cite{Giustino2017}.
The diagonal matrix elements of the SE in the KS basis set are given by
\begin{equation}
\begin{split}
    \Sigma^\FM_{n\kb}(\omega,\ef,T) =
                & \sum_{m,\nu} \int_\BZ \frac{d\qb}{\Omega_\BZ} |\gkkp|^2 \\
                & \times \left[
                    \frac{n_\qnu(T) + f_{m\kb+\qb}(\ef,T)}
                         {\omega - \emkq  + \wnuq + i \eta} \right.\\
                & \left. +
                    \frac{n_\qnu(T) + 1 - f_{m\kb+\qb}(\ef,T)}
                         {\omega - \emkq  - \wnuq + i \eta} \right] ,
\end{split}
\label{eq:fan_selfen}
\end{equation}
where $f_{m\kb+\qb}(\ef,T)$ and $n_\qnu(T)$ correspond to the Fermi-Dirac and Bose-Einstein occupation functions
with $T$ the temperature and $\ef$ the Fermi level.
For the sake of simplicity, the temperature and Fermi level are considered as parameters, and the dependence on $T$ and $\ef$ will be omitted in the following.
The integral in Eq.~\eqref{eq:fan_selfen} is performed over the $\qb$ points in the BZ of volume $\Omega_\BZ$
and $\eta$ is a positive real infinitesimal.

In the $\eta \rightarrow 0^+$ limit, the imaginary part of the SE (Eq.~\eqref{eq:fan_selfen}) 
evaluated at the KS energy is given by~\cite{Giustino2017}
    \begin{equation}
    \begin{split}
        \lim_{\eta \rightarrow 0^+} & \Im\{\Sigma^\FM_{n\kb}(\enk)\} =
                    \pi \sum_{m,\nu} \int_\BZ \frac{d\qb}{\Omega_\BZ} |\gkkp|^2\\
                    & \times \left[ (n_\qnu + f_{m\kb+\qb})
                                    \delta(\enk - \emkq  + \wnuq) \right.\\
                    & \left. + (n_\qnu + 1 - f_{m\kb+\qb})
                                    \delta(\enk - \emkq  - \wnuq ) \right]
    \end{split}
    \label{eq:imagfanks_selfen}
    \end{equation}
and corresponds to the linewidth of the electron state $n\kb$ due to the scattering with phonons.
Finally, the electron lifetime $\tau_{n\mathbf{k}}$ is
inversely proportional to the linewidth of the SE evaluated at the KS energy~\cite{Ponce2016,Ponce2018}:
    \begin{align}
    \frac{1}{\tau_{n\kb}} =
        2 \lim_{\eta \rightarrow 0^+} \Im\{\Sigma^\FM_{n\kb}(\enk)\}.
    \label{eq:fanlifetime}
    \end{align}
These lifetimes play an important role in different physical properties, such as optical absorption~\cite{Sangalli2019} or transport properties~\cite{Madsen2018}. 
In this work, we focus on the accurate computation of 
phonon-induced lifetimes and transport properties, without taking into account other scattering processes due to defects, impurities, grain boundaries or other electrons.
As we include only one of the possible scattering mechanisms,
our computed mobilities are expected 
to overestimate the experimental results~\footnote{
Scattering by defects, impurities, grain boundaries, or other electrons
can be described either using semi-empirical models~\cite{Calnan2010,Liu2017}
or first-principles computations~\cite{Bernardi2014,Lu2019}.
The computation of these effects is however outside the scope of the present work.
}.

\subsection{Carrier mobility}

In this article, we focus on the solution of the linearized Boltzmann transport formulation~\cite{Ashcroft1976}
within the relaxation time approximation (RTA).
The generalized transport coefficients are given by~\cite{Madsen2018}
\begin{equation}
    \begin{split}
    \Lc^{(m)}_{\alpha\beta} = 
    - \sum_n \int \frac{d\kb}{\Omega_\BZ} \vnka \vnkb\, \tau_{n\kb} & \\
    \times (\enk-\ef)^m 
    \left.\frac{\partial f}{\partial\varepsilon}\right|_{\enk}
    \label{eq:transport_lc}
    \end{split}
\end{equation}
where $\vnka$ is the $\alpha$-th component of the matrix element $\vnk$ of the electron velocity operator. These quantities can be obtained from interpolation methods such as
Wannier~\cite{Pizzi2014} or Shankland-Koelling-Wood (SKW)~\cite{Shankland1971,Euwema1969,Koelling1986,Madsen2006}
whose form allows one to analytically compute the derivatives of the eigenvalues with respect to $\kb$.
Alternatively, one can obtain $\vnk$ using finite differences between shifted
grids~\cite{Sangalli2019}.
In our implementation, we prefer to compute the velocity matrix elements without any approximation
using the commutator of the Hamiltonian with the position operator~\cite{DelSole1993}:
    \begin{align}
    \vnk = \dfrac{\partial \enk}{\partial \kb} = \braket{\psi_{n\kb}|\hat{\mathbf p} + \im\,[\Vnl, \rr] |\psi_{n\kb}}
    \label{eq:velocities}
    \end{align}
with $\hat{\mathbf p}$ the momentum operator, as done in DFPT calculations of the response to an external electric field~\cite{Gonze1997}.
We note, in passing, that the group velocities can be used to compute transport lifetimes within the so-called momentum relaxation time approximation~\cite{Li2015,Ma2018}.

The generalized transport coefficients can be used to obtain different transport properties such as
the electrical conductivity, Peltier and Seebeck coefficients, and charge carrier contribution to the
thermal conductivity tensors~\cite{Madsen2018}. The electrical conductivity tensor is given by
\begin{equation}
    \sigma_{\alpha\beta} =
    \frac{1}{\Omega} \Lc_{\alpha\beta}^{(0)} \label{eq:transport_sigma}
\end{equation}
and can be divided into hole and electron contributions~\footnote{We consider the conductivity related to the electric field only, not the Hall mobility~\cite{Ricci2017}.}:
    \begin{equation}
    \sigma = n_e \mu_{e} + n_h \mu_{h} \label{eq:mobility}
    \end{equation}
where $n_e$ and $n_h$ are the electron and hole concentrations in the conduction and valence bands respectively,
and $\mu_e$ and $\mu_h$ are the electron and hole mobilities, 
which can be obtained by selecting the conduction or valences states $n$ in Eq.~\eqref{eq:transport_lc}.
For electrons,
\begin{equation}
\begin{split}
    n_e = \sum_{n\in \text{CB}} \int \dfrac{d\kb}{\Omega_\BZ} f_{n\kb}, \\
    \mu_e = \dfrac{1}{n_e \Omega}\, \Lc_{n\in \text{CB}}^{(0)}
\end{split}
\end{equation}
where $n\in\text{CB}$ denotes states in the conduction bands. Similar expressions hold for holes.
At zero total carrier concentration, the Fermi level $\ef$ is located inside the band gap so that $n_e = n_h$.

The transport coefficients in Eq.~\eqref{eq:transport_lc}
depend both on the temperature and the doping level (through $\ef$).
Formally, $\tau_{n\kb}$ also depends on $\ef$ through the Fermi-Dirac occupations in
Eq.~\eqref{eq:imagfanks_selfen}. 
A commonly used approximation consists in neglecting the variation of $\tau_{n\kb}$ with $\ef$.
Under this assumption, Eq.~\eqref{eq:transport_lc} can be solved for different $\ef$ 
at almost no additional cost.
This approach is valid when the transport coefficients are obtained
for values of $\ef$ similar to the ones used to compute $\tau_{n\kb}$.

\section{Efficient computation of e-ph quantities}
\label{sec:implementation}

\subsection{Use of symmetries}

As mentioned in the introduction, our implementation exploits crystal symmetries to reduce the computational cost and the memory requirements.
As concerns the KS wave functions, the NSCF calculation can be restricted to the $\kk$ points in the IBZ of the unperturbed system as the Bloch states in the full BZ can be reconstructed by symmetry using the equations given in 
Appendix~\ref{app:symmetry_properties_wfs}.
At the level of the scattering potentials, the DFPT computations need to be performed only for $\qq$ points in the IBZ and, for each $\qq$ point, only an appropriate set of irreducible perturbations needs to be computed explicitly.
As shown in Appendix~\ref{app:symmetry_properties_dfpt}, all the scattering potentials can be reconstructed from this irreducible set.
These symmetry properties are used, for instance, to compute the sum over $\qq$ points in the full BZ in Eq.~\eqref{eq:dfpt_pot_realspace}.

Symmetries are also used in the computation of the electron lifetimes.
First of all, the SE operator is invariant under the action of all the operations of the space group of the crystal.
As a consequence, the computation of $\tau_{n\kb}$ can be restricted to the $\kk$ points in IBZ as $\tau_{n\kb}$ transforms as the KS eigenvalues $\ee_\nk$.
Last but not least, for a given $\kk$ point,
the integral in $\qq$ space in Eq.~\eqref{eq:imagfanks_selfen} 
can be restricted to
an appropriate irreducible wedge ($\text{IBZ}_\kk$) defined by the operations of the little group of $\kb$, $\ie$
the set of point group operations that leave the $\kb$ point invariant modulo a reciprocal lattice vector $\GG$.

\subsection{Tetrahedron method}

The evaluation of Eq.~\eqref{eq:imagfanks_selfen} requires an accurate integration of the Dirac $\delta$ distribution,
especially in the regions around the band edges where lifetimes usually show strong variations and anisotropic behavior.
For practical applications, it is customary to replace the $\delta$ distribution by Lorentzian or Gaussian functions 
with a small but finite width, also called the broadening parameter~$\eta$.
According to previous studies~\cite{Zhou2016}, Gaussian functions lead to a faster convergence of the $\qb$-point integral in Eq.~\eqref{eq:imagfanks_selfen}.
From a theoretical perspective, the use of Lorentzian functions is more appropriate because, in Eq.~\eqref{eq:fan_selfen},
the $\delta$ distribution is obtained as the limit of a Lorentzian for $\eta \rightarrow 0$.
It should be stressed, however, that both the Lorentzian and Gaussian methods require careful convergence studies: 
for a given value of $\eta$, physical quantities should be converged by increasing the density of $\qb$ points
then one should monitor the behavior of the converged values for $\eta \rightarrow 0$ and select a broadening
for which the results are relatively stable.

To avoid this additional convergence study, we prefer to compute Eq.~\eqref{eq:imagfanks_selfen} with the linear tetrahedron method.
The implementation closely follows the algorithm proposed by Blochl~\cite{Blochl1994}.
The full BZ is first partitioned into tetrahedra, then symmetries are used to identify a set of irreducible tetrahedra with the corresponding corners and multiplicity.
The presence of $\ee_{m\kk+\qq}$ in the argument of the two $\delta$ functions implies that only the symmetries of the little group of $\kk$ can be used to define the irreducible tetrahedra covering the $\text{IBZ}_\kk$.
Once the irreducible tetrahedra have been identified for a given $\kk$, 
the squared modulus of the e-ph matrix element as well as
the $F_{m\nu}^\pm(\qq) = \ee_{m{\kk+\qq}} \pm \omega_{\qnu}$ functions are linearly interpolated in $\qq$ space using the values at the corners of each tetrahedron. 
Within the linear approximation, the isosurfaces $F_{m\nu}^\pm(\qq) = \ee_{\nk}$ become planes that may cut the tetrahedron depending on the value of $\ee_\nk$ and the integration inside each tetrahedron is performed analytically.
Finally, as discussed in Ref.~\cite{Blochl1994},
the sum over tetrahedra is converted into a weighted sum over the $\qq$ points in the $\text{IBZ}_\kk$ 
with different weights $w^\pm(n, m, \nu, \qq)$
for absorption and emission processes.
The weights are nonzero only for $\qq$ points associated to tetrahedra intersecting one of the possible isosurfaces.
This is the key to the filtering algorithm in $\qq$ space discussed in more details in the next section. 

\subsection{$\kb$- and $\qb$-points filtering}
\label{sect:BZ_filtering}

\begin{figure}
    \centering
    \includegraphics[clip,trim=0.2cm 0.3cm 0.2cm 0.2cm,width=.48\textwidth]{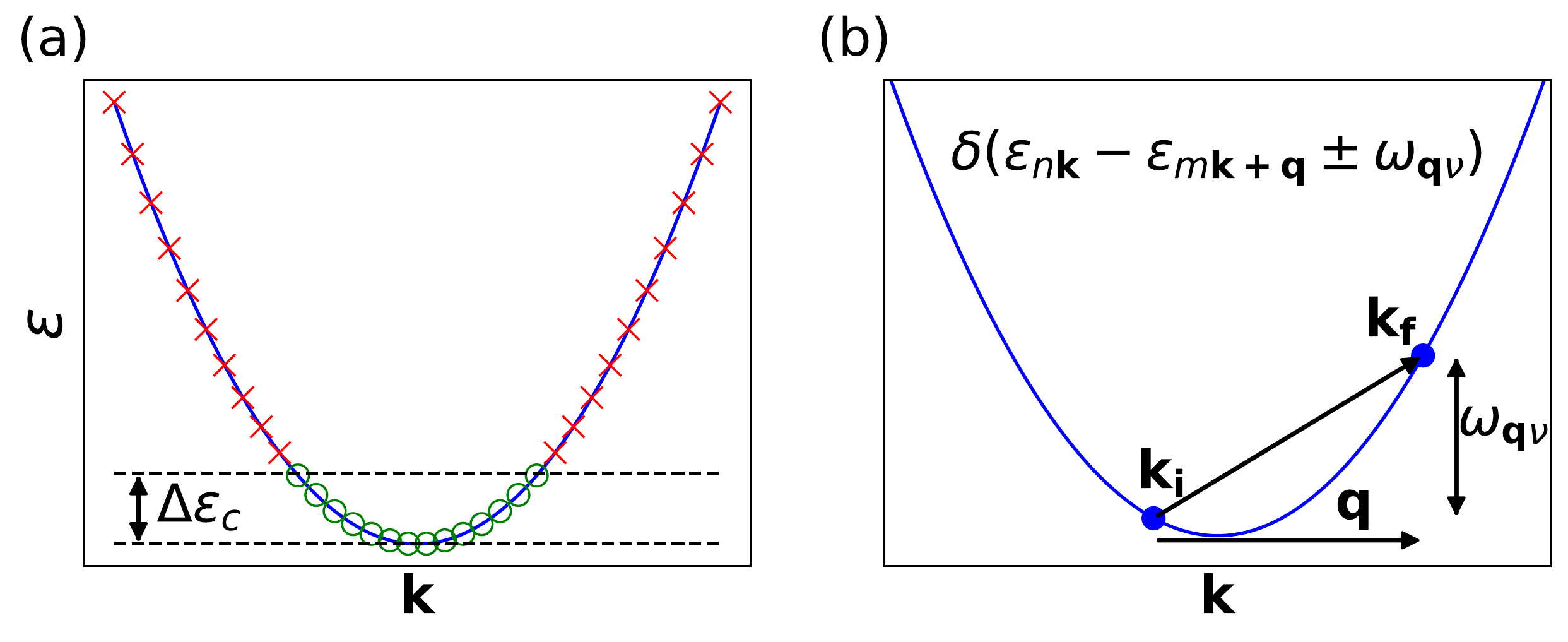}
    \caption{Schematic representation of the filtering of (a) $\kb$  and (b) $\qb$ points for the computation of the SE in the case of a single parabolic conduction band at the $\Gamma$ point. 
    (a) States represented by green dots (red crosses) are included (excluded) in (from) the transport computation 
    according to the input variable $\dec$.
    (b) For initial and final electron states $\kb_\mathbf{i}$ and $\kb_\mathbf{f}$, the wave vector $\qb = \kb_\mathbf{f} - \kb_\mathbf{i}$ requires a phonon mode with frequency $\wnuq = \varepsilon_{\kb_\mathbf{f}} - \varepsilon_{\kb_\mathbf{i}}$. If this condition is not fulfilled, the $\qq$ point is ignored.
    }
    \label{fig:filtering}
    \end{figure}

An initial filtering is achieved by noticing that
the derivative of the Fermi occupation $f$ in Eq.~\eqref{eq:transport_lc} 
is practically nonzero only for electronic states close to the Fermi level.
Since $\varepsilon_F$ is usually inside the band gap,
only electronic states with $\kb$ and $\enk$ close to the band edge(s) contribute to Eq.~\eqref{eq:transport_lc}~\cite{Sohier2018}.
In our implementation, we use an input variable that defines the energy range for electrons ($\dec$)
and holes ($\dev$): if 
$\enk - \varepsilon_{\text{CBM}} > \dec$ or 
$\varepsilon_{\text{VBM}} - \enk > \dev$, the lifetime $\tau_{n\mathbf{k}}$ is not computed because this state 
is assumed to give negligible contribution
to Eq.~\eqref{eq:transport_lc}.
This selection algorithm is depicted in Fig.~\ref{fig:filtering}(a), where $\tau_\nk$ is computed for the states represented by green circles but not for the red crosses. 

The achievable speedup depends on the band dispersion, the position of the Fermi level and the temperature. In most cases, the effective number of $\kk$ points is much smaller than the initial number of wave vectors in the IBZ associated to the homogeneous mesh.
In Sec.~\ref{sec:ph_limited_mobility},
it is shown that converged electron mobilities in Si and GaP can be obtained with around 29 and 60 irreducible $\kk$ points, respectively, representing only 1\% of the IBZ. 
An extreme scenario is represented by band dispersions with small effective masses for which very high resolution in $\kk$ space is needed to sample the small electron (hole) pockets.
For instance, in GaAs, due to the very small electron effective mass ($\sim$0.053~$m_0$ with $m_0$ the free electron mass), the effective number of $\kk$ points 
reduces to around 0.05\% of the IBZ. 
This class of systems is challenging also for Wannier-based approaches as is apparent from the large discrepancy in the results for the electron mobility in GaAs reported by different authors~\cite{Liu2017,Ma2018,Zhou2016}.
In our implementation, we can treat systems with small effective masses with the following procedure. 
We start with a NSCF calculation on a reasonably dense $\kk$ mesh to determine the position of the band edges within a certain tolerance.
Then we use the SKW method~\cite{Shankland1971,Euwema1969,Koelling1986}
to interpolate electron energies on a much denser $\kk$ grid.
This step allows us to identify the $\kk$ points lying inside a predefined energy window around the band edges.
Finally, a second NSCF calculation for this restricted set of wave vectors is performed and the resulting (exact) KS wave functions and energies are used to compute lifetimes and transport properties~\footnote{
The set of $\kk$ points obtained with this procedure belongs to a homogeneous mesh hence both symmetries and tetrahedron method can be used without any modification.
Small errors in the tetrahedron weights may be introduced by the linear interpolation because $\kk$ points that are slightly outside of the energy window are interpolated with SKW but this minor issue can be easily fixed by enlarging the energy window.
This techniques permits to reach very high resolution in $\kk$ space of the order of \grid{300}.
}.
We note that all the calculations reported in this work have been performed without this procedure because mobility computations in Si and GaP do not need particularly dense $\kk$ grids to converge. 
We were also able to reach converged computations in the worst-case scenario of GaAs without employing this technique despite the increased computational cost of the NSCF calculation.
It is however evident that 
e-ph calculations
in more challenging systems characterized by small effective masses and/or larger unit cells will benefit from this SKW-based approach.

The reduction in the number of $\kk$ points already leads to a large speed-up but 
another, distinct, optimization can be implemented when computing the imaginary part of the SE.
For a fixed $\kb$ point, indeed, not all $\qb$ points are compatible with energy conservation~\cite{Sohier2018}.
This selection rule is schematically depicted in Fig.~\ref{fig:filtering}(b).
Thanks to the tetrahedron method, it is possible to identify, for a given $\kb$ point,
the subset of $\qb$ points in the $\text{IBZ}_\kk$ contributing to Eq.~\eqref{eq:imagfanks_selfen}~\footnote{
In principle a similar optimization can be implemented for the Lorentzian and Gaussian broadening techniques,
however this requires choosing a threshold beyond which the broadened $\delta$ is treated as zero.
}.
This filtering technique restricts the computation of the matrix elements $\gkkp$ to a small set of $\qq$ points in the $\text{IBZ}_\kk$. 
The achievable reduction depends on the band structure and phonon dispersion, but in most cases only a few percents of the total number of $\qb$ points pass this filter.
In Si, for instance, less than 2\% of the $\text{IBZ}_\kk$
needs to be taken into account when evaluating the imaginary part of the SE thus leading to an additional significant speedup.
In GaAs, the reduction is even larger thanks to the highly-dispersive band centered at $\Gamma$ that allows for only a few intra-valley transitions. In this case, less than 0.2\% of the $\text{IBZ}_\kk$ has to be considered for each $\kk$ point.

\subsection{Double-grid technique}
\label{sect:double_grid}

\begin{figure}
    \centering
    \includegraphics[clip,trim=3.9cm 1.6cm 3.4cm 1.05cm,width=.34\textwidth]{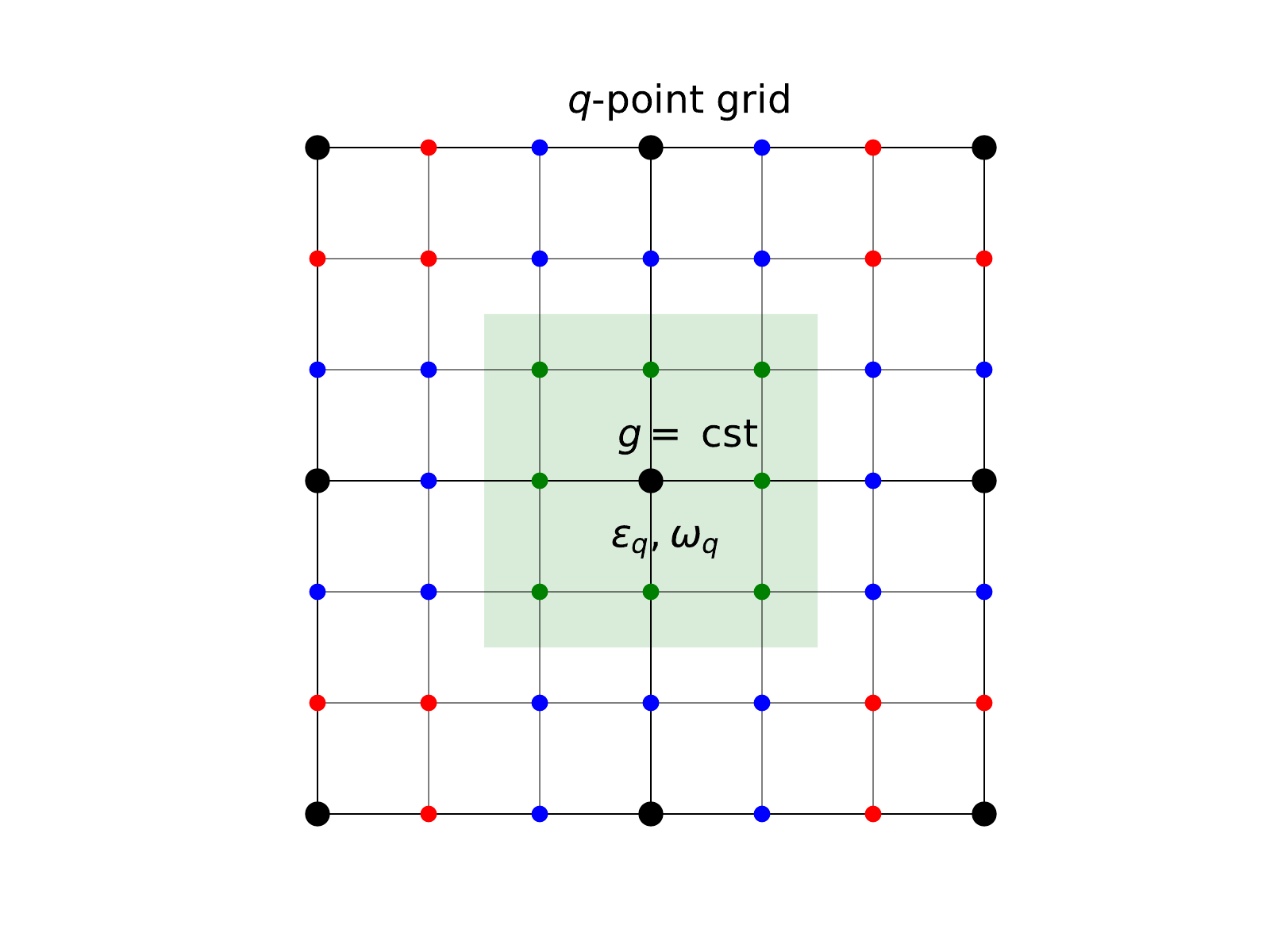}
    \caption{Schematic representation of the DG technique. 
    Black dots represent $\qb$ points belonging to the coarse grid 
    where the e-ph matrix elements $\gkkp$ are explicitly computed
    using our interpolation procedure.
    Colored small dots are $\qb$ points of the fine grid 
    where the $\gkkp$ are assumed to be constant 
    in the region surrounding the black dot (the shaded green area).}
    \label{fig:double_grid}
\end{figure}

The evaluation of $\ee_\nk$ and $\omega_\qnu$ is computationally less demanding than that of $\gkkp$.
This is evident in our implementation where the evaluation of
Eq.~\eqref{eq:elphon_mel} requires the electron wave functions and the derivative of the potential to
be stored in memory to apply the first-order KS Hamiltonian.
To improve the convergence rate without increasing the computational cost,
we implemented the possibility to use different $\qb$ meshes: one coarse mesh for $\gkkp$ and a finer one to describe the absorption and emission terms in Eq.~\eqref{eq:imagfanks_selfen}
where only $\emkq$
and $\omega_\qnu$ are needed.
This method, already implemented in the context of Bethe-Salpeter calculations~\cite{Kammerlander2012,Gillet2016,Sangalli2019} 
and used by Fiorentini \etal~\cite{Fiorentini2016} in the context of e-ph computations,
is commonly referred to as a double-grid (DG) technique
and is explained schematically in Fig.~\ref{fig:double_grid}.
The rationale behind such a technique is that, apart from the Fr\"ohlich divergence, 
the e-ph matrix elements
are expected to vary smoothly when compared to the Dirac $\delta$ functions in Eq.~\eqref{eq:imagfanks_selfen}.
Obviously this approximation breaks down in the region around $\qq=\Gamma$ in polar semiconductors,
where a high-resolution sampling would be required to capture the divergence.
Theoretically, it would be possible to address this issue by analytically integrating 
the diverging matrix elements close to $\Gamma$, or to compute the LR part of the matrix elements 
(as defined in Ref.~\cite{Verdi2015})
on the fine mesh used for the DG technique.
These extensions of the DG method are left for future studies.
For reasons of computational performance and reliability of the results, we only allow
commensurate grids. The phonon frequencies are obtained on the fine grid by
interpolating the dynamical matrix in $\qq$ space at negligible cost. The electron energies $\emkq$
on the fine grid can be obtained either
from a NSCF calculation or interpolated using the SKW 
method~\cite{Shankland1971,Euwema1969,Koelling1986}. 
This technique offers enough efficiency and flexibility to perform fast approximate computations of the
electron lifetimes for screening purposes with the additional benefit that the
accuracy can be systematically improved by densifying the coarse grid for $\gkkp$.
The following convention is used throughout this work: the $\qq$ mesh used for the DFPT computations
is called the initial mesh, while the meshes onto which the Fourier interpolation and the NSCF computations are performed are called the dense meshes.
When referring to results obtained with the DG technique, the mesh used for the KS wave functions and the e-ph matrix elements is called the coarse mesh while the mesh used for the $\emkq$ energies and the 
$\wqnu$ frequencies is denoted as the fine mesh.

\subsection{Parallelization}
\label{sec:parallel}

Our implementation uses five different MPI levels to distribute both the workload and the most memory-demanding data structures.
By default, the code distributes the $\qq$ points in the dense IBZ to reduce the memory allocated for the scattering potentials and the computational cost for the integrals in $\qq$ space. 
This parallelization level is quite efficient but tends to saturate when the number of MPI processes becomes comparable to 
the typical number of $\qq$ points used to integrate the self-energy.
In this regime, indeed, the wall-time required to compute the matrix elements becomes negligible and nonscalable parts such as the computation of the tetrahedron weights and symmetry tables begin to dominate.
In this case, one can activate the parallelization over $\kk$ points and spins to achieve better parallel efficiency. Calculations of the SE at different $\kk$ points and spins are completely independent hence these two MPI levels are embarrassingly parallel with excellent scalability, albeit they do not lead to any additional decrease in the memory requirements. 
If memory is of concern, one can employ the MPI distribution over perturbations to parallelize the computation of the e-ph matrix elements over the $(\kappa,\alpha)$ index and distribute the corresponding scattering potentials.
Finally, an additional distribution scheme over the band index $m$ is available when the real part of the SE is computed by summing over empty states.

\section{Results}
\label{sec:results}

\subsection{Computational details}
\begin{figure}
    \centering
    \includegraphics[clip,trim=0.2cm 0.3cm 0.2cm 0.2cm,width=.48\textwidth]{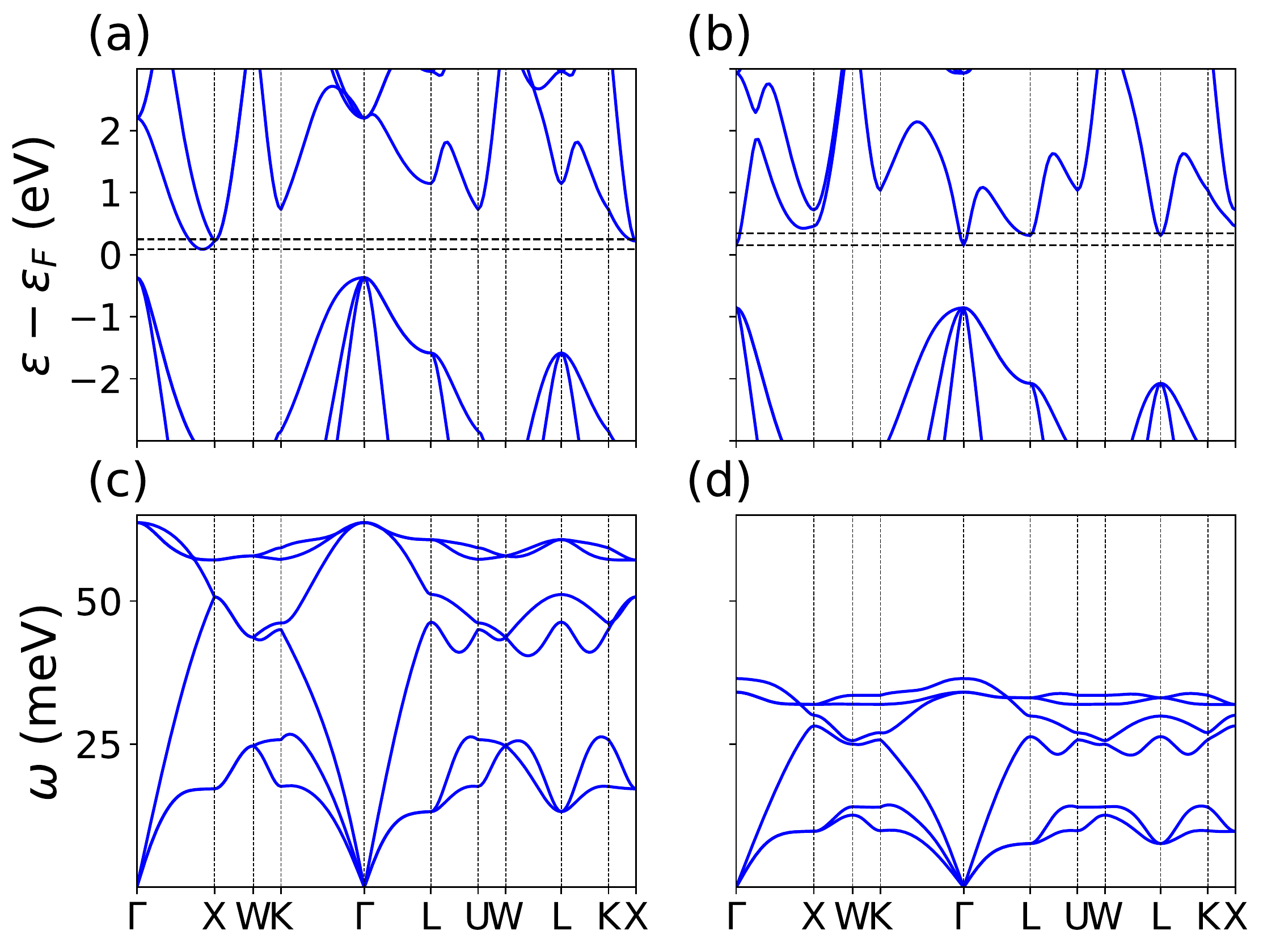}
    \caption{Band structure of (a) Si and (b) GaAs. The Fermi level is located so that $n_e =$ $10^{18}$ and $10^{15}$ cm$^{-3}$ for Si and GaAs, respectively. The energy window used to filter the $\kb$ points for transport computations is represented by horizontal dashed lines.
    Phonon dispersions of (c) Si and (d) GaAs.
    }
    \label{fig:SiGaAs_bs}
    \end{figure}

For all computations, we used norm-conserving pseudopotentials of the Troullier-Martins type~\cite{Troullier1991} 
in the local-density approximation (LDA) from Perdew and Wang~\cite{Perdew1992} parametrized by Ceperley and 
Alder~\cite{Ceperley1980}, with a plane-wave kinetic energy cutoff of 20~Ha for Si and 30~Ha for GaP and GaAs. 
The use of LDA pseudopotentials without nonlinear core corrections is imposed by the current implementation of the dynamical quadrupoles. 
The theoretical values of $Q^{\beta\gamma}_{\kappa\alpha}$ used in this work
are given in the Supplemental Material~\cite{supplemental_prb}.
The relaxed lattice parameter is 5.38~\AA~for Si, 5.32~\AA~for GaP and 5.53~\AA~for GaAs.
To obtain an accurate Fourier interpolation of the e-ph scattering potentials, we use \grid{9} DFPT $\qb$ grids for Si and GaP and \grid{6} for GaAs. 
The choice of the initial DFPT grid is based on the study reported in Ref.~\cite{Brunin2019prl}.
In the next sections, we discuss our results for Si and GaAs,
the detailed results for GaP can be found 
in the Supplemental Material~\cite{supplemental_prb}.
The \abinit band structures of Si and GaAs, are represented in Figs.~\ref{fig:SiGaAs_bs}(a) and (b), respectively, while the phonon dispersions are shown in panels (c) and (d).
The AbiPy python library~\cite{abipy-website}
has been used to automate part of the calculations as well as the post-processing of the results.

\subsection{Carrier linewidths}

\begin{figure}
    \centering
    \includegraphics[width=.45\textwidth]{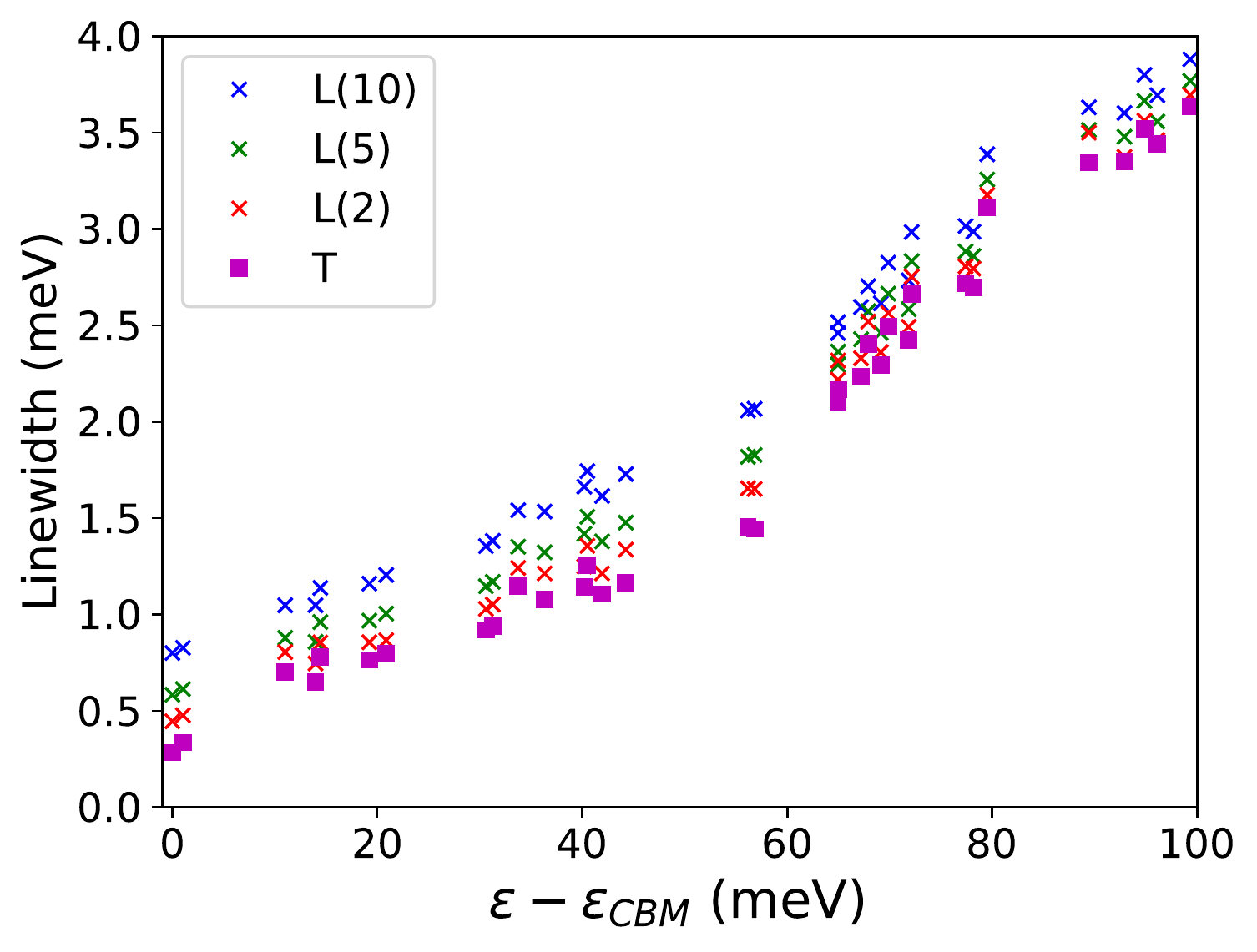}
    \caption{Electron linewidths near the CBM of Si obtained with a \grid{60} $\kb$-point grid at $T=300$\,K. 
    A very dense \grid{180} $\qb$-point grid is used for all curves to achieve well-converged results. 
    Results are obtained with Lorentzian (L)
    and tetrahedron (T) methods. 
    The broadening parameter is given in parentheses in meV.}
    \label{fig:intmethcomp}
    \end{figure}

In Fig.~\ref{fig:intmethcomp}, we report the linewidths 
for states within 90~meV from the conduction band minimum (CBM) of Si 
obtained with different integration schemes.
Close to the CBM, for energies  ${\varepsilon - \varepsilon_{\text{CBM}} < \omega_{\text{LO}}}$ 
(the frequency of the longitudinal optical phonon), 
the linewidths 
are mall as few scattering channels are available.
For higher energies, optical-phonon emission becomes possible and leads to an increase in the linewidths
around ${\varepsilon = \varepsilon_{\text{CBM}} + \omega_{\text{LO}}}$, which corresponds to 64~meV above the CBM of Si.
As expected, the values obtained with the Lorentzian
function approach the
tetrahedron results when the broadening is reduced provided the $\qq$ mesh is dense enough.
A similar study with Gaussian broadening is performed in the Supplemental Material where similar conclusions are reached~\cite{supplemental_prb}.

\begin{figure}
	\centering
	\includegraphics[clip,trim=0.2cm 0.2cm 0.2cm 0.2cm,width=.4\textwidth]{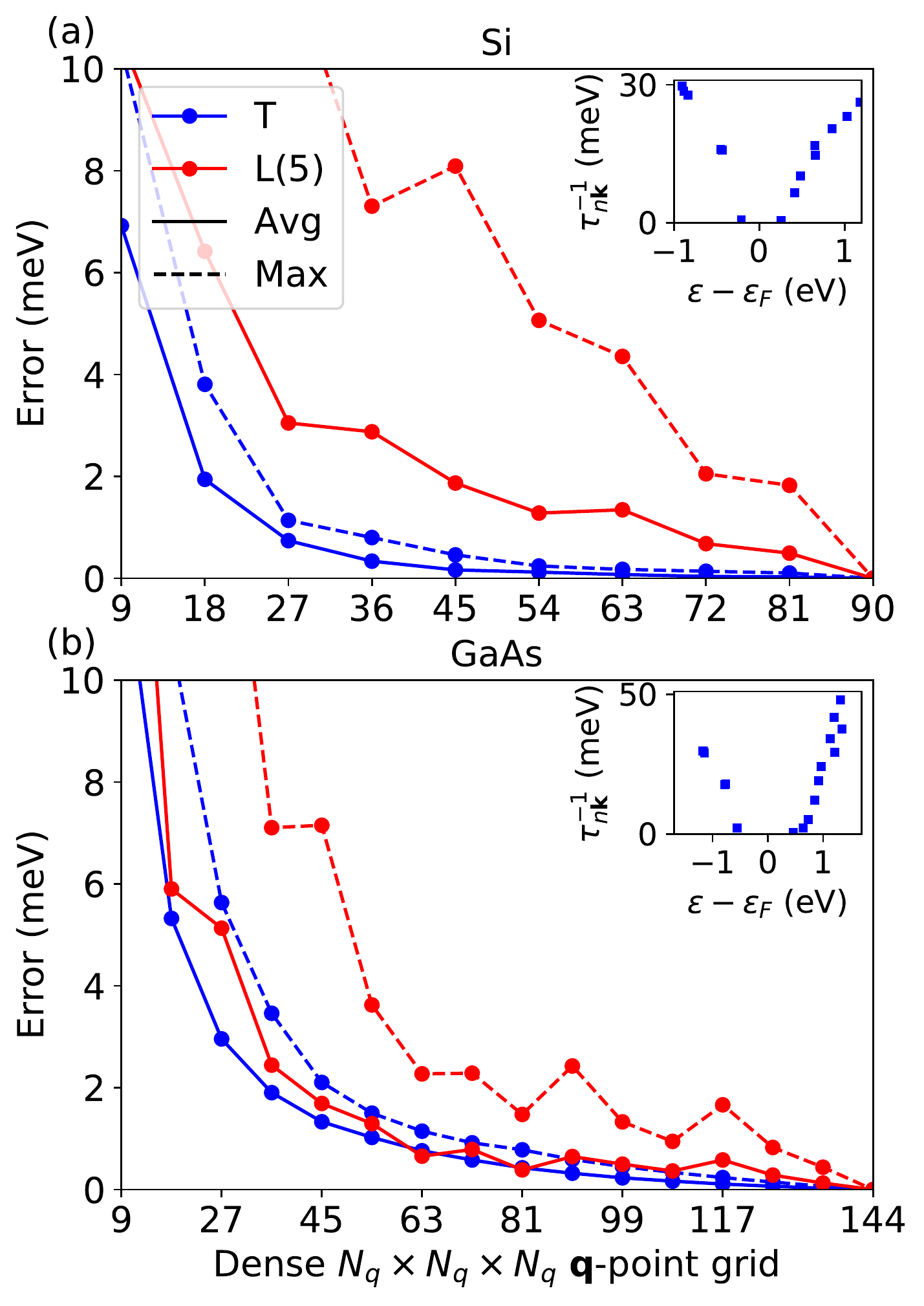}
	\caption{Average (solid line) and maximum (dashed line) error on the linewidths evaluated on a \grid{9} $\kb$-point grid in (a) Si and (b) GaAs as a function of the $\qb$ grid used for the integration. 
	Only states within 1~eV from the CBM and VBM are considered. 
	Results have been obtained with the tetrahedron method (T, blue) and a Lorentzian (L, red) broadening of 5~meV.
	The reference linewidths for each curve are the results obtained with the 
    densest $\qb$ grid and the corresponding integration technique.
	The insets show the linewidths obtained with the tetrahedron method and the densest $\qb$ grid.}
	\label{fig:errors_SiGaAs}
    \end{figure}

In Fig.~\ref{fig:errors_SiGaAs}, we compare the convergence rate of the tetrahedron integration scheme with the one obtained with a Lorentzian broadening of 5~meV.
The plot shows the average and maximum errors on the linewidths in Si and GaAs as a function of the $\qb$ mesh used to evaluate Eq.~\eqref{eq:imagfanks_selfen}.
The $\kk$ mesh for which the linewidths are obtained is fixed, and the dense $\qq$ mesh used for integrating
Eq.~\eqref{eq:fan_selfen} is progressively increased until convergence is reached.
In order to do this, we perform a NSCF computation for each dense mesh in order to 
have access to $\emkq$ and $\gkkp$ on the dense mesh, and select the $n\kk$ states belonging
to the \grid{9} mesh to compute the linewidths.
Two conclusions can be drawn from these results.
(i) In both materials, the tetrahedron method outperforms the Lorentzian broadening
since convergence is reached with less dense $\qb$-point grids.
(ii) The convergence in GaAs is slower than in Si because of the integrable Fr\"ohlich singularity for $\qb \rightarrow \zero$ present in polar semiconductors.
This is somehow expected as the numerical integration of the singularity requires dense $\qb$ meshes to sample enough points in the region around $\Gamma$.

\begin{figure}
    \centering
    \includegraphics[clip,trim=0.2cm 0.2cm 0.2cm 0.2cm,width=.4\textwidth]{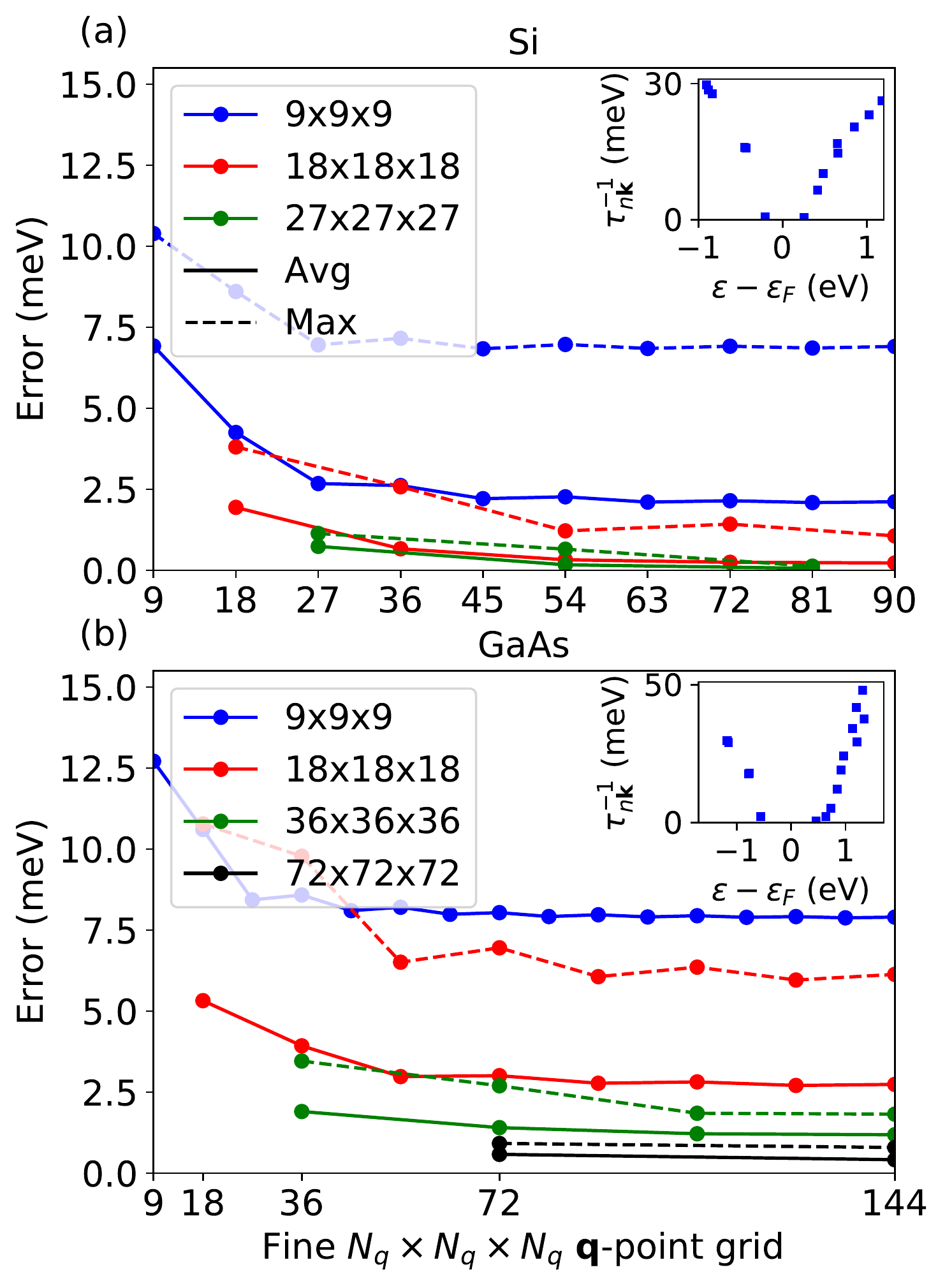}
    \caption{Average (solid line) and maximum (dashed line) error on the linewidths on a \grid{9} $\kb$-point grid in (a) Si and 
    (b) GaAs as a function of the fine $\qb$-point grid used for the integration. 
    The matrix elements are obtained on (a) coarse \grid{9} (blue), \grid{18} (red) 
    and \grid{27} (green) grids for Si and (b) coarse \grid{9} (blue), \grid{18} (red) 
    \grid{36} (green) and \grid{72} (black) grids for GaAs. 
    KS energies are computed on the fine grids.
    Only states within 1~eV from the CBM and VBM are considered. 
    The reference linewidths (insets) are the results obtained with a \grid{90} (\grid{144}) $\qb$ point grid for both matrix elements and 
    energies for Si (GaAs).}
    \label{fig:errors_dg_SiGaAs}
    \end{figure}

At this point, it is worth comparing the convergence rate of the standard tetrahedron scheme with that obtained with the DG technique presented in Sec.~\ref{sect:double_grid}.
Figure~\ref{fig:errors_dg_SiGaAs} gives the error on the linewidths in Si and GaAs,
similarly to Fig.~\ref{fig:errors_SiGaAs}, but now obtained with the
tetrahedron method together with the DG technique~\footnote{In all the calculations using the DG method, the KS eigenvalues on the fine $\kk$ mesh have been computed exactly by performing a NSCF computation.}.
We observe that, in Si, using a \grid{18} coarse $\qb$ grid for the e-ph matrix elements and a fine \grid{36} $\qb$ grid 
for the KS energies and phonon frequencies reduces the average error below 1~meV with a computational cost very similar to that required by \grid{18} $\kb$- and $\qb$-point meshes. 
Increasing the density of the coarse grid further reduces the error down to practically zero.
In the case of GaAs, a denser coarse $\qb$-point grid is required to reduce the error
because of the singularity around $\Gamma$.
The density of $\kk$ points used in these tests is not large enough to converge transport properties (as shown in the next Section) but it is clear that a sufficiently fine $\qq$ mesh is important to achieve accurate linewidths whose quality will affect the final results for the mobility.

\subsection{Phonon-limited mobility}
\label{sec:ph_limited_mobility}

For the computation of lifetimes and mobility, we consider an electron concentration of $10^{18}$ cm$^{-3}$ both in Si and GaP, following the approach used 
for effective mass computations 
in previous works~\cite{Hautier2013,Bhatia2016,Ricci2017}. This choice, indeed, helps avoiding numerical noise and allows one to reach faster convergence when compared to calculations done with lower concentrations. Note, however, that the computed mobility does not correspond to the experimental one measured for $10^{18}$ electrons per cm$^{3}$, because such concentrations require impurities in the crystal that would increase the scattering rates and decrease the mobility. The results are representative of the intrinsic mobility 
as long as the Fermi level remains inside the band gap and far enough from the band edges (see also the tests reported in Ref.~\cite{supplemental_prb}).
In GaAs, we used an electron concentration of $10^{15}$ cm$^{-3}$ because the density of states in the conduction band is very small.

\begin{figure}
    \centering
	\includegraphics[width=.49\textwidth]{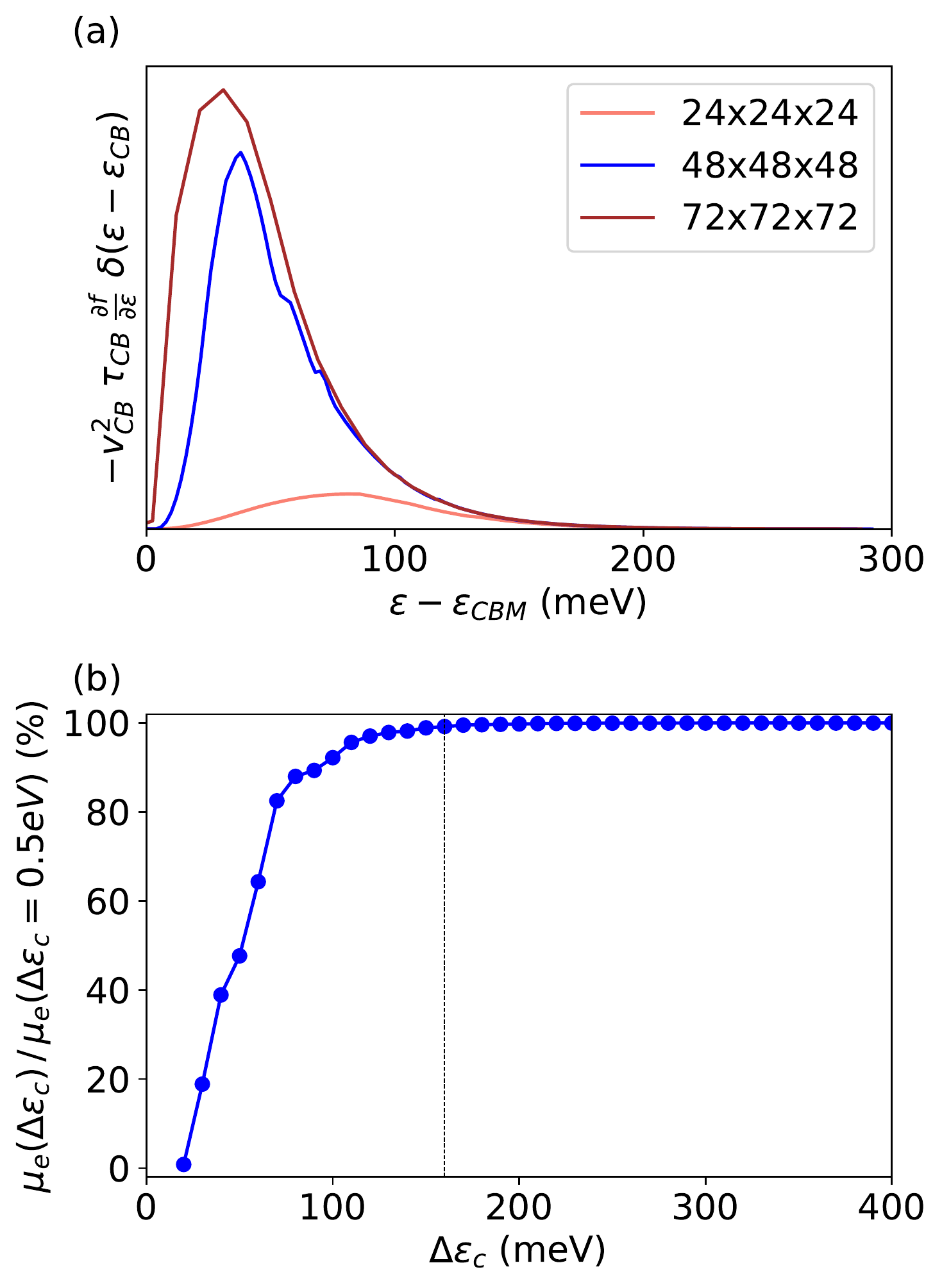}
    \caption{(a) $xx$ component of the integrand in Eq.~\eqref{eq:transport_lc} convoluted with $\delta(\varepsilon-\enk)$ 
    in the conduction band of Si, for \grid{24} (black), \grid{48} (red) and \grid{72} (green) $\kb$-point and \grid{144} $\qb$-point grids at $T = 300$K.
    (b) Electron mobility of Si as a function of the energy window used for the computation 
    of the lifetimes in Eq.~\eqref{eq:transport_lc}. 
    \grid{72} $\kb$-point and \grid{144} $\qb$-point grids have been used.
    For these grids, the converged value is 1509~cm$^2$V$^{-1}$s$^{-1}$.
    A 160~meV energy range is enough to reach a relative error in the mobility lower than 1\%.}
    \label{fig:Si_kernelerange}
	\end{figure}

In our implementation, as mentioned in Sec.~\ref{sect:BZ_filtering},
lifetimes and group velocities are computed only for the KS states
that contribute to Eq.~\eqref{eq:transport_lc} by introducing
an energy window around the Fermi level. 
Figure~\ref{fig:Si_kernelerange}(a) shows the convergence of the integrand of Eq.~\eqref{eq:transport_lc} 
(convoluted by $\delta(\varepsilon-\enk)$ as discussed in Appendix~\ref{app:boltzmann_transport}) 
for increasing $\kb$-point densities,
in the case of electrons in the conduction band of Si. 
Integrating this function directly gives the electron mobility.
Figure~\ref{fig:Si_kernelerange}(a) shows that 
the integrand quickly vanishes for energies far from the Fermi level (and, therefore, far from the CBM).
As a consequence, it is possible 
to find an optimal energy window for the electrons
that leads to a considerable computational saving without affecting the quality of the calculation.
This is demonstrated in Fig.~\ref{fig:Si_kernelerange}(b) where we report the mobility obtained for different values of the window. 
In Si, an energy range of 0.16~eV is sufficient to reach less than 1\% relative error on the mobility while
a slightly larger value of 0.19 (0.18)~eV is needed for GaAs (GaP).
We therefore use these values for the energy window and, hereafter, we focus on the convergence of the electron mobility with respect to the $\kk$  and $\qq$ meshes, including a detailed analysis of the effect of the DG integration scheme on the convergence rate.

\begin{figure}
    \centering
    \includegraphics[width=0.49\textwidth]{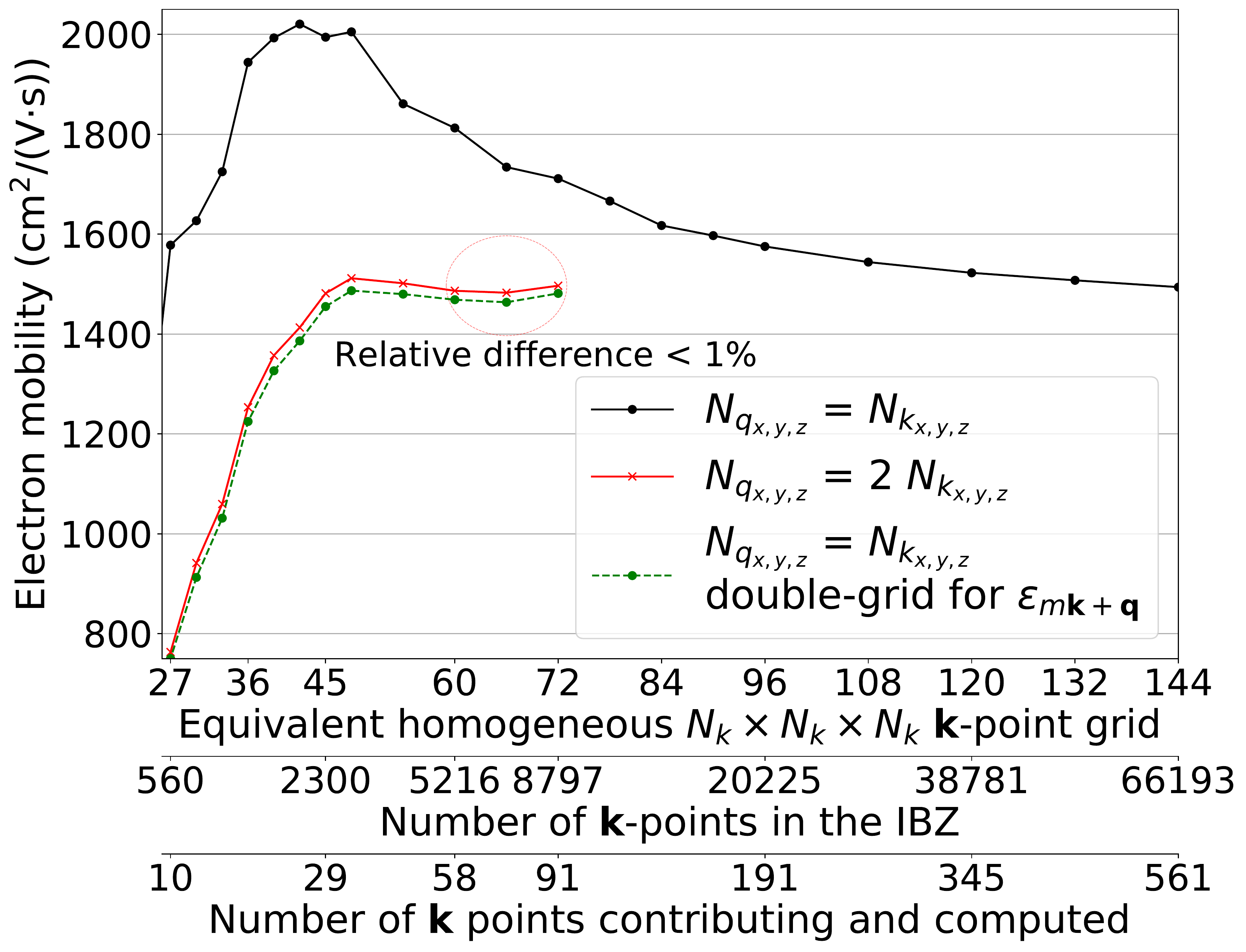}
    \caption{Electron mobility in Si ($T=300$\,K) as a function of the $\kb$- and $\qb$-point grids.
    The density of $\qb$ points is the same as the density of $\kb$ points (black), or twice as dense (red).
    The results obtained with DG technique are also reported (green, dashed). 
    In this case, the densities of the $\kb$- and $\qb$-point grids 
    are the same for matrix elements and lifetimes, but a grid twice as dense is used for the energies 
    in Eq.~\eqref{eq:imagfanks_selfen}.
    The equivalent homogeneous $\kb$ mesh is first reported, together with the corresponding
    number of points in the IBZ.
    The effective number of $\kb$ points 
    included in the computation of $\tau_\nk$ is also given.
    }
    \label{fig:Si_mobility}
	\end{figure}

Figure~\ref{fig:Si_mobility} shows the dependence of the electron mobility in Si on the equivalent $\kb$ grid used for the integration of Eq.~\eqref{eq:transport_lc}, as well as on the dense $\qb$ mesh used for the evaluation of the lifetimes (Eq.~\eqref{eq:fanlifetime}). 
The mesh of $\qb$ points is the same as the $\kb$ grid (black dots), or twice as dense in each direction (red crosses). 
For a given $\kb$ grid, increasing the $\qb$-point density 
(compare vertically-aligned data points in
Fig.~\ref{fig:Si_mobility}) systematically decreases the mobility as more scattering channels are included. 
We need to include a sufficient number of $\kb$ wave vectors to evaluate Eq.~\eqref{eq:transport_lc} and enough $\qb$ points to evaluate Eq.~\eqref{eq:fanlifetime}.
All the convergence curves shown in Fig.~\ref{fig:Si_mobility} follow the same trend:
at low $\kb$-point densities, the (unconverged) mobility increases with the $\kb$-point density.
This is because the $\kk$-point sampling of the conduction band becomes denser close to the CBM where the integrand of Eq.~\eqref{eq:transport_lc} reaches its maximum 
(see Fig.~\ref{fig:Si_kernelerange}(a)).
Once the number of $\kb$ points included in the computation of the mobility is large enough, the mobility starts to decrease because the density of $\qb$ points increases and more scattering channels are 
properly captured.

The total number of $\kb$ points in the IBZ is reported on the second $x$ axis of Fig.~\ref{fig:Si_mobility}, together with the effective number of $\kk$ points included in the integration of Eq.~\eqref{eq:transport_lc} (third $x$ axis). Note that only around 1\% of the IBZ needs to be computed. 
The red curve (obtained with a $\qb$ mesh twice as dense in each direction as the $\kb$ mesh) shows that convergence within 1\% is achieved
with a \grid{60} $\kb$ mesh together with a \grid{120} $\qb$ grid. 
Using a \grid{45} grid for electrons 
and a \grid{90} grid for phonons
already leads to an error lower than 5\%.
We conclude that, in Si, a $\qb$ mesh twice as dense in each direction as the $\kb$ mesh is required to accelerate convergence.
If the mobility is computed with the same mesh for electrons and phonons, indeed, 
the convergence is much slower
as we include an unnecessarily large number of $\kb$ points whose lifetimes are still far from convergence.

The mobility obtained with the DG technique for the lifetimes is also reported (dashed green line). 
In this case, the density of $\kb$  and $\qb$ points is the same for the lifetimes and the e-ph matrix elements, 
but now the $\qb$-point density is doubled in each direction for the energies in Eq.~\eqref{eq:imagfanks_selfen}, 
thus allowing a better description of phonon absorption and emission processes.
The results are almost identical to those obtained by explicitly computing the e-ph matrix elements on the same grid as the energies (red).
To appreciate the efficiency of the DG integration scheme, it is worthwhile to compare the wall-time required by two calculations done with the same sampling. 
The computation time of a standard e-ph calculation in Si with a \grid{45} $\kk$ grid and a \grid{90} $\qq$ mesh using a single CPU is around 2 hours 
and decreases down to 40 minutes when the DG technique is used.
Obviously, the wall-time can be easily decreased by running on multiple CPUs using the MPI implementation discussed in Sec.~\ref{sec:parallel}.

Our converged value for the electron mobility in Si is 1509~cm$^2$/(V$\cdot$s) that
compares well with experimental data comprised between 1300 and 1450 cm$^2$/(V$\cdot$s)~\cite{Ponce2018}.
Our results are consistent with other theoretical SERTA calculations reported in the literature that range from 1555 and 1872 cm$^2$/(V$\cdot$s)~\cite{Ponce2018,Li2015,Ma2018}.
%
%
For the sake of completeness, 
one should also stress that mobilities are quite sensitive to the computational parameters.
Ponc\'e \etal showed that, in Si, the choice of the XC functional can lead to a difference of about 10\% in the mobility~\cite{Ponce2018}. The lattice parameter also has a large effect as well as the formalism used to compute the electronic dispersion (for example, LDA vs $GW$). 

\begin{figure}
    \centering
    \includegraphics[width=0.49\textwidth]{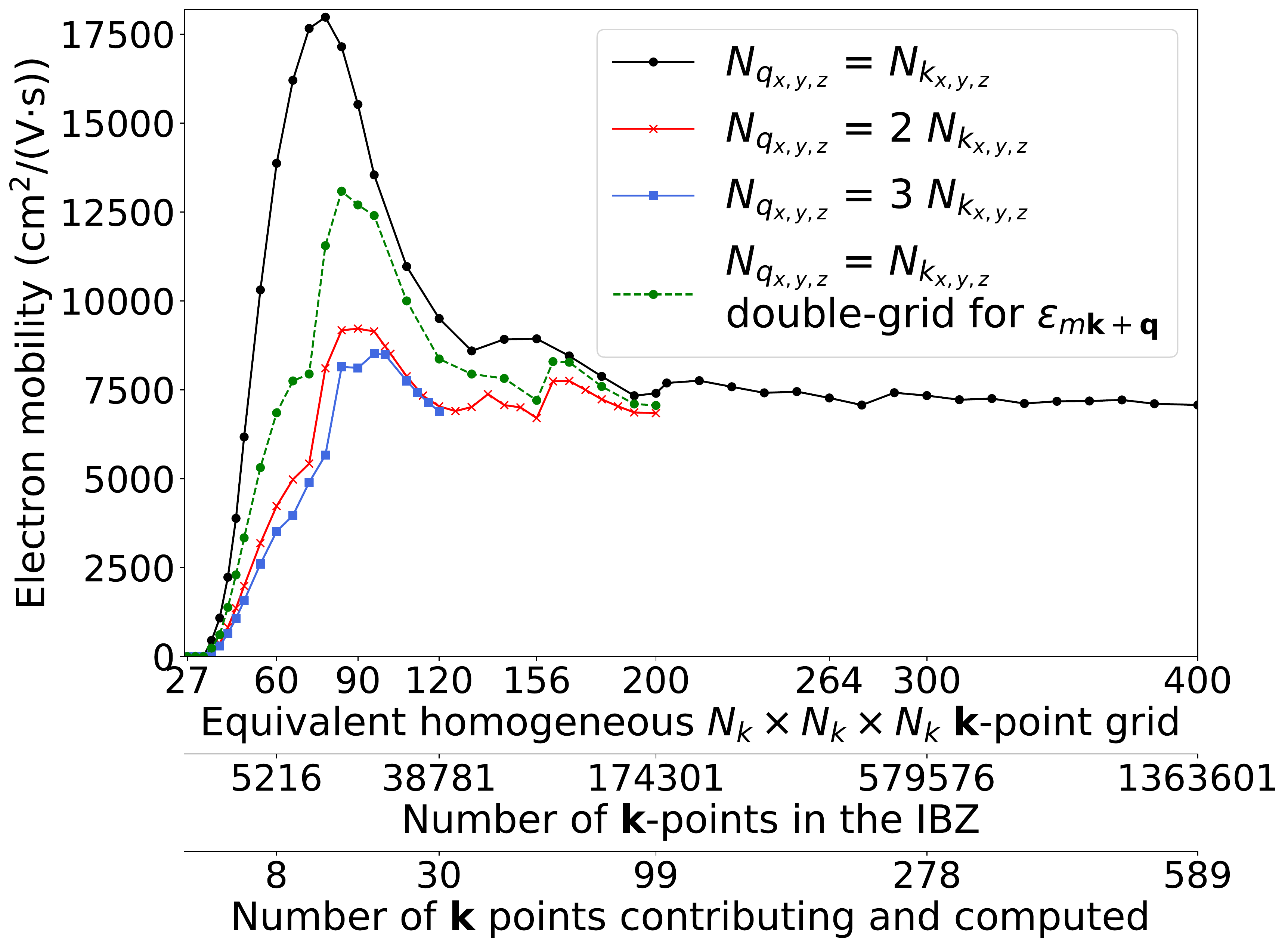}
    \caption{Electron mobility in GaAs ($T=300$\,K) as a function of the $\kb$- and $\qb$-point grids.
    The density of $\qb$ points is the same as the density of $\kb$ points (black), twice (red), 
    or three times (blue) as dense in each direction.
    The DG technique results are also reported (green, dashed). In this case, the densities of 
    the $\kb$- and $\qb$-point grids are the same for e-ph matrix elements and lifetimes, 
    but a grid twice as dense in all directions is used for the energies in Eq.~\eqref{eq:imagfanks_selfen}.
    The equivalent homogeneous $\kb$ mesh is reported, together with the corresponding number of points in the IBZ.
    The effective number of $\kb$ points 
    included in the computation of $\tau_\nk$ is also given.
    }
    \label{fig:GaAs_mobility}
	\end{figure}

We now discuss our results
for the electron mobility in GaAs that are summarized in Fig.~\ref{fig:GaAs_mobility}. 
First of all, we note that, despite the very dense homogeneous $\kk$ mesh employed,
a relatively small number of effective $\kk$ points contribute 
and
therefore need to be computed.
As explained in Sec.~\ref{sect:BZ_filtering}, this is due to the small effective mass (high dispersion) of the CB in GaAs.
The blue curve in Fig.~\ref{fig:GaAs_mobility} has been obtained with a $\qq$ mesh that is three times denser than the $\kk$ mesh in each direction. For the \grid{108} $\kk$ grid, the red and blue curves give very similar results thus confirming that a \grid{216} $\qq$ mesh is dense enough for an accurate integration of the linewidths. Note, however, that a \grid{216} $\kk$ mesh for electrons is not sufficient to converge the mobility.
Convergence within 5\% is indeed reached with a \grid{264} $\kk$ mesh (and the same $\qq$ mesh). 
The mobility obtained using the DG technique is also reported in Fig.~\ref{fig:GaAs_mobility} (green dashed line).
The results with the double grid are systematically improved (closer to red crosses, compared to black dots). 
Unfortunately, the convergence of the double grid is not as smooth and fast as the one observed in Si, likely due to the divergence of the e-ph matrix elements around $\Gamma$.
Our final value for the electron mobility in GaAs
obtained with
a \grid{400} grid both for electrons and phonons
is
7075~cm$^2$/(V$\cdot$s).
This result is in the range of reported electron mobilities for GaAs, roughly between 7000 and 12000~cm$^2$/(V$\cdot$s)~\cite{Ma2018,Liu2017,Agapito2018}.
The large range of computed mobilities has been analyzed in Ref.~\cite{Ma2018}. The lattice parameter has a very important role, as well as the formalism used both for the electronic band structure (GGA, $GW$, etc.) and the transport computation (iterative BTE, momentum-relaxation time approximation, etc.). In addition, we use the tetrahedron integration method and treat dynamical quadrupoles in the interpolation of the scattering potentials.

\begin{figure}
    \centering
    \includegraphics[width=0.49\textwidth]{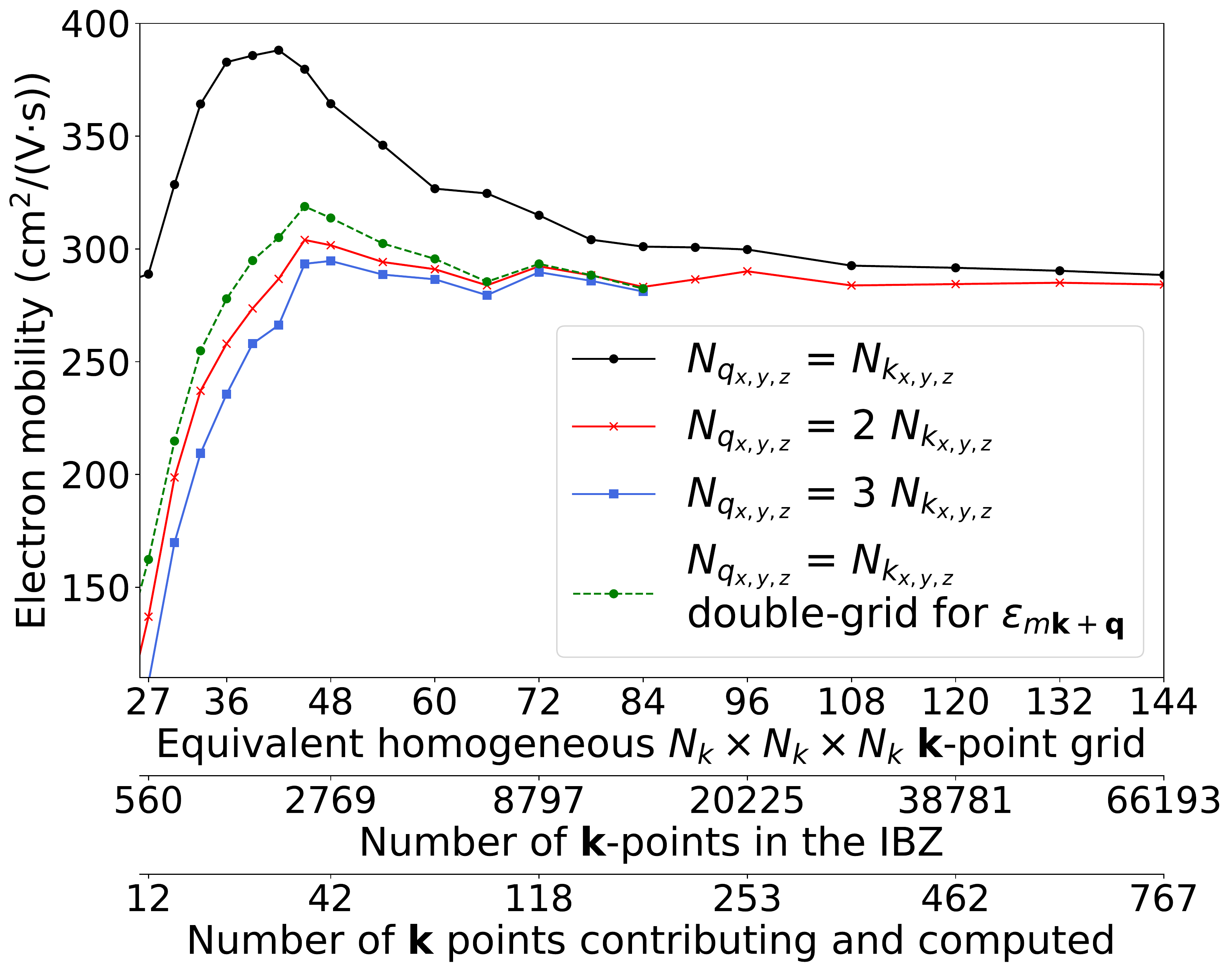}
    \caption{Electron mobility in GaP ($T=300$\,K) as a function of the $\kb$- and $\qb$-point grids.
    The density of $\qb$ points is the same as the density of $\kb$ points (black), twice (red), 
    or three times (blue) as dense in each direction.
    The DG technique results are also reported (green, dashed). In this case, the densities of 
    the $\kb$- and $\qb$-point grids are the same for matrix elements and lifetimes, 
    but a grid twice as dense in all directions is used for the energies in Eq.~\eqref{eq:imagfanks_selfen}.
    The equivalent homogeneous $\kb$ mesh is reported, together with the corresponding number of points in the IBZ.
    The effective number of $\kb$ points 
    included in the computation of $\tau_\nk$ is also given.
    }
    \label{fig:GaP_mobility}
	\end{figure}

Finally, in Fig.~\ref{fig:GaP_mobility}, we report
the electron mobility in GaP.
This system is less problematic than GaAs and convergence within 5\% is reached with \grid{54} $\kb$-point and \grid{108} $\qb$-point meshes. 
Oscillations are still observed for $\kb$ meshes denser than \grid{78}, but not larger than 2\%. For these grids, the $\qb$-point mesh is well converged, as shown by the blue curve ($\qb$ mesh three times as dense in each direction as the $\kb$ mesh). 
The mobility obtained with the DG technique 
is also reported in the same figure (green dashed line).
Once the $\qb$ mesh is dense enough to correctly describe this divergence (\grid{66} mesh), the DG results are very similar to a full computation taking into account the variation of the e-ph matrix elements on a $\qb$ mesh twice as dense as the $\kb$ mesh (red crosses).
There are few experimental data for the electron mobility in GaP 
with values 
between 200 and 330~cm$^2$/(V$\cdot$s)~\cite{Darnon2015} but experiments are usually performed on polycrystalline samples.
A more accurate description of transport properties in GaP 
would therefore require the inclusion of grain-boundary scattering effects
that are beyond the scope of the present work.
It is clear that additional theoretical work and more advanced \abinitio techniques including
self-consistency in the Boltzmann transport equation,
many-body effects in the electron-electron interaction as well as additional scattering processes are needed in order 
to improve the agreement between theory and experiment. 
Nonetheless,
our convergence studies for the phonon-limited electron mobility in Si and GaP indicate that the $\qq$ mesh should be at least as dense as the $\kk$ mesh in order to capture enough scattering channels with small momentum transfer. 
In GaAs, a very fine $\kk$ mesh is required due to the high dispersion of the conduction band. 
In this case, it is not as interesting to use a denser $\qq$ mesh.

\section{Conclusion}

We present an efficient method 
based on plane waves and Bloch states
for the computation of the e-ph self-energy and carrier mobilities in the self-energy relaxation time approximation.
Our approach takes advantage of symmetries and advanced integration techniques such as the linear tetrahedron and double-grid methods to achieve accurate results with a computational cost that is competitive with state-of-the-art implementations based on localized orbitals.
The number of explicit DFPT calculations is significantly reduced by interpolating the scattering potentials using a real-space representation in conjunction with Fourier transforms while Bloch states 
are treated exactly.
A systematic analysis of the convergence behavior of electron linewidths and mobilities in Si, GaAs, and GaP as a function of the initial \abinitio $\qq$ mesh reveals that the proper treatment of the long-range fields due to dynamical quadrupoles is crucial for an accurate and efficient interpolation in the region around $\Gamma$~\cite{Brunin2019prl}.
We discussed how to treat quadrupolar terms in e-ph calculations in a fully \abinitio way using the recently proposed first-principles theory of spatial dispersion~\cite{Royo2019}.
Our approach is implemented as direct post-processing of standard DFPT calculations and the e-ph results can be easily converged by monitoring a small set of parameters defining the BZ sampling and an energy window around the Fermi level.
Smart sampling techniques and highly parallel algorithms
make it possible to obtain converged carrier mobilities for relatively small crystalline systems in a few hours with commodity-class clusters.
Our work will hopefully pave the way towards systematic and high-throughput
studies of materials properties related to e-ph interaction both in polar and nonpolar semiconductors.

\section{Acknowledgment}

G.~B., G.-M.~R. and G.~H.~acknowledge financial support from F.R.S.-FNRS.
H.~P.~C.~M. acknowledges financial support from F.R.S.-FNRS through the PDR Grants HTBaSE (T.1071.15).
M.~G., M.~J.~V and X.~G.~acknowledge financial support from F.R.S.-FNRS through the PDR Grants AIXPHO (T.0238.13) and ALPS (T.0103.19). M.~J.~V thanks FNRS and ULiege for a sabbatical grant in ICN2.
M.~S. and M.~R. acknowledge the support of Ministerio de Economia,
Industria y Competitividad (MINECO-Spain) through
Grants  No.~MAT2016-77100-C2-2-P  and  No.~SEV-2015-0496;
of Generalitat de Catalunya (Grant No.~2017 SGR1506); and
of the European Research Council (ERC) under the European Union's
Horizon 2020 research and innovation program (Grant
Agreement No.~724529).
The present research benefited from computational resources made available on the {Tier-1} 
supercomputer of the F\'ed\'eration Wallonie-Bruxelles, infrastructure funded by the Walloon 
Region under grant agreement n\textsuperscript{o}1117545.

\vspace{2mm}
\textit{Note added} -- Recently, we became aware of a related work
by another group that reaches similar conclusions about the
importance of the dynamical quadrupole term to obtain an
accurate physical description of e-ph interactions~\cite{Jhalani2020,Park2020}.
\vspace{2mm}

\appendix
\section{Long-wavelength limit in semiconductors}
\label{app:lr_limit}

\subsection{DFPT potentials for $\qq\rightarrow \zero$}

In the e-ph scattering potentials, there are two different nonanalytic terms  in $\qq$ space.
The derivative of the local part of the pseudopotential, $\partial_{\kappa\alpha,\qb} \vloc$, behaves like~\cite{Gonze1997a,Stengel2013,Royo2019}:
\begin{equation}
\partial_{\kappa\alpha,\qb}\vloc(\Gb=\zero)
=i\frac{4 \pi}{\Omega}
 \frac{q_{\alpha}}{q^2}Z_{\kappa}
 + {\cal{O}}(q_{\alpha}),
\end{equation}
with $Z_\kappa$ the charge of the (pseudo) ion $\kappa$.
The second term is the derivative of the Hartree potential
$\partial_{\kappa\alpha,\qb} \vH 
$
that is linked to 
the density variation by a simple formula in reciprocal space:
\begin{equation}
\partial_{\kappa\alpha,\qb}\vH(\Gb)
=\frac{4 \pi}{|\qG|^2}
 \partial_{\kappa\alpha,\qb}n(\Gb).
\label{eq:Hartree_change}
\end{equation}
The $\Gb=\zero$ component of the density change is therefore amplified by a diverging $\frac{4 \pi}{q^2}$ factor. 
This is handled in DFPT by the separate treatment of the screened electric field as discussed in Ref.~\cite{Gonze1997a,Stengel2013,Ponce2015}.
This term is subtracted in auxiliary DFPT computations, as explained in Sec.~\ref{sec:lr_limit}.
In principle there is also a $1/q^2$ contribution coming from the (exact) XC kernel but approximated functionals such as LDA or GGA fail to reproduce such behavior~\cite{Gonze1997a,Gonze1995}.
The nonlocal part of the pseudopotential is short-ranged in real space and 
does not pose additional challenges. 

\subsection{Derivations of Eqs.~\eqref{eq:wfNA_q_rightarrow_0} and \eqref{eq:wfNA_q}}

Equation~\eqref{eq:wfNA_q_rightarrow_0} can be obtained by the combination of different equations coming from Appendix A of Ref.~\cite{Ponce2015}.
Namely, one should use Eqs.~(A5), (A6), (A26), (A51), (A54), (A73) and (A66), rewritten here with the notation of this paper.
The first-order derivatives of the wavefunctions with respect to collective oscillations for $\qq\rightarrow\zero$ are obtained as
\begin{equation}
    |u_{n\kk,\qq\rightarrow\zero}^{\tau_{\kappa\alpha}}\rangle = -{\bf A}^{-1} {\bf w} + \frac{-b + a({\bf u}^\dagger {\bf A}^{-1} {\bf w}) }{1 + a({\bf u}^\dagger {\bf A}^{-1} {\bf u}) }{\bf A}^{-1} {\bf u},
    \label{eq:x1}
\end{equation}
where the definitions of ${\bf A}$, ${\bf w}$, ${\bf u}$, $a$ and $b$ can be found in Ref.~\cite{Ponce2015}.
The denominator in the previous equation is  
the $\qq$-dependent macroscopic dielectric function:
\begin{align}
    \epsilon_M(\qq) & = 1 + a({\bf u}^\dagger {\bf A}^{-1} {\bf u}). \label{eq:epsilon_macro}
\end{align}
The numerator is given by extending Eq.~(A73) of Ref.~\cite{Ponce2015} to the next order in $q$:
\begin{equation}
    - b + a({\bf u}^\dagger {\bf A}^{-1} {\bf w}) = -\frac{4\pi i q_\gamma}{\Omega q^2} Z^*_{\kappa\alpha,\gamma}  e^{-iq_\eta \tau_{\kappa\eta}}.
    \label{eq:x1_numerator}
\end{equation}
The first term in Eq.~\eqref{eq:x1} is the first-order wavefunction due to collective atomic displacements at $\qq=\zero$, where the $\Gb=\zero$ term of the Hartree potential change has been removed:
\begin{equation}
    -{\bf A}^{-1} {\bf w} = |u^{\tau_{\kappa\alpha}^\prime}_{n\kb,\qb = \zero}\rangle,
    \label{eq:mam1w}
\end{equation}
and ${\bf A}^{-1} {\bf u}$ is the same first-order wavefunction but with an electric field perturbation:
\begin{equation}
    {\bf A}^{-1} {\bf u} = -i {\cal{Q}} q_{\lambda} |u^{\cal{E}_{\lambda}^\prime}_{n\kb}\rangle.
    \label{eq:am1u}
\end{equation}
Inserting Eqs.~\eqref{eq:epsilon_macro}--\eqref{eq:am1u} into Eq.~\eqref{eq:x1} gives Eq.~\eqref{eq:wfNA_q_rightarrow_0} of the main text.
Note that the electric field perturbation in Eqs.~\eqref{eq:wfNA_q_rightarrow_0} and \eqref{eq:am1u} has been defined following the convention of Ref.~\cite{Royo2019}, that differs from the convention of Ref.~\cite{Gonze1997a}. The difference is the charge of the electron, that has already been included in the latter. In Ref.~\cite{Gonze1997a}, the perturbation might be better referred to as the $\cal{Q}\cal{E}_{\alpha}$ perturbation. 
So, in this sense, \abinit implements the $\cal{Q}\cal{E}_{\alpha}$ perturbation and not the 
$\cal{E}_{\alpha}$ perturbation.

Equation~\eqref{eq:wfNA_q} is obtained using Eq.~\eqref{eq:epsilon_macro} and extending
\begin{equation}
    - b + a({\bf u}^\dagger {\bf A}^{-1} {\bf w})
\end{equation}
in Eq.~\eqref{eq:x1_numerator} to the cell-integrated charge response to a monochromatic atomic displacement from Eq.~\eqref{eq:Qqka_expansion}.

\section{Symmetry properties of the wavefunctions}
\label{app:symmetry_properties_wfs}

In this appendix, we discuss how to use symmetries to reconstruct KS eigenvalues and wavefunctions in the full BZ from the IBZ.
We focus on scalar wavefunctions,
the generalization to two-component spinors is discussed in Ref.~\cite{Bassani1975}.

A generic element of the crystalline space group will be denoted in the following with the $\oSv$ symbol where $\mcS$ is a real orthonormal matrix corresponding to
a proper or improper rotation and $\bvec$ the associated fractional translation in Cartesian coordinates.
The application of the symmetry operator $\oSv$ to the atomic position $\btau$ is defined by
\begin{equation}
\oSv\, \btau \equiv \mcS \btau + \bvec.
\end{equation}
%
%
%
The inverse of $\oSv$ has rotational part $\mcS^{-1}$ and fractional translation $-\mcS^{-1}\bvec$.
The application of the symmetry operation $\oSv$ on a generic function $F(\rr)$ 
of the three spatial coordinates $\rr$ is conventionally defined by:
\begin{equation}
\oSv\, F (\rr) \equiv F(\Si (\rr-\bvec)).  
\end{equation}
Since $\oSv$ commutes with the KS Hamiltonian $\HH$ of the crystal, it readily
follows that, given $\psi_{n\kk}(\rr)$ eigenstate of $\HH$ with eigenvalue $\ee_\nk$
also $\oSv\, \psi_{n\kk}(\rr)$ is eigenstate of the Schr\"odinger problem with the same eigenvalue:
\begin{equation}
\HH\oSv\, \psi_{n\kk}= 
 \oSv\HH\,\psi_{n\kk} = \ee_{n\kk}\,\oSv\psi_{n\kk}.
\end{equation}
Although $\oSv\,\psi_{n\kk}(\rr)$ has the same eigenenergy as $\psi_{n\kk}(\rr)$, its 
crystalline momentum is different.
The operator $\oSv$ 
transforms a Bloch eigenstate $e^{i \kk\cdot\rr} u_\nk(\rr)$
with vector $\kk$
into a new Bloch state of crystalline momentum $\mcS \kk$.
This important property can be seen as follows:
\begin{equation}
\label{eq:Rotation_of_psi}
 \begin{split}
 \Bigl[ \omcS_\bvec\psi_\nk \Bigr] (\rr+\RR) = &
 \quad \psi_\nk \bigl( \Si (\rr+\RR-\bvec) \bigr) \\ = 
  & \quad e^{i \kk\cdot\Si (\rr+\RR-\bvec)}\, u_\nk \bigl( \Si(\rr-\bvec) \bigr) \\ = 
  & \quad e^{i\ltrans\Si \kk\cdot\RR}\, \psi_\nk \bigl( \Si (\rr-\bvec)\bigr ) \\ =
  & \quad 
  e^{i\,\mcS \kk \cdot \RR}\, \omcS_\bvec \psi_\nk(\rr),
 \end{split}
\end{equation}
where the invariance under lattice translation of the periodic part of 
the Bloch wave function $u_\nk(\rr)$  has been exploited.
The set of equations below summarizes the relationships employed to reconstruct the wavefunctions in the BZ~\footnote{
In our implementation, we prefer to perform the symmetrization of the periodic part of the wavefunction in $\GG$ space because the fractional translation can be easily taken into account by multiplying by a phase factor.
The symmetrization in real space, on the other hand, requires a real-space FFT grid compatible with all the fractional translations of the space group.
}
\begin{eqnarray}
\label{eq:space_group_symmetry}
\begin{cases}
\ee_{\mcS\kk}    & =  \quad \ee_{\kk} 
\\
u_{\mcS\kk}(\rr) & =  \quad e^{-i \mcS\kk \cdot \bvec}\, u_{\kk}\bigl(\Si(\rr - \bvec)\bigr),
\\
u_{\mcS\kk}(\GG) & =  \quad e^{-i(\mcS\kk+\GG)\cdot\bvec}\, u_{\kk}(\Si\GG)
\end{cases}
\end{eqnarray}
where we used the following conventions for the Fourier series of periodic lattice quantities
\begin{eqnarray}\label{eq:FT_1point_convention}
 u(\GG) & = \frac{1}{\Omega} \int_\Omega u(\rr)e^{-i\GG\cdot\rr}\dd\rr, \\
 u(\rr) & = \sum_\GG u(\GG)e^{i\GG\cdot\rr}.
\end{eqnarray}
The time invariance of the Hamiltonian leads to the additional symmetries
\begin{eqnarray}
\label{eq:time_reversal_symmetry}
\begin{cases}
\ee_{-\kk}   & =  \quad \ee_{\kk}, 
\\
u_{-\kk}(\rr) & =  \quad u_\kk^*(\rr), 
\\
u_{-\kk}(-\GG) & = \quad u_\kk^*(\GG)
\end{cases}
\end{eqnarray}
that may be used to halve the number of $\kk$ points even if the system is not invariant under spatial inversion.
Finally, wave functions with $\kk$ wave vector outside the first Brillouin zone are obtained with the gauge
\begin{equation}
\label{eq:parallel_transport}
\psi_{n\kk+\GG}(\rr) = \psi_\nk(\rr).
\end{equation}
It is important to stress that in 
Eqs.~\eqref{eq:space_group_symmetry}, ~\eqref{eq:time_reversal_symmetry} 
and~\eqref{eq:parallel_transport},
we assumed nondegenerate eigenstates.
In the presence of degeneracy, indeed, it is always possible to construct new orthonormal eigenstates of the Hamiltonian by performing a unitary transformation within the degenerate subspace.
However, electron lifetimes (and physical observables in general), do not depend on this gauge since
all the degenerate states are summed over in Eq.~\eqref{eq:fanlifetime}
hence the presence of this gauge is irrelevant in our context.

\section{Symmetry properties of the scattering potentials}
\label{app:symmetry_properties_dfpt}

Symmetry properties can also be used to reconstruct the e-ph scattering potentials from an appropriate set of atomic perturbations with wavevector $\qq$ in the IBZ of the unperturbed crystal.
For the local external potential, the derivation of the symmetry properties is particularly simple as the lattice-periodic part of the first order derivative can be easily expressed in reciprocal space using~\cite{Gonze1997a}
\begin{equation}
    \partial_{\kappa\alpha,\qq} v^{\text{loc}}(\GG) = -\dfrac{i}{\Omega} (\qq+\GG)_\alpha e^{-i(\qq+\GG) \cdot \btau_\kappa}
    v^{\text{loc}}_\kappa(|\qq+\GG|)
\end{equation}
where $v^{\text{loc}}_\kappa(|\qq+\GG|)$ is the Fourier transform of the local part of the pseudopotential associated to atom $\kappa$.
Starting from this formula, one obtains that the Fourier component at the rotated $(\mcS\qq, \mcS\GG)$ is given by a linear combination of the symmetry-related $(\kappa'\beta)$ terms with wavevector $\qq$ in the IBZ

\begin{equation}
\label{eq:v1loc_qg_symmetry}
\begin{split}
    \partial_{\kappa\alpha,\mcS\qq}v^{\text{loc}}(\mcS\GG) = \sum_\beta 
    \mcS_{\alpha\beta}\, \partial_{\kappa'\beta,\qq}\, v^{\text{loc}}(\GG) \\   \times e^{-i(\qq+\GG) \cdot (\LL_0 + \mcS^{-1}\bvec )}
\end{split}
\end{equation}
where $\btau_{\kappa'}$ denotes the position of the symmetric atom in the first unit cell and $\LL_0$ is a, possibly null, real-space lattice vector.
The two symmetric atoms at $\btau_{\kappa}$ and 
$\btau_{\kappa'}$ are related by the inverse of  $\oSv$:
\begin{equation}
\mcS^{-1} (\btau_\kappa - \bvec) = 
    \btau_{\kappa'} + \LL_0.
\end{equation}
%
We now derive the symmetry properties of the SCF part of the DFPT potential using Eq.~\eqref{eq:v1loc_qg_symmetry} 
and well-known results for the inverse dielectric matrix $\epsilon^{-1}$.
The first-order variation of the self-consistent potential is indeed related to the first-order change of the external potential by 
%
\begin{equation}
\label{Eq:vtot_from_em1}
\delta V^{\text{scf}}(\rr) = \int \dd\rr' \epsilon^{-1}(\rr,\rr') \delta V^{\text{ext}}(\rr')
\end{equation}
where $\epsilon^{-1}$ is invariant 
under all the symmetry operations of the crystal:
\begin{equation}
\epsilon^{-1}(\rr + \RR, \rr' + \RR) = \epsilon^{-1}(\rr, \rr'),
\end{equation}
\begin{equation}
\label{eq:em1_rotsym}
\epsilon^{-1}(\oSv\rr, \oSv\rr') = \epsilon^{-1}(\rr, \rr').
\end{equation}
In Fourier space, Eq.~\eqref{eq:em1_rotsym} leads to the following symmetry property~\cite{Hybertsen1987}
\begin{equation}
\label{eq:epsilon_qgg_symmetry}
\epsilon^{-1}_{\mcS\GG_1,\mcS\GG_2}(\mcS\qq) = e^{+i\mcS(\GG_2-\GG_1) \cdot \bvec}\,\epsilon^{-1}_{\GG_1,\GG_2}(\qq)
\end{equation}
where the Fourier transform of a two-point function is defined by
\begin{equation}\label{eq:FT_2points_convention}
\begin{split}
 f_{\GG_1,\GG_2}(\qq) = \frac{1}{V} \iint_V 
 e^{-i(\qq+\GG_1) \cdot \rr_1}\,f(\rr_1, \rr_2) \\ 
 \times e^{i (\qq+\GG_2) \cdot \rr_2}\dd\rr_1\dd\rr_2,
\end{split}
\end{equation}
and $V$ is the volume of the 
Born-von K\'arm\'an supercell.
In $\GG$ space, the screened potential is related to the external perturbation by:
\begin{equation}
\label{eq:v1scf_em1vloc}
\partial_{\kappa\alpha,\qq} \vscf(\GG) =
\sum_{\GG'} \epsilon^{-1}_{\GG,\GG'}(\qq)
\partial_{\kappa\alpha,\qq} v^{\text{loc}}(\GG')
\end{equation}
Using Eqs.~\eqref{eq:v1loc_qg_symmetry} and
\eqref{eq:epsilon_qgg_symmetry} 
in Eq.~\eqref{eq:v1scf_em1vloc}, after some algebra to rearrange the different terms, 
one obtains that the Fourier components of 
the SCF part of the DFPT potential transform similarly to the local part of the pseudopotential (compare with Eq.~\eqref{eq:v1loc_qg_symmetry})
\begin{equation}
\begin{split}
    \partial_{\kappa\alpha,\mcS\qq}\vscf(\mcS\GG) = \sum_\beta 
    \mcS_{\alpha\beta}\, \partial_{\kappa'\beta,\qq}\, \vscf(\GG) \\   \times e^{-i(\qq+\GG) \cdot (\LL_0 + \mcS^{-1}\bvec )}.
\end{split}
\end{equation}
%
%
%
%
When $\mcS\qq = \qq$, the equation provides a connection among atomic perturbations with the same wavevector $\qq$ that can be used to reconstruct by symmetry all the $3 N_{\text{atom}}$ potentials starting from an irreducible set.
Finally, we note that
the contribution associated to the nonlocal part of the pseudopotential can be computed explicitly for arbitrary $\qq$ using the equations given in~\cite{Gonze1997a} hence no symmetrization is required for this term.
%
%
%
%
%
%

\section{Boltzmann transport equation}
\label{app:boltzmann_transport}

An equivalent formulation of the linearized Boltzmann transport equation in the RTA uses the transport distribution function~\cite{Madsen2018}:
    \begin{align}
    L_{\alpha\beta}(\omega) =
    \sum_{n} \int \frac{d\kb}{\Omega_{BZ}} \vnka \vnkb \tau_{n\kb} \delta(\omega-\enk)
    \label{eq:transport_l}
    \end{align}    
that can be used to express the generalized transport coefficients as:
    \begin{align}
    \Lc^{(m)}_{\alpha\beta} =
    -\int L_{\alpha\beta}(\omega)(\omega-\ef)^m \frac{\partial f(\omega)}{\partial\omega}d\omega.
    \label{eq:transport_lc_delta}
    \end{align}
The expression given in the main text (Eq.~\eqref{eq:transport_lc}) is obtained by replacing Eq.~\eqref{eq:transport_l} 
in Eq.~\eqref{eq:transport_lc_delta} and integrating the $\delta$ function in frequency space. The integrand represented in Fig.~\textcolor{blue}{7(a)} 
of the main text is
\begin{equation}
-\vnka \vnkb \tau_{n\kb} \frac{\partial f(\omega)}{\partial\omega} \delta(\omega-\enk),
\end{equation}
which corresponds to the integrand in Eq.~\eqref{eq:transport_lc_delta} for ${m=0}$.

\nocite{Petretto2018,Hamann2013}
\bibliographystyle{apsrev4-1}

\begin{thebibliography}{99}%
\makeatletter
\providecommand \@ifxundefined [1]{%
 \@ifx{#1\undefined}
}%
\providecommand \@ifnum [1]{%
 \ifnum #1\expandafter \@firstoftwo
 \else \expandafter \@secondoftwo
 \fi
}%
\providecommand \@ifx [1]{%
 \ifx #1\expandafter \@firstoftwo
 \else \expandafter \@secondoftwo
 \fi
}%
\providecommand \natexlab [1]{#1}%
\providecommand \enquote  [1]{``#1''}%
\providecommand \bibnamefont  [1]{#1}%
\providecommand \bibfnamefont [1]{#1}%
\providecommand \citenamefont [1]{#1}%
\providecommand \href@noop [0]{\@secondoftwo}%
\providecommand \href [0]{\begingroup \@sanitize@url \@href}%
\providecommand \@href[1]{\@@startlink{#1}\@@href}%
\providecommand \@@href[1]{\endgroup#1\@@endlink}%
\providecommand \@sanitize@url [0]{\catcode `\\12\catcode `\$12\catcode
  `\&12\catcode `\#12\catcode `\^12\catcode `\_12\catcode `\%12\relax}%
\providecommand \@@startlink[1]{}%
\providecommand \@@endlink[0]{}%
\providecommand \url  [0]{\begingroup\@sanitize@url \@url }%
\providecommand \@url [1]{\endgroup\@href {#1}{\urlprefix }}%
\providecommand \urlprefix  [0]{URL }%
\providecommand \Eprint [0]{\href }%
\providecommand \doibase [0]{http://dx.doi.org/}%
\providecommand \selectlanguage [0]{\@gobble}%
\providecommand \bibinfo  [0]{\@secondoftwo}%
\providecommand \bibfield  [0]{\@secondoftwo}%
\providecommand \translation [1]{[#1]}%
\providecommand \BibitemOpen [0]{}%
\providecommand \bibitemStop [0]{}%
\providecommand \bibitemNoStop [0]{.\EOS\space}%
\providecommand \EOS [0]{\spacefactor3000\relax}%
\providecommand \BibitemShut  [1]{\csname bibitem#1\endcsname}%
\let\auto@bib@innerbib\@empty
\bibitem [{\citenamefont {Giustino}(2017)}]{Giustino2017}%
  \BibitemOpen
  \bibfield  {author} {\bibinfo {author} {\bibfnamefont {F.}~\bibnamefont
  {Giustino}},\ }\href {\doibase 10.1103/RevModPhys.89.015003} {\bibfield
  {journal} {\bibinfo  {journal} {Rev.~Mod.~Phys.}\ }\textbf {\bibinfo {volume}
  {89}},\ \bibinfo {pages} {015003} (\bibinfo {year} {2017})}\BibitemShut
  {NoStop}%
\bibitem [{\citenamefont {McMillan}(1968)}]{Mcmillan1968}%
  \BibitemOpen
  \bibfield  {author} {\bibinfo {author} {\bibfnamefont {W.~L.}\ \bibnamefont
  {McMillan}},\ }\href {\doibase 10.1103/PhysRev.167.331} {\bibfield  {journal}
  {\bibinfo  {journal} {Phys.~Rev.}\ }\textbf {\bibinfo {volume} {167}},\
  \bibinfo {pages} {331} (\bibinfo {year} {1968})}\BibitemShut {NoStop}%
\bibitem [{\citenamefont {Choi}\ \emph {et~al.}(2002)\citenamefont {Choi},
  \citenamefont {Roundy}, \citenamefont {Sun}, \citenamefont {Cohen},\ and\
  \citenamefont {Louie}}]{Choi2002}%
  \BibitemOpen
  \bibfield  {author} {\bibinfo {author} {\bibfnamefont {H.~J.}\ \bibnamefont
  {Choi}}, \bibinfo {author} {\bibfnamefont {D.}~\bibnamefont {Roundy}},
  \bibinfo {author} {\bibfnamefont {H.}~\bibnamefont {Sun}}, \bibinfo {author}
  {\bibfnamefont {M.~L.}\ \bibnamefont {Cohen}}, \ and\ \bibinfo {author}
  {\bibfnamefont {S.~G.}\ \bibnamefont {Louie}},\ }\href {\doibase
  10.1038/nature00898} {\bibfield  {journal} {\bibinfo  {journal} {Nature}\
  }\textbf {\bibinfo {volume} {418}},\ \bibinfo {pages} {758} (\bibinfo {year}
  {2002})}\BibitemShut {NoStop}%
\bibitem [{\citenamefont {L{\"u}ders}\ \emph {et~al.}(2005)\citenamefont
  {L{\"u}ders}, \citenamefont {Marques}, \citenamefont {Lathiotakis},
  \citenamefont {Floris}, \citenamefont {Profeta}, \citenamefont {Fast},
  \citenamefont {Continenza}, \citenamefont {Massidda},\ and\ \citenamefont
  {Gross}}]{Luders2005}%
  \BibitemOpen
  \bibfield  {author} {\bibinfo {author} {\bibfnamefont {M.}~\bibnamefont
  {L{\"u}ders}}, \bibinfo {author} {\bibfnamefont {M.~A.~L.}\ \bibnamefont
  {Marques}}, \bibinfo {author} {\bibfnamefont {N.~N.}\ \bibnamefont
  {Lathiotakis}}, \bibinfo {author} {\bibfnamefont {A.}~\bibnamefont {Floris}},
  \bibinfo {author} {\bibfnamefont {G.}~\bibnamefont {Profeta}}, \bibinfo
  {author} {\bibfnamefont {L.}~\bibnamefont {Fast}}, \bibinfo {author}
  {\bibfnamefont {A.}~\bibnamefont {Continenza}}, \bibinfo {author}
  {\bibfnamefont {S.}~\bibnamefont {Massidda}}, \ and\ \bibinfo {author}
  {\bibfnamefont {E.~K.~U.}\ \bibnamefont {Gross}},\ }\href {\doibase
  10.1103/PhysRevB.72.024545} {\bibfield  {journal} {\bibinfo  {journal}
  {Phys.~Rev.~B}\ }\textbf {\bibinfo {volume} {72}},\ \bibinfo {pages} {024545}
  (\bibinfo {year} {2005})}\BibitemShut {NoStop}%
\bibitem [{\citenamefont {Marques}\ \emph {et~al.}(2005)\citenamefont
  {Marques}, \citenamefont {L{\"u}ders}, \citenamefont {Lathiotakis},
  \citenamefont {Profeta}, \citenamefont {Floris}, \citenamefont {Fast},
  \citenamefont {Continenza}, \citenamefont {Gross},\ and\ \citenamefont
  {Massidda}}]{Marques2005}%
  \BibitemOpen
  \bibfield  {author} {\bibinfo {author} {\bibfnamefont {M.~A.~L.}\
  \bibnamefont {Marques}}, \bibinfo {author} {\bibfnamefont {M.}~\bibnamefont
  {L{\"u}ders}}, \bibinfo {author} {\bibfnamefont {N.~N.}\ \bibnamefont
  {Lathiotakis}}, \bibinfo {author} {\bibfnamefont {G.}~\bibnamefont
  {Profeta}}, \bibinfo {author} {\bibfnamefont {A.}~\bibnamefont {Floris}},
  \bibinfo {author} {\bibfnamefont {L.}~\bibnamefont {Fast}}, \bibinfo {author}
  {\bibfnamefont {A.}~\bibnamefont {Continenza}}, \bibinfo {author}
  {\bibfnamefont {E.~K.~U.}\ \bibnamefont {Gross}}, \ and\ \bibinfo {author}
  {\bibfnamefont {S.}~\bibnamefont {Massidda}},\ }\href {\doibase
  10.1103/PhysRevB.72.024546} {\bibfield  {journal} {\bibinfo  {journal}
  {Phys.~Rev.~B}\ }\textbf {\bibinfo {volume} {72}},\ \bibinfo {pages} {024546}
  (\bibinfo {year} {2005})}\BibitemShut {NoStop}%
\bibitem [{\citenamefont {Margine}\ and\ \citenamefont
  {Giustino}(2014)}]{Margine2014}%
  \BibitemOpen
  \bibfield  {author} {\bibinfo {author} {\bibfnamefont {E.~R.}\ \bibnamefont
  {Margine}}\ and\ \bibinfo {author} {\bibfnamefont {F.}~\bibnamefont
  {Giustino}},\ }\href {\doibase 10.1103/PhysRevB.90.014518} {\bibfield
  {journal} {\bibinfo  {journal} {Phys.~Rev.~B}\ }\textbf {\bibinfo {volume}
  {90}},\ \bibinfo {pages} {014518} (\bibinfo {year} {2014})}\BibitemShut
  {NoStop}%
\bibitem [{\citenamefont {Margine}\ \emph {et~al.}(2016)\citenamefont
  {Margine}, \citenamefont {Lambert},\ and\ \citenamefont
  {Giustino}}]{Margine2016}%
  \BibitemOpen
  \bibfield  {author} {\bibinfo {author} {\bibfnamefont {E.~R.}\ \bibnamefont
  {Margine}}, \bibinfo {author} {\bibfnamefont {H.}~\bibnamefont {Lambert}}, \
  and\ \bibinfo {author} {\bibfnamefont {F.}~\bibnamefont {Giustino}},\ }\href
  {\doibase 10.1038/srep21414} {\bibfield  {journal} {\bibinfo  {journal}
  {Sci.~Rep.}\ }\textbf {\bibinfo {volume} {6}},\ \bibinfo {pages} {21414}
  (\bibinfo {year} {2016})}\BibitemShut {NoStop}%
\bibitem [{\citenamefont {Kioupakis}\ \emph {et~al.}(2010)\citenamefont
  {Kioupakis}, \citenamefont {Rinke}, \citenamefont {Schleife}, \citenamefont
  {Bechstedt},\ and\ \citenamefont {Van~de Walle}}]{Kioupakis2010}%
  \BibitemOpen
  \bibfield  {author} {\bibinfo {author} {\bibfnamefont {E.}~\bibnamefont
  {Kioupakis}}, \bibinfo {author} {\bibfnamefont {P.}~\bibnamefont {Rinke}},
  \bibinfo {author} {\bibfnamefont {A.}~\bibnamefont {Schleife}}, \bibinfo
  {author} {\bibfnamefont {F.}~\bibnamefont {Bechstedt}}, \ and\ \bibinfo
  {author} {\bibfnamefont {C.~G.}\ \bibnamefont {Van~de Walle}},\ }\href
  {\doibase 10.1103/PhysRevB.81.241201} {\bibfield  {journal} {\bibinfo
  {journal} {Phys.~Rev.~B}\ }\textbf {\bibinfo {volume} {81}},\ \bibinfo
  {pages} {241201} (\bibinfo {year} {2010})}\BibitemShut {NoStop}%
\bibitem [{\citenamefont {Noffsinger}\ \emph {et~al.}(2012)\citenamefont
  {Noffsinger}, \citenamefont {Kioupakis}, \citenamefont {Van~de Walle},
  \citenamefont {Louie},\ and\ \citenamefont {Cohen}}]{Noffsinger2012}%
  \BibitemOpen
  \bibfield  {author} {\bibinfo {author} {\bibfnamefont {J.}~\bibnamefont
  {Noffsinger}}, \bibinfo {author} {\bibfnamefont {E.}~\bibnamefont
  {Kioupakis}}, \bibinfo {author} {\bibfnamefont {C.~G.}\ \bibnamefont {Van~de
  Walle}}, \bibinfo {author} {\bibfnamefont {S.~G.}\ \bibnamefont {Louie}}, \
  and\ \bibinfo {author} {\bibfnamefont {M.~L.}\ \bibnamefont {Cohen}},\ }\href
  {\doibase 10.1103/PhysRevLett.108.167402} {\bibfield  {journal} {\bibinfo
  {journal} {Phys.~Rev.~Lett.}\ }\textbf {\bibinfo {volume} {108}},\ \bibinfo
  {pages} {167402} (\bibinfo {year} {2012})}\BibitemShut {NoStop}%
\bibitem [{\citenamefont {Peelaers}\ \emph {et~al.}(2012)\citenamefont
  {Peelaers}, \citenamefont {Kioupakis},\ and\ \citenamefont {Van~de
  Walle}}]{Peelaers2012}%
  \BibitemOpen
  \bibfield  {author} {\bibinfo {author} {\bibfnamefont {H.}~\bibnamefont
  {Peelaers}}, \bibinfo {author} {\bibfnamefont {E.}~\bibnamefont {Kioupakis}},
  \ and\ \bibinfo {author} {\bibfnamefont {C.~G.}\ \bibnamefont {Van~de
  Walle}},\ }\href {\doibase 10.1063/1.3671162} {\bibfield  {journal} {\bibinfo
   {journal} {Appl.~Phys.~Lett.}\ }\textbf {\bibinfo {volume} {100}},\ \bibinfo
  {pages} {011914} (\bibinfo {year} {2012})}\BibitemShut {NoStop}%
\bibitem [{\citenamefont {Allen}\ and\ \citenamefont
  {Heine}(1976)}]{Allen1976}%
  \BibitemOpen
  \bibfield  {author} {\bibinfo {author} {\bibfnamefont {P.~B.}\ \bibnamefont
  {Allen}}\ and\ \bibinfo {author} {\bibfnamefont {V.}~\bibnamefont {Heine}},\
  }\href {\doibase 10.1088/0022-3719/9/12/013} {\bibfield  {journal} {\bibinfo
  {journal} {J.~Phys.~C}\ }\textbf {\bibinfo {volume} {9}},\ \bibinfo {pages}
  {2305} (\bibinfo {year} {1976})}\BibitemShut {NoStop}%
\bibitem [{\citenamefont {Allen}\ and\ \citenamefont
  {Cardona}(1983)}]{Allen1983}%
  \BibitemOpen
  \bibfield  {author} {\bibinfo {author} {\bibfnamefont {P.~B.}\ \bibnamefont
  {Allen}}\ and\ \bibinfo {author} {\bibfnamefont {M.}~\bibnamefont
  {Cardona}},\ }\href {\doibase 10.1103/PhysRevB.27.4760} {\bibfield  {journal}
  {\bibinfo  {journal} {Phys.~Rev.~B}\ }\textbf {\bibinfo {volume} {27}},\
  \bibinfo {pages} {4760} (\bibinfo {year} {1983})}\BibitemShut {NoStop}%
\bibitem [{\citenamefont {Ponc{\'e}}\ \emph {et~al.}(2014)\citenamefont
  {Ponc{\'e}}, \citenamefont {Antonius}, \citenamefont {Gillet}, \citenamefont
  {Boulanger}, \citenamefont {Laflamme~Janssen}, \citenamefont {Marini},
  \citenamefont {C{\^o}t{\'e}},\ and\ \citenamefont {Gonze}}]{Ponce2014}%
  \BibitemOpen
  \bibfield  {author} {\bibinfo {author} {\bibfnamefont {S.}~\bibnamefont
  {Ponc{\'e}}}, \bibinfo {author} {\bibfnamefont {G.}~\bibnamefont {Antonius}},
  \bibinfo {author} {\bibfnamefont {Y.}~\bibnamefont {Gillet}}, \bibinfo
  {author} {\bibfnamefont {P.}~\bibnamefont {Boulanger}}, \bibinfo {author}
  {\bibfnamefont {J.}~\bibnamefont {Laflamme~Janssen}}, \bibinfo {author}
  {\bibfnamefont {A.}~\bibnamefont {Marini}}, \bibinfo {author} {\bibfnamefont
  {M.}~\bibnamefont {C{\^o}t{\'e}}}, \ and\ \bibinfo {author} {\bibfnamefont
  {X.}~\bibnamefont {Gonze}},\ }\href {\doibase 10.1103/PhysRevB.90.214304}
  {\bibfield  {journal} {\bibinfo  {journal} {Phys.~Rev.~B}\ }\textbf {\bibinfo
  {volume} {90}},\ \bibinfo {pages} {214304} (\bibinfo {year}
  {2014})}\BibitemShut {NoStop}%
\bibitem [{\citenamefont {Antonius}\ \emph {et~al.}(2015)\citenamefont
  {Antonius}, \citenamefont {Ponc{\'e}}, \citenamefont {Lantagne-Hurtubise},
  \citenamefont {Auclair}, \citenamefont {Gonze},\ and\ \citenamefont
  {C{\^o}t{\'e}}}]{Antonius2015}%
  \BibitemOpen
  \bibfield  {author} {\bibinfo {author} {\bibfnamefont {G.}~\bibnamefont
  {Antonius}}, \bibinfo {author} {\bibfnamefont {S.}~\bibnamefont {Ponc{\'e}}},
  \bibinfo {author} {\bibfnamefont {E.}~\bibnamefont {Lantagne-Hurtubise}},
  \bibinfo {author} {\bibfnamefont {G.}~\bibnamefont {Auclair}}, \bibinfo
  {author} {\bibfnamefont {X.}~\bibnamefont {Gonze}}, \ and\ \bibinfo {author}
  {\bibfnamefont {M.}~\bibnamefont {C{\^o}t{\'e}}},\ }\href {\doibase
  10.1103/PhysRevB.92.085137} {\bibfield  {journal} {\bibinfo  {journal}
  {Phys.~Rev.~B}\ }\textbf {\bibinfo {volume} {92}},\ \bibinfo {pages} {085137}
  (\bibinfo {year} {2015})}\BibitemShut {NoStop}%
\bibitem [{\citenamefont {Ponc{\'e}}\ \emph {et~al.}(2015)\citenamefont
  {Ponc{\'e}}, \citenamefont {Gillet}, \citenamefont {Laflamme~Janssen},
  \citenamefont {Marini}, \citenamefont {Verstraete},\ and\ \citenamefont
  {Gonze}}]{Ponce2015}%
  \BibitemOpen
  \bibfield  {author} {\bibinfo {author} {\bibfnamefont {S.}~\bibnamefont
  {Ponc{\'e}}}, \bibinfo {author} {\bibfnamefont {Y.}~\bibnamefont {Gillet}},
  \bibinfo {author} {\bibfnamefont {J.}~\bibnamefont {Laflamme~Janssen}},
  \bibinfo {author} {\bibfnamefont {A.}~\bibnamefont {Marini}}, \bibinfo
  {author} {\bibfnamefont {M.}~\bibnamefont {Verstraete}}, \ and\ \bibinfo
  {author} {\bibfnamefont {X.}~\bibnamefont {Gonze}},\ }\href {\doibase
  10.1063/1.4927081} {\bibfield  {journal} {\bibinfo  {journal}
  {J.~Chem.~Phys.}\ }\textbf {\bibinfo {volume} {143}},\ \bibinfo {pages}
  {102813} (\bibinfo {year} {2015})}\BibitemShut {NoStop}%
\bibitem [{\citenamefont {Broido}\ \emph {et~al.}(2007)\citenamefont {Broido},
  \citenamefont {Malorny}, \citenamefont {Birner}, \citenamefont {Mingo},\ and\
  \citenamefont {Stewart}}]{Broido2007}%
  \BibitemOpen
  \bibfield  {author} {\bibinfo {author} {\bibfnamefont {D.~A.}\ \bibnamefont
  {Broido}}, \bibinfo {author} {\bibfnamefont {M.}~\bibnamefont {Malorny}},
  \bibinfo {author} {\bibfnamefont {G.}~\bibnamefont {Birner}}, \bibinfo
  {author} {\bibfnamefont {N.}~\bibnamefont {Mingo}}, \ and\ \bibinfo {author}
  {\bibfnamefont {D.~A.}\ \bibnamefont {Stewart}},\ }\href {\doibase
  10.1063/1.2822891} {\bibfield  {journal} {\bibinfo  {journal}
  {Appl.~Phys.~Lett.}\ }\textbf {\bibinfo {volume} {91}},\ \bibinfo {pages}
  {231922} (\bibinfo {year} {2007})}\BibitemShut {NoStop}%
\bibitem [{\citenamefont {Liao}\ \emph
  {et~al.}(2015{\natexlab{a}})\citenamefont {Liao}, \citenamefont {Qiu},
  \citenamefont {Zhou}, \citenamefont {Huberman}, \citenamefont {Esfarjani},\
  and\ \citenamefont {Chen}}]{Liao2015}%
  \BibitemOpen
  \bibfield  {author} {\bibinfo {author} {\bibfnamefont {B.}~\bibnamefont
  {Liao}}, \bibinfo {author} {\bibfnamefont {B.}~\bibnamefont {Qiu}}, \bibinfo
  {author} {\bibfnamefont {J.}~\bibnamefont {Zhou}}, \bibinfo {author}
  {\bibfnamefont {S.}~\bibnamefont {Huberman}}, \bibinfo {author}
  {\bibfnamefont {K.}~\bibnamefont {Esfarjani}}, \ and\ \bibinfo {author}
  {\bibfnamefont {G.}~\bibnamefont {Chen}},\ }\href {\doibase
  10.1103/PhysRevLett.114.115901} {\bibfield  {journal} {\bibinfo  {journal}
  {Phys.~Rev.~Lett.}\ }\textbf {\bibinfo {volume} {114}},\ \bibinfo {pages}
  {115901} (\bibinfo {year} {2015}{\natexlab{a}})}\BibitemShut {NoStop}%
\bibitem [{\citenamefont {Kim}\ \emph {et~al.}(2016)\citenamefont {Kim},
  \citenamefont {Park},\ and\ \citenamefont {Marzari}}]{Kim2016}%
  \BibitemOpen
  \bibfield  {author} {\bibinfo {author} {\bibfnamefont {T.~Y.}\ \bibnamefont
  {Kim}}, \bibinfo {author} {\bibfnamefont {C.-H.}\ \bibnamefont {Park}}, \
  and\ \bibinfo {author} {\bibfnamefont {N.}~\bibnamefont {Marzari}},\ }\href
  {\doibase 10.1021/acs.nanolett.5b05288} {\bibfield  {journal} {\bibinfo
  {journal} {Nano Lett.}\ }\textbf {\bibinfo {volume} {16}},\ \bibinfo {pages}
  {2439} (\bibinfo {year} {2016})}\BibitemShut {NoStop}%
\bibitem [{\citenamefont {Xu}\ and\ \citenamefont {Verstraete}(2014)}]{Xu2014}%
  \BibitemOpen
  \bibfield  {author} {\bibinfo {author} {\bibfnamefont {B.}~\bibnamefont
  {Xu}}\ and\ \bibinfo {author} {\bibfnamefont {M.~J.}\ \bibnamefont
  {Verstraete}},\ }\href {\doibase 10.1103/PhysRevLett.112.196603} {\bibfield
  {journal} {\bibinfo  {journal} {Phys.~Rev.~Lett.}\ }\textbf {\bibinfo
  {volume} {112}},\ \bibinfo {pages} {196603} (\bibinfo {year}
  {2014})}\BibitemShut {NoStop}%
\bibitem [{\citenamefont {Bernardi}\ \emph {et~al.}(2014)\citenamefont
  {Bernardi}, \citenamefont {Vigil-Fowler}, \citenamefont {Lischner},
  \citenamefont {Neaton},\ and\ \citenamefont {Louie}}]{Bernardi2014}%
  \BibitemOpen
  \bibfield  {author} {\bibinfo {author} {\bibfnamefont {M.}~\bibnamefont
  {Bernardi}}, \bibinfo {author} {\bibfnamefont {D.}~\bibnamefont
  {Vigil-Fowler}}, \bibinfo {author} {\bibfnamefont {J.}~\bibnamefont
  {Lischner}}, \bibinfo {author} {\bibfnamefont {J.~B.}\ \bibnamefont
  {Neaton}}, \ and\ \bibinfo {author} {\bibfnamefont {S.~G.}\ \bibnamefont
  {Louie}},\ }\href {\doibase 10.1103/PhysRevLett.112.257402} {\bibfield
  {journal} {\bibinfo  {journal} {Phys.~Rev.~Lett.}\ }\textbf {\bibinfo
  {volume} {112}},\ \bibinfo {pages} {257402} (\bibinfo {year}
  {2014})}\BibitemShut {NoStop}%
\bibitem [{\citenamefont {Krishnaswamy}\ \emph {et~al.}(2017)\citenamefont
  {Krishnaswamy}, \citenamefont {Himmetoglu}, \citenamefont {Kang},
  \citenamefont {Janotti},\ and\ \citenamefont {Van~de
  Walle}}]{Krishnaswamy2017}%
  \BibitemOpen
  \bibfield  {author} {\bibinfo {author} {\bibfnamefont {K.}~\bibnamefont
  {Krishnaswamy}}, \bibinfo {author} {\bibfnamefont {B.}~\bibnamefont
  {Himmetoglu}}, \bibinfo {author} {\bibfnamefont {Y.}~\bibnamefont {Kang}},
  \bibinfo {author} {\bibfnamefont {A.}~\bibnamefont {Janotti}}, \ and\
  \bibinfo {author} {\bibfnamefont {C.~G.}\ \bibnamefont {Van~de Walle}},\
  }\href {\doibase 10.1103/PhysRevB.95.205202} {\bibfield  {journal} {\bibinfo
  {journal} {Phys.~Rev.~B}\ }\textbf {\bibinfo {volume} {95}},\ \bibinfo
  {pages} {205202} (\bibinfo {year} {2017})}\BibitemShut {NoStop}%
\bibitem [{\citenamefont {Ma}\ \emph {et~al.}(2018)\citenamefont {Ma},
  \citenamefont {Nissimagoudar},\ and\ \citenamefont {Li}}]{Ma2018}%
  \BibitemOpen
  \bibfield  {author} {\bibinfo {author} {\bibfnamefont {J.}~\bibnamefont
  {Ma}}, \bibinfo {author} {\bibfnamefont {A.~S.}\ \bibnamefont
  {Nissimagoudar}}, \ and\ \bibinfo {author} {\bibfnamefont {W.}~\bibnamefont
  {Li}},\ }\href {\doibase 10.1103/PhysRevB.97.045201} {\bibfield  {journal}
  {\bibinfo  {journal} {Phys.~Rev.~B}\ }\textbf {\bibinfo {volume} {97}},\
  \bibinfo {pages} {045201} (\bibinfo {year} {2018})}\BibitemShut {NoStop}%
\bibitem [{\citenamefont {Li}(2015)}]{Li2015}%
  \BibitemOpen
  \bibfield  {author} {\bibinfo {author} {\bibfnamefont {W.}~\bibnamefont
  {Li}},\ }\href {\doibase 10.1103/PhysRevB.92.075405} {\bibfield  {journal}
  {\bibinfo  {journal} {Phys.~Rev.~B}\ }\textbf {\bibinfo {volume} {92}},\
  \bibinfo {pages} {075405} (\bibinfo {year} {2015})}\BibitemShut {NoStop}%
\bibitem [{\citenamefont {Mustafa}\ \emph {et~al.}(2016)\citenamefont
  {Mustafa}, \citenamefont {Bernardi}, \citenamefont {Neaton},\ and\
  \citenamefont {Louie}}]{Mustafa2016}%
  \BibitemOpen
  \bibfield  {author} {\bibinfo {author} {\bibfnamefont {J.~I.}\ \bibnamefont
  {Mustafa}}, \bibinfo {author} {\bibfnamefont {M.}~\bibnamefont {Bernardi}},
  \bibinfo {author} {\bibfnamefont {J.~B.}\ \bibnamefont {Neaton}}, \ and\
  \bibinfo {author} {\bibfnamefont {S.~G.}\ \bibnamefont {Louie}},\ }\href
  {\doibase 10.1103/PhysRevB.94.155105} {\bibfield  {journal} {\bibinfo
  {journal} {Phys.~Rev.~B}\ }\textbf {\bibinfo {volume} {94}},\ \bibinfo
  {pages} {155105} (\bibinfo {year} {2016})}\BibitemShut {NoStop}%
\bibitem [{\citenamefont {Liao}\ \emph
  {et~al.}(2015{\natexlab{b}})\citenamefont {Liao}, \citenamefont {Zhou},
  \citenamefont {Qiu}, \citenamefont {Dresselhaus},\ and\ \citenamefont
  {Chen}}]{Liao2015b}%
  \BibitemOpen
  \bibfield  {author} {\bibinfo {author} {\bibfnamefont {B.}~\bibnamefont
  {Liao}}, \bibinfo {author} {\bibfnamefont {J.}~\bibnamefont {Zhou}}, \bibinfo
  {author} {\bibfnamefont {B.}~\bibnamefont {Qiu}}, \bibinfo {author}
  {\bibfnamefont {M.~S.}\ \bibnamefont {Dresselhaus}}, \ and\ \bibinfo {author}
  {\bibfnamefont {G.}~\bibnamefont {Chen}},\ }\href {\doibase
  10.1103/PhysRevB.91.235419} {\bibfield  {journal} {\bibinfo  {journal}
  {Phys.~Rev.~B}\ }\textbf {\bibinfo {volume} {91}},\ \bibinfo {pages} {235419}
  (\bibinfo {year} {2015}{\natexlab{b}})}\BibitemShut {NoStop}%
\bibitem [{\citenamefont {Ponc{\'e}}\ \emph {et~al.}(2018)\citenamefont
  {Ponc{\'e}}, \citenamefont {Margine},\ and\ \citenamefont
  {Giustino}}]{Ponce2018}%
  \BibitemOpen
  \bibfield  {author} {\bibinfo {author} {\bibfnamefont {S.}~\bibnamefont
  {Ponc{\'e}}}, \bibinfo {author} {\bibfnamefont {E.~R.}\ \bibnamefont
  {Margine}}, \ and\ \bibinfo {author} {\bibfnamefont {F.}~\bibnamefont
  {Giustino}},\ }\href {\doibase 10.1103/PhysRevB.97.121201} {\bibfield
  {journal} {\bibinfo  {journal} {Phys.~Rev.~B}\ }\textbf {\bibinfo {volume}
  {97}},\ \bibinfo {pages} {121201} (\bibinfo {year} {2018})}\BibitemShut
  {NoStop}%
\bibitem [{\citenamefont {Zeng}\ \emph {et~al.}(2018)\citenamefont {Zeng},
  \citenamefont {Zhao}, \citenamefont {Li},\ and\ \citenamefont
  {Yang}}]{Zeng2018}%
  \BibitemOpen
  \bibfield  {author} {\bibinfo {author} {\bibfnamefont {X.}~\bibnamefont
  {Zeng}}, \bibinfo {author} {\bibfnamefont {S.}~\bibnamefont {Zhao}}, \bibinfo
  {author} {\bibfnamefont {Z.}~\bibnamefont {Li}}, \ and\ \bibinfo {author}
  {\bibfnamefont {J.}~\bibnamefont {Yang}},\ }\href {\doibase
  10.1103/PhysRevB.98.155443} {\bibfield  {journal} {\bibinfo  {journal}
  {Phys.~Rev.~B}\ }\textbf {\bibinfo {volume} {98}},\ \bibinfo {pages} {155443}
  (\bibinfo {year} {2018})}\BibitemShut {NoStop}%
\bibitem [{\citenamefont {Zhou}\ and\ \citenamefont
  {Bernardi}(2016)}]{Zhou2016}%
  \BibitemOpen
  \bibfield  {author} {\bibinfo {author} {\bibfnamefont {J.-J.}\ \bibnamefont
  {Zhou}}\ and\ \bibinfo {author} {\bibfnamefont {M.}~\bibnamefont
  {Bernardi}},\ }\href {\doibase 10.1103/PhysRevB.94.201201} {\bibfield
  {journal} {\bibinfo  {journal} {Phys.~Rev.~B}\ }\textbf {\bibinfo {volume}
  {94}},\ \bibinfo {pages} {201201} (\bibinfo {year} {2016})}\BibitemShut
  {NoStop}%
\bibitem [{\citenamefont {Gonze}\ \emph {et~al.}(2011)\citenamefont {Gonze},
  \citenamefont {Boulanger},\ and\ \citenamefont {C{\^o}t{\'e}}}]{Gonze2011}%
  \BibitemOpen
  \bibfield  {author} {\bibinfo {author} {\bibfnamefont {X.}~\bibnamefont
  {Gonze}}, \bibinfo {author} {\bibfnamefont {P.}~\bibnamefont {Boulanger}}, \
  and\ \bibinfo {author} {\bibfnamefont {M.}~\bibnamefont {C{\^o}t{\'e}}},\
  }\href {\doibase 10.1002/andp.201000100} {\bibfield  {journal} {\bibinfo
  {journal} {Ann.~Phys.}\ }\textbf {\bibinfo {volume} {523}},\ \bibinfo {pages}
  {168} (\bibinfo {year} {2011})}\BibitemShut {NoStop}%
\bibitem [{\citenamefont {Gonze}\ \emph {et~al.}(2016)\citenamefont {Gonze},
  \citenamefont {Jollet}, \citenamefont {Araujo}, \citenamefont {Adams},
  \citenamefont {Amadon}, \citenamefont {Applencourt}, \citenamefont {Audouze},
  \citenamefont {Beuken}, \citenamefont {Bieder}, \citenamefont {Bokhanchuk}
  \emph {et~al.}}]{Gonze2016}%
  \BibitemOpen
  \bibfield  {author} {\bibinfo {author} {\bibfnamefont {X.}~\bibnamefont
  {Gonze}}, \bibinfo {author} {\bibfnamefont {F.}~\bibnamefont {Jollet}},
  \bibinfo {author} {\bibfnamefont {F.~A.}\ \bibnamefont {Araujo}}, \bibinfo
  {author} {\bibfnamefont {D.}~\bibnamefont {Adams}}, \bibinfo {author}
  {\bibfnamefont {B.}~\bibnamefont {Amadon}}, \bibinfo {author} {\bibfnamefont
  {T.}~\bibnamefont {Applencourt}}, \bibinfo {author} {\bibfnamefont
  {C.}~\bibnamefont {Audouze}}, \bibinfo {author} {\bibfnamefont {J.-M.}\
  \bibnamefont {Beuken}}, \bibinfo {author} {\bibfnamefont {J.}~\bibnamefont
  {Bieder}}, \bibinfo {author} {\bibfnamefont {A.}~\bibnamefont {Bokhanchuk}},
  \emph {et~al.},\ }\href {\doibase 10.1016/j.cpc.2016.04.003} {\bibfield
  {journal} {\bibinfo  {journal} {Comput.~Phys.~Commun.}\ }\textbf {\bibinfo
  {volume} {205}},\ \bibinfo {pages} {106} (\bibinfo {year}
  {2016})}\BibitemShut {NoStop}%
\bibitem [{\citenamefont {Gonze}\ \emph {et~al.}(2019)\citenamefont {Gonze},
  \citenamefont {Amadon}, \citenamefont {Antonius}, \citenamefont {Arnardi},
  \citenamefont {Baguet}, \citenamefont {Beuken}, \citenamefont {Bieder},
  \citenamefont {Bottin}, \citenamefont {Bouchet}, \citenamefont {Bousquet}
  \emph {et~al.}}]{Gonze2019}%
  \BibitemOpen
  \bibfield  {author} {\bibinfo {author} {\bibfnamefont {X.}~\bibnamefont
  {Gonze}}, \bibinfo {author} {\bibfnamefont {B.}~\bibnamefont {Amadon}},
  \bibinfo {author} {\bibfnamefont {G.}~\bibnamefont {Antonius}}, \bibinfo
  {author} {\bibfnamefont {F.}~\bibnamefont {Arnardi}}, \bibinfo {author}
  {\bibfnamefont {L.}~\bibnamefont {Baguet}}, \bibinfo {author} {\bibfnamefont
  {J.-M.}\ \bibnamefont {Beuken}}, \bibinfo {author} {\bibfnamefont
  {J.}~\bibnamefont {Bieder}}, \bibinfo {author} {\bibfnamefont
  {F.}~\bibnamefont {Bottin}}, \bibinfo {author} {\bibfnamefont
  {J.}~\bibnamefont {Bouchet}}, \bibinfo {author} {\bibfnamefont
  {E.}~\bibnamefont {Bousquet}},  \emph {et~al.},\ }\href {\doibase
  https://doi.org/10.1016/j.cpc.2019.107042} {\bibfield  {journal} {\bibinfo
  {journal} {Comput.~Phys.~Commun.}\ }\textbf {\bibinfo {volume} {248}},\
  \bibinfo {pages} {107042} (\bibinfo {year} {2019})}\BibitemShut {NoStop}%
\bibitem [{\citenamefont {Marini}(2008)}]{Marini2008}%
  \BibitemOpen
  \bibfield  {author} {\bibinfo {author} {\bibfnamefont {A.}~\bibnamefont
  {Marini}},\ }\href {\doibase 10.1103/PhysRevLett.101.106405} {\bibfield
  {journal} {\bibinfo  {journal} {Phys.~Rev.~Lett.}\ }\textbf {\bibinfo
  {volume} {101}},\ \bibinfo {pages} {106405} (\bibinfo {year}
  {2008})}\BibitemShut {NoStop}%
\bibitem [{\citenamefont {Marini}\ \emph {et~al.}(2009)\citenamefont {Marini},
  \citenamefont {Hogan}, \citenamefont {Gr{\"u}ning},\ and\ \citenamefont
  {Varsano}}]{Marini2009}%
  \BibitemOpen
  \bibfield  {author} {\bibinfo {author} {\bibfnamefont {A.}~\bibnamefont
  {Marini}}, \bibinfo {author} {\bibfnamefont {C.}~\bibnamefont {Hogan}},
  \bibinfo {author} {\bibfnamefont {M.}~\bibnamefont {Gr{\"u}ning}}, \ and\
  \bibinfo {author} {\bibfnamefont {D.}~\bibnamefont {Varsano}},\ }\href
  {\doibase 10.1016/j.cpc.2009.02.003} {\bibfield  {journal} {\bibinfo
  {journal} {Comput.~Phys.~Commun.}\ }\textbf {\bibinfo {volume} {180}},\
  \bibinfo {pages} {1392} (\bibinfo {year} {2009})}\BibitemShut {NoStop}%
\bibitem [{\citenamefont {Giannozzi}\ \emph {et~al.}(2009)\citenamefont
  {Giannozzi}, \citenamefont {Baroni}, \citenamefont {Bonini}, \citenamefont
  {Calandra}, \citenamefont {Car}, \citenamefont {Cavazzoni}, \citenamefont
  {Ceresoli}, \citenamefont {Chiarotti}, \citenamefont {Cococcioni},
  \citenamefont {Dabo} \emph {et~al.}}]{Giannozzi2009}%
  \BibitemOpen
  \bibfield  {author} {\bibinfo {author} {\bibfnamefont {P.}~\bibnamefont
  {Giannozzi}}, \bibinfo {author} {\bibfnamefont {S.}~\bibnamefont {Baroni}},
  \bibinfo {author} {\bibfnamefont {N.}~\bibnamefont {Bonini}}, \bibinfo
  {author} {\bibfnamefont {M.}~\bibnamefont {Calandra}}, \bibinfo {author}
  {\bibfnamefont {R.}~\bibnamefont {Car}}, \bibinfo {author} {\bibfnamefont
  {C.}~\bibnamefont {Cavazzoni}}, \bibinfo {author} {\bibfnamefont
  {D.}~\bibnamefont {Ceresoli}}, \bibinfo {author} {\bibfnamefont {G.~L.}\
  \bibnamefont {Chiarotti}}, \bibinfo {author} {\bibfnamefont {M.}~\bibnamefont
  {Cococcioni}}, \bibinfo {author} {\bibfnamefont {I.}~\bibnamefont {Dabo}},
  \emph {et~al.},\ }\href {\doibase 10.1088/0953-8984/21/39/395502} {\bibfield
  {journal} {\bibinfo  {journal} {J.~Phys.:~Condens.~Matter}\ }\textbf
  {\bibinfo {volume} {21}},\ \bibinfo {pages} {395502} (\bibinfo {year}
  {2009})}\BibitemShut {NoStop}%
\bibitem [{\citenamefont {Ponc{\'e}}\ \emph {et~al.}(2016)\citenamefont
  {Ponc{\'e}}, \citenamefont {Margine}, \citenamefont {Verdi},\ and\
  \citenamefont {Giustino}}]{Ponce2016}%
  \BibitemOpen
  \bibfield  {author} {\bibinfo {author} {\bibfnamefont {S.}~\bibnamefont
  {Ponc{\'e}}}, \bibinfo {author} {\bibfnamefont {E.~R.}\ \bibnamefont
  {Margine}}, \bibinfo {author} {\bibfnamefont {C.}~\bibnamefont {Verdi}}, \
  and\ \bibinfo {author} {\bibfnamefont {F.}~\bibnamefont {Giustino}},\ }\href
  {\doibase 10.1016/j.cpc.2016.07.028} {\bibfield  {journal} {\bibinfo
  {journal} {Comput.~Phys.~Commun.}\ }\textbf {\bibinfo {volume} {209}},\
  \bibinfo {pages} {116} (\bibinfo {year} {2016})}\BibitemShut {NoStop}%
\bibitem [{\citenamefont {Marzari}\ \emph {et~al.}(2012)\citenamefont
  {Marzari}, \citenamefont {Mostofi}, \citenamefont {Yates}, \citenamefont
  {Souza},\ and\ \citenamefont {Vanderbilt}}]{Marzari2012}%
  \BibitemOpen
  \bibfield  {author} {\bibinfo {author} {\bibfnamefont {N.}~\bibnamefont
  {Marzari}}, \bibinfo {author} {\bibfnamefont {A.~A.}\ \bibnamefont
  {Mostofi}}, \bibinfo {author} {\bibfnamefont {J.~R.}\ \bibnamefont {Yates}},
  \bibinfo {author} {\bibfnamefont {I.}~\bibnamefont {Souza}}, \ and\ \bibinfo
  {author} {\bibfnamefont {D.}~\bibnamefont {Vanderbilt}},\ }\href {\doibase
  10.1103/RevModPhys.84.1419} {\bibfield  {journal} {\bibinfo  {journal}
  {Rev.~Mod.~Phys.}\ }\textbf {\bibinfo {volume} {84}},\ \bibinfo {pages}
  {1419} (\bibinfo {year} {2012})}\BibitemShut {NoStop}%
\bibitem [{Note1()}]{Note1}%
  \BibitemOpen
  \bibinfo {note} {Considerable efforts have been made in recent years to
  develop more robust algorithms to localize Wannier orbitals using few tunable
  parameters. See, for example, the SCDM method proposed in Ref.~\cite
  {Damle2018}.}\BibitemShut {Stop}%
\bibitem [{\citenamefont {Van~Setten}\ \emph {et~al.}(2018)\citenamefont
  {Van~Setten}, \citenamefont {Giantomassi}, \citenamefont {Bousquet},
  \citenamefont {Verstraete}, \citenamefont {Hamann}, \citenamefont {Gonze},\
  and\ \citenamefont {Rignanese}}]{Vansetten2018}%
  \BibitemOpen
  \bibfield  {author} {\bibinfo {author} {\bibfnamefont {M.~J.}\ \bibnamefont
  {Van~Setten}}, \bibinfo {author} {\bibfnamefont {M.}~\bibnamefont
  {Giantomassi}}, \bibinfo {author} {\bibfnamefont {E.}~\bibnamefont
  {Bousquet}}, \bibinfo {author} {\bibfnamefont {M.~J.}\ \bibnamefont
  {Verstraete}}, \bibinfo {author} {\bibfnamefont {D.~R.}\ \bibnamefont
  {Hamann}}, \bibinfo {author} {\bibfnamefont {X.}~\bibnamefont {Gonze}}, \
  and\ \bibinfo {author} {\bibfnamefont {G.-M.}\ \bibnamefont {Rignanese}},\
  }\href {\doibase 10.1016/j.cpc.2018.01.012} {\bibfield  {journal} {\bibinfo
  {journal} {Comput.~Phys.~Commun.}\ }\textbf {\bibinfo {volume} {226}},\
  \bibinfo {pages} {39} (\bibinfo {year} {2018})}\BibitemShut {NoStop}%
\bibitem [{\citenamefont {Agapito}\ and\ \citenamefont
  {Bernardi}(2018)}]{Agapito2018}%
  \BibitemOpen
  \bibfield  {author} {\bibinfo {author} {\bibfnamefont {L.~A.}\ \bibnamefont
  {Agapito}}\ and\ \bibinfo {author} {\bibfnamefont {M.}~\bibnamefont
  {Bernardi}},\ }\href {\doibase 10.1103/PhysRevB.97.235146} {\bibfield
  {journal} {\bibinfo  {journal} {Phys.~Rev.~B}\ }\textbf {\bibinfo {volume}
  {97}},\ \bibinfo {pages} {235146} (\bibinfo {year} {2018})}\BibitemShut
  {NoStop}%
\bibitem [{\citenamefont {Gonze}\ \emph {et~al.}(2009)\citenamefont {Gonze},
  \citenamefont {Amadon}, \citenamefont {Anglade}, \citenamefont {Beuken},
  \citenamefont {Bottin}, \citenamefont {Boulanger}, \citenamefont {Bruneval},
  \citenamefont {Caliste}, \citenamefont {Caracas}, \citenamefont
  {C{\^o}t{\'e}} \emph {et~al.}}]{Gonze2009}%
  \BibitemOpen
  \bibfield  {author} {\bibinfo {author} {\bibfnamefont {X.}~\bibnamefont
  {Gonze}}, \bibinfo {author} {\bibfnamefont {B.}~\bibnamefont {Amadon}},
  \bibinfo {author} {\bibfnamefont {P.-M.}\ \bibnamefont {Anglade}}, \bibinfo
  {author} {\bibfnamefont {J.-M.}\ \bibnamefont {Beuken}}, \bibinfo {author}
  {\bibfnamefont {F.}~\bibnamefont {Bottin}}, \bibinfo {author} {\bibfnamefont
  {P.}~\bibnamefont {Boulanger}}, \bibinfo {author} {\bibfnamefont
  {F.}~\bibnamefont {Bruneval}}, \bibinfo {author} {\bibfnamefont
  {D.}~\bibnamefont {Caliste}}, \bibinfo {author} {\bibfnamefont
  {R.}~\bibnamefont {Caracas}}, \bibinfo {author} {\bibfnamefont
  {M.}~\bibnamefont {C{\^o}t{\'e}}},  \emph {et~al.},\ }\href {\doibase
  10.1016/j.cpc.2009.07.007} {\bibfield  {journal} {\bibinfo  {journal}
  {Comput.~Phys.~Commun.}\ }\textbf {\bibinfo {volume} {180}},\ \bibinfo
  {pages} {2582} (\bibinfo {year} {2009})}\BibitemShut {NoStop}%
\bibitem [{\citenamefont {Eiguren}\ and\ \citenamefont
  {Ambrosch-Draxl}(2008)}]{Eiguren2008}%
  \BibitemOpen
  \bibfield  {author} {\bibinfo {author} {\bibfnamefont {A.}~\bibnamefont
  {Eiguren}}\ and\ \bibinfo {author} {\bibfnamefont {C.}~\bibnamefont
  {Ambrosch-Draxl}},\ }\href {\doibase 10.1103/PhysRevB.78.045124} {\bibfield
  {journal} {\bibinfo  {journal} {Phys.~Rev.~B}\ }\textbf {\bibinfo {volume}
  {78}},\ \bibinfo {pages} {045124} (\bibinfo {year} {2008})}\BibitemShut
  {NoStop}%
\bibitem [{\citenamefont {Brunin}\ \emph {et~al.}(2020)\citenamefont {Brunin},
  \citenamefont {Miranda}, \citenamefont {Giantomassi}, \citenamefont {Royo},
  \citenamefont {Stengel}, \citenamefont {Verstraete}, \citenamefont {Gonze},
  \citenamefont {Rignanese},\ and\ \citenamefont {Hautier}}]{Brunin2019prl}%
  \BibitemOpen
  \bibfield  {author} {\bibinfo {author} {\bibfnamefont {G.}~\bibnamefont
  {Brunin}}, \bibinfo {author} {\bibfnamefont {H.~P.~C.}\ \bibnamefont
  {Miranda}}, \bibinfo {author} {\bibfnamefont {M.}~\bibnamefont
  {Giantomassi}}, \bibinfo {author} {\bibfnamefont {M.}~\bibnamefont {Royo}},
  \bibinfo {author} {\bibfnamefont {M.}~\bibnamefont {Stengel}}, \bibinfo
  {author} {\bibfnamefont {M.~J.}\ \bibnamefont {Verstraete}}, \bibinfo
  {author} {\bibfnamefont {X.}~\bibnamefont {Gonze}}, \bibinfo {author}
  {\bibfnamefont {G.-M.}\ \bibnamefont {Rignanese}}, \ and\ \bibinfo {author}
  {\bibfnamefont {G.}~\bibnamefont {Hautier}},\ }\href@noop {} {\bibfield
  {journal} {\bibinfo  {journal} {Unknown}\ }\textbf {\bibinfo {volume} {X}},\
  \bibinfo {pages} {XYZ} (\bibinfo {year} {2020})}\BibitemShut {NoStop}%
\bibitem [{\citenamefont {Mahan}(2014)}]{Mahan2014}%
  \BibitemOpen
  \bibfield  {author} {\bibinfo {author} {\bibfnamefont {G.~D.}\ \bibnamefont
  {Mahan}},\ }\href {\doibase https://doi.org/10.1007/978-1-4757-5714-9} {\emph
  {\bibinfo {title} {Many-Particle Physics}}}\ (\bibinfo  {publisher}
  {Springer},\ \bibinfo {year} {2014})\BibitemShut {NoStop}%
\bibitem [{\citenamefont {Marini}\ \emph {et~al.}(2015)\citenamefont {Marini},
  \citenamefont {Ponc\'e},\ and\ \citenamefont {Gonze}}]{Marini2015}%
  \BibitemOpen
  \bibfield  {author} {\bibinfo {author} {\bibfnamefont {A.}~\bibnamefont
  {Marini}}, \bibinfo {author} {\bibfnamefont {S.}~\bibnamefont {Ponc\'e}}, \
  and\ \bibinfo {author} {\bibfnamefont {X.}~\bibnamefont {Gonze}},\ }\href
  {\doibase 10.1103/PhysRevB.91.224310} {\bibfield  {journal} {\bibinfo
  {journal} {Phys.~Rev.~B}\ }\textbf {\bibinfo {volume} {91}},\ \bibinfo
  {pages} {224310} (\bibinfo {year} {2015})}\BibitemShut {NoStop}%
\bibitem [{\citenamefont {Gonze}\ and\ \citenamefont {Lee}(1997)}]{Gonze1997}%
  \BibitemOpen
  \bibfield  {author} {\bibinfo {author} {\bibfnamefont {X.}~\bibnamefont
  {Gonze}}\ and\ \bibinfo {author} {\bibfnamefont {C.}~\bibnamefont {Lee}},\
  }\href {\doibase 10.1103/PhysRevB.55.10355} {\bibfield  {journal} {\bibinfo
  {journal} {Phys.~Rev.~B}\ }\textbf {\bibinfo {volume} {55}},\ \bibinfo
  {pages} {10355} (\bibinfo {year} {1997})}\BibitemShut {NoStop}%
\bibitem [{\citenamefont {Baroni}\ \emph {et~al.}(2001)\citenamefont {Baroni},
  \citenamefont {de~Gironcoli}, \citenamefont {Dal~Corso},\ and\ \citenamefont
  {Giannozzi}}]{Baroni2001}%
  \BibitemOpen
  \bibfield  {author} {\bibinfo {author} {\bibfnamefont {S.}~\bibnamefont
  {Baroni}}, \bibinfo {author} {\bibfnamefont {S.}~\bibnamefont
  {de~Gironcoli}}, \bibinfo {author} {\bibfnamefont {A.}~\bibnamefont
  {Dal~Corso}}, \ and\ \bibinfo {author} {\bibfnamefont {P.}~\bibnamefont
  {Giannozzi}},\ }\href {\doibase 10.1103/revmodphys.73.515} {\bibfield
  {journal} {\bibinfo  {journal} {Rev.~Mod.~Phys.}\ }\textbf {\bibinfo {volume}
  {73}},\ \bibinfo {pages} {515} (\bibinfo {year} {2001})}\BibitemShut
  {NoStop}%
\bibitem [{Note2()}]{Note2}%
  \BibitemOpen
  \bibinfo {note} {We assume norm-conserving pseudopotentials.}\BibitemShut
  {Stop}%
\bibitem [{\citenamefont {Payne}\ \emph {et~al.}(1992)\citenamefont {Payne},
  \citenamefont {Teter}, \citenamefont {Allan}, \citenamefont {Arias},\ and\
  \citenamefont {Joannopoulos}}]{Payne1992}%
  \BibitemOpen
  \bibfield  {author} {\bibinfo {author} {\bibfnamefont {M.~C.}\ \bibnamefont
  {Payne}}, \bibinfo {author} {\bibfnamefont {M.~P.}\ \bibnamefont {Teter}},
  \bibinfo {author} {\bibfnamefont {D.~C.}\ \bibnamefont {Allan}}, \bibinfo
  {author} {\bibfnamefont {T.~A.}\ \bibnamefont {Arias}}, \ and\ \bibinfo
  {author} {\bibfnamefont {J.~D.}\ \bibnamefont {Joannopoulos}},\ }\href
  {\doibase 10.1103/RevModPhys.64.1045} {\bibfield  {journal} {\bibinfo
  {journal} {Rev.~Mod.~Phys.}\ }\textbf {\bibinfo {volume} {64}},\ \bibinfo
  {pages} {1045} (\bibinfo {year} {1992})}\BibitemShut {NoStop}%
\bibitem [{\citenamefont {Giustino}\ \emph {et~al.}(2007)\citenamefont
  {Giustino}, \citenamefont {Cohen},\ and\ \citenamefont
  {Louie}}]{Giustino2007}%
  \BibitemOpen
  \bibfield  {author} {\bibinfo {author} {\bibfnamefont {F.}~\bibnamefont
  {Giustino}}, \bibinfo {author} {\bibfnamefont {M.~L.}\ \bibnamefont {Cohen}},
  \ and\ \bibinfo {author} {\bibfnamefont {S.~G.}\ \bibnamefont {Louie}},\
  }\href {\doibase 10.1103/PhysRevB.76.165108} {\bibfield  {journal} {\bibinfo
  {journal} {Phys.~Rev.~B}\ }\textbf {\bibinfo {volume} {76}},\ \bibinfo
  {pages} {165108} (\bibinfo {year} {2007})}\BibitemShut {NoStop}%
\bibitem [{\citenamefont {Kleinman}\ and\ \citenamefont
  {Bylander}(1982)}]{Kleinman1982}%
  \BibitemOpen
  \bibfield  {author} {\bibinfo {author} {\bibfnamefont {L.}~\bibnamefont
  {Kleinman}}\ and\ \bibinfo {author} {\bibfnamefont {D.~M.}\ \bibnamefont
  {Bylander}},\ }\href {\doibase 10.1103/PhysRevLett.48.1425} {\bibfield
  {journal} {\bibinfo  {journal} {Phys.~Rev.~Lett.}\ }\textbf {\bibinfo
  {volume} {48}},\ \bibinfo {pages} {1425} (\bibinfo {year}
  {1982})}\BibitemShut {NoStop}%
\bibitem [{\citenamefont {Kohn}(1959)}]{Kohn1959}%
  \BibitemOpen
  \bibfield  {author} {\bibinfo {author} {\bibfnamefont {W.}~\bibnamefont
  {Kohn}},\ }\href {\doibase 10.1103/PhysRevLett.2.393} {\bibfield  {journal}
  {\bibinfo  {journal} {Phys.~Rev.~Lett.}\ }\textbf {\bibinfo {volume} {2}},\
  \bibinfo {pages} {393} (\bibinfo {year} {1959})}\BibitemShut {NoStop}%
\bibitem [{\citenamefont {Vogl}(1976)}]{Vogl1976}%
  \BibitemOpen
  \bibfield  {author} {\bibinfo {author} {\bibfnamefont {P.}~\bibnamefont
  {Vogl}},\ }\href {\doibase 10.1103/PhysRevB.13.694} {\bibfield  {journal}
  {\bibinfo  {journal} {Phys.~Rev.~B}\ }\textbf {\bibinfo {volume} {13}},\
  \bibinfo {pages} {694} (\bibinfo {year} {1976})}\BibitemShut {NoStop}%
\bibitem [{\citenamefont {Stengel}(2013)}]{Stengel2013}%
  \BibitemOpen
  \bibfield  {author} {\bibinfo {author} {\bibfnamefont {M.}~\bibnamefont
  {Stengel}},\ }\href {\doibase 10.1103/PhysRevB.88.174106} {\bibfield
  {journal} {\bibinfo  {journal} {Phys.~Rev.~B}\ }\textbf {\bibinfo {volume}
  {88}},\ \bibinfo {pages} {174106} (\bibinfo {year} {2013})}\BibitemShut
  {NoStop}%
\bibitem [{\citenamefont {Verdi}\ and\ \citenamefont
  {Giustino}(2015)}]{Verdi2015}%
  \BibitemOpen
  \bibfield  {author} {\bibinfo {author} {\bibfnamefont {C.}~\bibnamefont
  {Verdi}}\ and\ \bibinfo {author} {\bibfnamefont {F.}~\bibnamefont
  {Giustino}},\ }\href {\doibase 10.1103/PhysRevLett.115.176401} {\bibfield
  {journal} {\bibinfo  {journal} {Phys.~Rev.~Lett.}\ }\textbf {\bibinfo
  {volume} {115}},\ \bibinfo {pages} {176401} (\bibinfo {year}
  {2015})}\BibitemShut {NoStop}%
\bibitem [{\citenamefont {Royo}\ and\ \citenamefont
  {Stengel}(2019)}]{Royo2019}%
  \BibitemOpen
  \bibfield  {author} {\bibinfo {author} {\bibfnamefont {M.}~\bibnamefont
  {Royo}}\ and\ \bibinfo {author} {\bibfnamefont {M.}~\bibnamefont {Stengel}},\
  }\href {\doibase 10.1103/PhysRevX.9.021050} {\bibfield  {journal} {\bibinfo
  {journal} {Phys.~Rev.~X}\ }\textbf {\bibinfo {volume} {9}},\ \bibinfo {pages}
  {021050} (\bibinfo {year} {2019})}\BibitemShut {NoStop}%
\bibitem [{\citenamefont {Sjakste}\ \emph {et~al.}(2015)\citenamefont
  {Sjakste}, \citenamefont {Vast}, \citenamefont {Calandra},\ and\
  \citenamefont {Mauri}}]{Sjakste2015}%
  \BibitemOpen
  \bibfield  {author} {\bibinfo {author} {\bibfnamefont {J.}~\bibnamefont
  {Sjakste}}, \bibinfo {author} {\bibfnamefont {N.}~\bibnamefont {Vast}},
  \bibinfo {author} {\bibfnamefont {M.}~\bibnamefont {Calandra}}, \ and\
  \bibinfo {author} {\bibfnamefont {F.}~\bibnamefont {Mauri}},\ }\href
  {\doibase 10.1103/PhysRevB.92.054307} {\bibfield  {journal} {\bibinfo
  {journal} {Phys.~Rev.~B}\ }\textbf {\bibinfo {volume} {92}},\ \bibinfo
  {pages} {054307} (\bibinfo {year} {2015})}\BibitemShut {NoStop}%
\bibitem [{Note3()}]{Note3}%
  \BibitemOpen
  \bibinfo {note} {The $\alpha $ parameter determines the separation between
  the long-range and the short-range parts of the interaction and is used to
  express Ewald sums~\cite {Pickett1989,Gonze1997} in terms of a sum in
  ${\protect \bf G}$ space (long-range part) and a sum in real space that,
  being short ranged, is not relevant for the definition of the LR model. The
  optimal value of $\alpha $ is material-dependent and should therefore be
  subject to convergence studies. In our systems, we observed small changes in
  the physical observables ($\sim $1\%) with $\alpha $. A value of
  $0.1$~Bohr$^{-2}$ is used in all calculations.}\BibitemShut {Stop}%
\bibitem [{\citenamefont {Sangalli}\ \emph {et~al.}(2019)\citenamefont
  {Sangalli}, \citenamefont {Ferretti}, \citenamefont {Miranda}, \citenamefont
  {Attaccalite}, \citenamefont {Marri}, \citenamefont {Cannuccia},
  \citenamefont {Melo}, \citenamefont {Marsili}, \citenamefont {Paleari},
  \citenamefont {Marrazzo} \emph {et~al.}}]{Sangalli2019}%
  \BibitemOpen
  \bibfield  {author} {\bibinfo {author} {\bibfnamefont {D.}~\bibnamefont
  {Sangalli}}, \bibinfo {author} {\bibfnamefont {A.}~\bibnamefont {Ferretti}},
  \bibinfo {author} {\bibfnamefont {H.}~\bibnamefont {Miranda}}, \bibinfo
  {author} {\bibfnamefont {C.}~\bibnamefont {Attaccalite}}, \bibinfo {author}
  {\bibfnamefont {I.}~\bibnamefont {Marri}}, \bibinfo {author} {\bibfnamefont
  {E.}~\bibnamefont {Cannuccia}}, \bibinfo {author} {\bibfnamefont
  {P.}~\bibnamefont {Melo}}, \bibinfo {author} {\bibfnamefont {M.}~\bibnamefont
  {Marsili}}, \bibinfo {author} {\bibfnamefont {F.}~\bibnamefont {Paleari}},
  \bibinfo {author} {\bibfnamefont {A.}~\bibnamefont {Marrazzo}},  \emph
  {et~al.},\ }\href {\doibase https://doi.org/10.1088/1361-648X/ab15d0}
  {\bibfield  {journal} {\bibinfo  {journal} {J.~Phys.~Condens.~Matter}\
  }\textbf {\bibinfo {volume} {31}},\ \bibinfo {pages} {325902} (\bibinfo
  {year} {2019})}\BibitemShut {NoStop}%
\bibitem [{\citenamefont {Madsen}\ \emph {et~al.}(2018)\citenamefont {Madsen},
  \citenamefont {Carrete},\ and\ \citenamefont {Verstraete}}]{Madsen2018}%
  \BibitemOpen
  \bibfield  {author} {\bibinfo {author} {\bibfnamefont {G.~K.~H.}\
  \bibnamefont {Madsen}}, \bibinfo {author} {\bibfnamefont {J.}~\bibnamefont
  {Carrete}}, \ and\ \bibinfo {author} {\bibfnamefont {M.~J.}\ \bibnamefont
  {Verstraete}},\ }\href {\doibase 10.1016/j.cpc.2018.05.010} {\bibfield
  {journal} {\bibinfo  {journal} {Comput.~Phys.~Commun.}\ }\textbf {\bibinfo
  {volume} {231}},\ \bibinfo {pages} {140} (\bibinfo {year}
  {2018})}\BibitemShut {NoStop}%
\bibitem [{Note4()}]{Note4}%
  \BibitemOpen
  \bibinfo {note} {Scattering by defects, impurities, grain boundaries, or
  other electrons can be described either using semi-empirical models~\cite
  {Calnan2010,Liu2017} or first-principles computations~\cite
  {Bernardi2014,Lu2019}. The computation of these effects is however outside
  the scope of the present work.}\BibitemShut {Stop}%
\bibitem [{\citenamefont {Ashcroft}\ and\ \citenamefont
  {Mermin}(1976)}]{Ashcroft1976}%
  \BibitemOpen
  \bibfield  {author} {\bibinfo {author} {\bibfnamefont {N.~W.}\ \bibnamefont
  {Ashcroft}}\ and\ \bibinfo {author} {\bibfnamefont {N.~D.}\ \bibnamefont
  {Mermin}},\ }\href@noop {} {\emph {\bibinfo {title} {{Solid state
  physics}}}},\ \bibinfo {edition} {1st}\ ed.\ (\bibinfo  {publisher} {Saunders
  College},\ \bibinfo {year} {1976})\BibitemShut {NoStop}%
\bibitem [{\citenamefont {Pizzi}\ \emph {et~al.}(2014)\citenamefont {Pizzi},
  \citenamefont {Volja}, \citenamefont {Kozinsky}, \citenamefont {Fornari},\
  and\ \citenamefont {Marzari}}]{Pizzi2014}%
  \BibitemOpen
  \bibfield  {author} {\bibinfo {author} {\bibfnamefont {G.}~\bibnamefont
  {Pizzi}}, \bibinfo {author} {\bibfnamefont {D.}~\bibnamefont {Volja}},
  \bibinfo {author} {\bibfnamefont {B.}~\bibnamefont {Kozinsky}}, \bibinfo
  {author} {\bibfnamefont {M.}~\bibnamefont {Fornari}}, \ and\ \bibinfo
  {author} {\bibfnamefont {N.}~\bibnamefont {Marzari}},\ }\href {\doibase
  10.1016/j.cpc.2013.09.015} {\bibfield  {journal} {\bibinfo  {journal}
  {Comput.~Phys.~Commun.}\ }\textbf {\bibinfo {volume} {185}},\ \bibinfo
  {pages} {422} (\bibinfo {year} {2014})}\BibitemShut {NoStop}%
\bibitem [{\citenamefont {Shankland}(1971)}]{Shankland1971}%
  \BibitemOpen
  \bibfield  {author} {\bibinfo {author} {\bibfnamefont {D.~G.}\ \bibnamefont
  {Shankland}},\ }\href {\doibase 10.1002/qua.560050857} {\bibfield  {journal}
  {\bibinfo  {journal} {Int.~J.~Quantum Chem.}\ }\textbf {\bibinfo {volume}
  {5}},\ \bibinfo {pages} {497} (\bibinfo {year} {1971})}\BibitemShut {NoStop}%
\bibitem [{\citenamefont {Euwema}\ \emph {et~al.}(1969)\citenamefont {Euwema},
  \citenamefont {Stukel}, \citenamefont {Collins}, \citenamefont {DeWitt},\
  and\ \citenamefont {Shankland}}]{Euwema1969}%
  \BibitemOpen
  \bibfield  {author} {\bibinfo {author} {\bibfnamefont {R.~N.}\ \bibnamefont
  {Euwema}}, \bibinfo {author} {\bibfnamefont {D.~J.}\ \bibnamefont {Stukel}},
  \bibinfo {author} {\bibfnamefont {T.~C.}\ \bibnamefont {Collins}}, \bibinfo
  {author} {\bibfnamefont {J.~S.}\ \bibnamefont {DeWitt}}, \ and\ \bibinfo
  {author} {\bibfnamefont {D.~G.}\ \bibnamefont {Shankland}},\ }\href {\doibase
  10.1103/PhysRev.178.1419} {\bibfield  {journal} {\bibinfo  {journal}
  {Phys.~Rev.}\ }\textbf {\bibinfo {volume} {178}},\ \bibinfo {pages} {1419}
  (\bibinfo {year} {1969})}\BibitemShut {NoStop}%
\bibitem [{\citenamefont {Koelling}\ and\ \citenamefont
  {Wood}(1986)}]{Koelling1986}%
  \BibitemOpen
  \bibfield  {author} {\bibinfo {author} {\bibfnamefont {D.~D.}\ \bibnamefont
  {Koelling}}\ and\ \bibinfo {author} {\bibfnamefont {J.~H.}\ \bibnamefont
  {Wood}},\ }\href {\doibase 10.1016/0021-9991(86)90261-5} {\bibfield
  {journal} {\bibinfo  {journal} {J.~Comput.~Phys.}\ }\textbf {\bibinfo
  {volume} {67}},\ \bibinfo {pages} {253} (\bibinfo {year} {1986})}\BibitemShut
  {NoStop}%
\bibitem [{\citenamefont {Madsen}\ and\ \citenamefont
  {Singh}(2006)}]{Madsen2006}%
  \BibitemOpen
  \bibfield  {author} {\bibinfo {author} {\bibfnamefont {G.~K.~H.}\
  \bibnamefont {Madsen}}\ and\ \bibinfo {author} {\bibfnamefont {D.~J.}\
  \bibnamefont {Singh}},\ }\href {\doibase 10.1016/j.cpc.2006.03.007}
  {\bibfield  {journal} {\bibinfo  {journal} {Comput.~Phys.~Commun.}\ }\textbf
  {\bibinfo {volume} {175}},\ \bibinfo {pages} {67} (\bibinfo {year}
  {2006})}\BibitemShut {NoStop}%
\bibitem [{\citenamefont {Del~Sole}\ and\ \citenamefont
  {Girlanda}(1993)}]{DelSole1993}%
  \BibitemOpen
  \bibfield  {author} {\bibinfo {author} {\bibfnamefont {R.}~\bibnamefont
  {Del~Sole}}\ and\ \bibinfo {author} {\bibfnamefont {R.}~\bibnamefont
  {Girlanda}},\ }\href {\doibase 10.1103/PhysRevB.48.11789} {\bibfield
  {journal} {\bibinfo  {journal} {Phys.~Rev.~B}\ }\textbf {\bibinfo {volume}
  {48}},\ \bibinfo {pages} {11789} (\bibinfo {year} {1993})}\BibitemShut
  {NoStop}%
\bibitem [{Note5()}]{Note5}%
  \BibitemOpen
  \bibinfo {note} {We consider the conductivity related to the electric field
  only, not the Hall mobility~\cite {Ricci2017}.}\BibitemShut {Stop}%
\bibitem [{\citenamefont {Bl{\"o}chl}\ \emph {et~al.}(1994)\citenamefont
  {Bl{\"o}chl}, \citenamefont {Jepsen},\ and\ \citenamefont
  {Andersen}}]{Blochl1994}%
  \BibitemOpen
  \bibfield  {author} {\bibinfo {author} {\bibfnamefont {P.~E.}\ \bibnamefont
  {Bl{\"o}chl}}, \bibinfo {author} {\bibfnamefont {O.}~\bibnamefont {Jepsen}},
  \ and\ \bibinfo {author} {\bibfnamefont {O.~K.}\ \bibnamefont {Andersen}},\
  }\href {\doibase 10.1103/PhysRevB.49.16223} {\bibfield  {journal} {\bibinfo
  {journal} {Phys.~Rev.~B}\ }\textbf {\bibinfo {volume} {49}},\ \bibinfo
  {pages} {16223} (\bibinfo {year} {1994})}\BibitemShut {NoStop}%
\bibitem [{\citenamefont {Sohier}\ \emph {et~al.}(2018)\citenamefont {Sohier},
  \citenamefont {Campi}, \citenamefont {Marzari},\ and\ \citenamefont
  {Gibertini}}]{Sohier2018}%
  \BibitemOpen
  \bibfield  {author} {\bibinfo {author} {\bibfnamefont {T.}~\bibnamefont
  {Sohier}}, \bibinfo {author} {\bibfnamefont {D.}~\bibnamefont {Campi}},
  \bibinfo {author} {\bibfnamefont {N.}~\bibnamefont {Marzari}}, \ and\
  \bibinfo {author} {\bibfnamefont {M.}~\bibnamefont {Gibertini}},\ }\href
  {\doibase 10.1103/PhysRevMaterials.2.114010} {\bibfield  {journal} {\bibinfo
  {journal} {Phys.~Rev.~Mat.}\ }\textbf {\bibinfo {volume} {2}},\ \bibinfo
  {pages} {114010} (\bibinfo {year} {2018})}\BibitemShut {NoStop}%
\bibitem [{\citenamefont {Liu}\ \emph {et~al.}(2017)\citenamefont {Liu},
  \citenamefont {Zhou}, \citenamefont {Liao}, \citenamefont {Singh},\ and\
  \citenamefont {Chen}}]{Liu2017}%
  \BibitemOpen
  \bibfield  {author} {\bibinfo {author} {\bibfnamefont {T.-H.}\ \bibnamefont
  {Liu}}, \bibinfo {author} {\bibfnamefont {J.}~\bibnamefont {Zhou}}, \bibinfo
  {author} {\bibfnamefont {B.}~\bibnamefont {Liao}}, \bibinfo {author}
  {\bibfnamefont {D.~J.}\ \bibnamefont {Singh}}, \ and\ \bibinfo {author}
  {\bibfnamefont {G.}~\bibnamefont {Chen}},\ }\href {\doibase
  10.1103/PhysRevB.95.075206} {\bibfield  {journal} {\bibinfo  {journal}
  {Phys.~Rev.~B}\ }\textbf {\bibinfo {volume} {95}},\ \bibinfo {pages} {075206}
  (\bibinfo {year} {2017})}\BibitemShut {NoStop}%
\bibitem [{Note6()}]{Note6}%
  \BibitemOpen
  \bibinfo {note} {The set of $\protect \mathbf {k}$ points obtained with this
  procedure belongs to a homogeneous mesh hence both symmetries and tetrahedron
  method can be used without any modification. Small errors in the tetrahedron
  weights may be introduced by the linear interpolation because $\protect
  \mathbf {k}$ points that are slightly outside of the energy window are
  interpolated with SKW but this minor issue can be easily fixed by enlarging
  the energy window. This techniques permits to reach very high resolution in
  $\protect \mathbf {k}$ space of the order of ${300}\times {300}\times
  {300}$.}\BibitemShut {Stop}%
\bibitem [{Note7()}]{Note7}%
  \BibitemOpen
  \bibinfo {note} {In principle a similar optimization can be implemented for
  the Lorentzian and Gaussian broadening techniques, however this requires
  choosing a threshold beyond which the broadened $\delta $ is treated as
  zero.}\BibitemShut {Stop}%
\bibitem [{\citenamefont {Kammerlander}\ \emph {et~al.}(2012)\citenamefont
  {Kammerlander}, \citenamefont {Botti}, \citenamefont {Marques}, \citenamefont
  {Marini},\ and\ \citenamefont {Attaccalite}}]{Kammerlander2012}%
  \BibitemOpen
  \bibfield  {author} {\bibinfo {author} {\bibfnamefont {D.}~\bibnamefont
  {Kammerlander}}, \bibinfo {author} {\bibfnamefont {S.}~\bibnamefont {Botti}},
  \bibinfo {author} {\bibfnamefont {M.~A.~L.}\ \bibnamefont {Marques}},
  \bibinfo {author} {\bibfnamefont {A.}~\bibnamefont {Marini}}, \ and\ \bibinfo
  {author} {\bibfnamefont {C.}~\bibnamefont {Attaccalite}},\ }\href {\doibase
  10.1103/PhysRevB.86.125203} {\bibfield  {journal} {\bibinfo  {journal}
  {Phys.~Rev.~B}\ }\textbf {\bibinfo {volume} {86}},\ \bibinfo {pages} {125203}
  (\bibinfo {year} {2012})}\BibitemShut {NoStop}%
\bibitem [{\citenamefont {Gillet}\ \emph {et~al.}(2016)\citenamefont {Gillet},
  \citenamefont {Giantomassi},\ and\ \citenamefont {Gonze}}]{Gillet2016}%
  \BibitemOpen
  \bibfield  {author} {\bibinfo {author} {\bibfnamefont {Y.}~\bibnamefont
  {Gillet}}, \bibinfo {author} {\bibfnamefont {M.}~\bibnamefont {Giantomassi}},
  \ and\ \bibinfo {author} {\bibfnamefont {X.}~\bibnamefont {Gonze}},\ }\href
  {\doibase 10.1016/j.cpc.2016.02.008} {\bibfield  {journal} {\bibinfo
  {journal} {Comput.~Phys.~Commun.}\ }\textbf {\bibinfo {volume} {203}},\
  \bibinfo {pages} {83} (\bibinfo {year} {2016})}\BibitemShut {NoStop}%
\bibitem [{\citenamefont {Fiorentini}\ and\ \citenamefont
  {Bonini}(2016)}]{Fiorentini2016}%
  \BibitemOpen
  \bibfield  {author} {\bibinfo {author} {\bibfnamefont {M.}~\bibnamefont
  {Fiorentini}}\ and\ \bibinfo {author} {\bibfnamefont {N.}~\bibnamefont
  {Bonini}},\ }\href {\doibase https://doi.org/10.1103/PhysRevB.94.085204}
  {\bibfield  {journal} {\bibinfo  {journal} {Phys.~Rev.~B}\ }\textbf {\bibinfo
  {volume} {94}},\ \bibinfo {pages} {085204} (\bibinfo {year}
  {2016})}\BibitemShut {NoStop}%
\bibitem [{\citenamefont {Troullier}\ and\ \citenamefont
  {Martins}(1991)}]{Troullier1991}%
  \BibitemOpen
  \bibfield  {author} {\bibinfo {author} {\bibfnamefont {N.}~\bibnamefont
  {Troullier}}\ and\ \bibinfo {author} {\bibfnamefont {J.~L.}\ \bibnamefont
  {Martins}},\ }\href {\doibase https://doi.org/10.1103/PhysRevB.43.1993}
  {\bibfield  {journal} {\bibinfo  {journal} {Phys.~Rev.~B}\ }\textbf {\bibinfo
  {volume} {43}},\ \bibinfo {pages} {1993} (\bibinfo {year}
  {1991})}\BibitemShut {NoStop}%
\bibitem [{\citenamefont {Perdew}\ and\ \citenamefont
  {Wang}(1992)}]{Perdew1992}%
  \BibitemOpen
  \bibfield  {author} {\bibinfo {author} {\bibfnamefont {J.~P.}\ \bibnamefont
  {Perdew}}\ and\ \bibinfo {author} {\bibfnamefont {Y.}~\bibnamefont {Wang}},\
  }\href {\doibase https://doi.org/10.1103/PhysRevB.45.13244} {\bibfield
  {journal} {\bibinfo  {journal} {Phys.~Rev.~B}\ }\textbf {\bibinfo {volume}
  {45}},\ \bibinfo {pages} {13244} (\bibinfo {year} {1992})}\BibitemShut
  {NoStop}%
\bibitem [{\citenamefont {Ceperley}\ and\ \citenamefont
  {Alder}(1980)}]{Ceperley1980}%
  \BibitemOpen
  \bibfield  {author} {\bibinfo {author} {\bibfnamefont {D.~M.}\ \bibnamefont
  {Ceperley}}\ and\ \bibinfo {author} {\bibfnamefont {B.}~\bibnamefont
  {Alder}},\ }\href {\doibase https://doi.org/10.1103/PhysRevLett.45.566}
  {\bibfield  {journal} {\bibinfo  {journal} {Phys.~Rev.~Lett.}\ }\textbf
  {\bibinfo {volume} {45}},\ \bibinfo {pages} {566} (\bibinfo {year}
  {1980})}\BibitemShut {NoStop}%
\bibitem [{\citenamefont {{See Supplemental Material for
  more information about computational details and
  results}}()}]{supplemental_prb}%
  \BibitemOpen
  \bibfield  {author} {\bibinfo {author} {\bibnamefont {{See Supplemental
  Material for more information about computational
  details and results}}},\ }\href@noop {} {}\BibitemShut {NoStop}%
\bibitem [{\citenamefont {Giantomassi}\ and\ \citenamefont {\textit{et
  al.}}()}]{abipy-website}%
  \BibitemOpen
  \bibfield  {author} {\bibinfo {author} {\bibfnamefont {M.}~\bibnamefont
  {Giantomassi}}\ and\ \bibinfo {author} {\bibnamefont {\textit{et al.}}},\
  }\href@noop {} {\enquote {\bibinfo {title} {Abipy project},}\ }\bibinfo
  {howpublished} {\url {https://github.com/abinit/abipy}}\BibitemShut {NoStop}%
\bibitem [{Note8()}]{Note8}%
  \BibitemOpen
  \bibinfo {note} {In all the calculations using the DG method, the KS
  eigenvalues on the fine $\protect \mathbf {k}$ mesh have been computed
  exactly by performing a NSCF computation.}\BibitemShut {Stop}%
\bibitem [{\citenamefont {Hautier}\ \emph {et~al.}(2013)\citenamefont
  {Hautier}, \citenamefont {Miglio}, \citenamefont {Ceder}, \citenamefont
  {Rignanese},\ and\ \citenamefont {Gonze}}]{Hautier2013}%
  \BibitemOpen
  \bibfield  {author} {\bibinfo {author} {\bibfnamefont {G.}~\bibnamefont
  {Hautier}}, \bibinfo {author} {\bibfnamefont {A.}~\bibnamefont {Miglio}},
  \bibinfo {author} {\bibfnamefont {G.}~\bibnamefont {Ceder}}, \bibinfo
  {author} {\bibfnamefont {G.~M.}\ \bibnamefont {Rignanese}}, \ and\ \bibinfo
  {author} {\bibfnamefont {X.}~\bibnamefont {Gonze}},\ }\href {\doibase
  10.1038/ncomms3292} {\bibfield  {journal} {\bibinfo  {journal}
  {Nat.~Commun.}\ }\textbf {\bibinfo {volume} {4}} (\bibinfo {year} {2013}),\
  10.1038/ncomms3292}\BibitemShut {NoStop}%
\bibitem [{\citenamefont {Bhatia}\ \emph {et~al.}(2016)\citenamefont {Bhatia},
  \citenamefont {Hautier}, \citenamefont {Nilgianskul}, \citenamefont {Miglio},
  \citenamefont {Sun}, \citenamefont {Kim}, \citenamefont {Kim}, \citenamefont
  {Chen}, \citenamefont {Rignanese}, \citenamefont {Gonze},\ and\ \citenamefont
  {Suntivich}}]{Bhatia2016}%
  \BibitemOpen
  \bibfield  {author} {\bibinfo {author} {\bibfnamefont {A.}~\bibnamefont
  {Bhatia}}, \bibinfo {author} {\bibfnamefont {G.}~\bibnamefont {Hautier}},
  \bibinfo {author} {\bibfnamefont {T.}~\bibnamefont {Nilgianskul}}, \bibinfo
  {author} {\bibfnamefont {A.}~\bibnamefont {Miglio}}, \bibinfo {author}
  {\bibfnamefont {J.}~\bibnamefont {Sun}}, \bibinfo {author} {\bibfnamefont
  {H.~J.}\ \bibnamefont {Kim}}, \bibinfo {author} {\bibfnamefont {K.~H.}\
  \bibnamefont {Kim}}, \bibinfo {author} {\bibfnamefont {S.}~\bibnamefont
  {Chen}}, \bibinfo {author} {\bibfnamefont {G.~M.}\ \bibnamefont {Rignanese}},
  \bibinfo {author} {\bibfnamefont {X.}~\bibnamefont {Gonze}}, \ and\ \bibinfo
  {author} {\bibfnamefont {J.}~\bibnamefont {Suntivich}},\ }\href {\doibase
  10.1021/acs.chemmater.5b03794} {\bibfield  {journal} {\bibinfo  {journal}
  {Chem.~Mater.}\ }\textbf {\bibinfo {volume} {28}},\ \bibinfo {pages} {30}
  (\bibinfo {year} {2016})}\BibitemShut {NoStop}%
\bibitem [{\citenamefont {Ricci}\ \emph {et~al.}(2017)\citenamefont {Ricci},
  \citenamefont {Chen}, \citenamefont {Aydemir}, \citenamefont {Snyder},
  \citenamefont {Rignanese}, \citenamefont {Jain},\ and\ \citenamefont
  {Hautier}}]{Ricci2017}%
  \BibitemOpen
  \bibfield  {author} {\bibinfo {author} {\bibfnamefont {F.}~\bibnamefont
  {Ricci}}, \bibinfo {author} {\bibfnamefont {W.}~\bibnamefont {Chen}},
  \bibinfo {author} {\bibfnamefont {U.}~\bibnamefont {Aydemir}}, \bibinfo
  {author} {\bibfnamefont {G.~J.}\ \bibnamefont {Snyder}}, \bibinfo {author}
  {\bibfnamefont {G.-M.}\ \bibnamefont {Rignanese}}, \bibinfo {author}
  {\bibfnamefont {A.}~\bibnamefont {Jain}}, \ and\ \bibinfo {author}
  {\bibfnamefont {G.}~\bibnamefont {Hautier}},\ }\href {\doibase
  10.1038/sdata.2017.85} {\bibfield  {journal} {\bibinfo  {journal}
  {Sci.~Data.}\ }\textbf {\bibinfo {volume} {4}},\ \bibinfo {pages} {170085}
  (\bibinfo {year} {2017})}\BibitemShut {NoStop}%
\bibitem [{\citenamefont {Darnon}\ \emph {et~al.}(2015)\citenamefont {Darnon},
  \citenamefont {Varache}, \citenamefont {Descazeaux}, \citenamefont {Quinci},
  \citenamefont {Martin}, \citenamefont {Baron},\ and\ \citenamefont
  {Mu{\~n}oz}}]{Darnon2015}%
  \BibitemOpen
  \bibfield  {author} {\bibinfo {author} {\bibfnamefont {M.}~\bibnamefont
  {Darnon}}, \bibinfo {author} {\bibfnamefont {R.}~\bibnamefont {Varache}},
  \bibinfo {author} {\bibfnamefont {M.}~\bibnamefont {Descazeaux}}, \bibinfo
  {author} {\bibfnamefont {T.}~\bibnamefont {Quinci}}, \bibinfo {author}
  {\bibfnamefont {M.}~\bibnamefont {Martin}}, \bibinfo {author} {\bibfnamefont
  {T.}~\bibnamefont {Baron}}, \ and\ \bibinfo {author} {\bibfnamefont
  {D.}~\bibnamefont {Mu{\~n}oz}},\ }\href {\doibase 10.1063/1.4931514}
  {\bibfield  {journal} {\bibinfo  {journal} {AIP Conf.~Proc.}\ }\textbf
  {\bibinfo {volume} {1679}},\ \bibinfo {pages} {040003} (\bibinfo {year}
  {2015})}\BibitemShut {NoStop}%
\bibitem [{\citenamefont {Jhalani}\ \emph {et~al.}(2020)\citenamefont
  {Jhalani}, \citenamefont {Zhou}, \citenamefont {Park}, \citenamefont
  {Dreyer},\ and\ \citenamefont {Bernardi}}]{Jhalani2020}%
  \BibitemOpen
  \bibfield  {author} {\bibinfo {author} {\bibfnamefont {V.~A.}\ \bibnamefont
  {Jhalani}}, \bibinfo {author} {\bibfnamefont {J.-J.}\ \bibnamefont {Zhou}},
  \bibinfo {author} {\bibfnamefont {J.}~\bibnamefont {Park}}, \bibinfo {author}
  {\bibfnamefont {C.~E.}\ \bibnamefont {Dreyer}}, \ and\ \bibinfo {author}
  {\bibfnamefont {M.}~\bibnamefont {Bernardi}},\ }\href {\doibase
  https://arxiv.org/abs/2002.08351} {\bibfield  {journal} {\bibinfo  {journal}
  {arXiv}\ } (\bibinfo {year} {2020}),\
  https://arxiv.org/abs/2002.08351}\BibitemShut {NoStop}%
\bibitem [{\citenamefont {Park}\ \emph {et~al.}(2020)\citenamefont {Park},
  \citenamefont {Zhou}, \citenamefont {Jhalani}, \citenamefont {Dreyer},\ and\
  \citenamefont {Bernardi}}]{Park2020}%
  \BibitemOpen
  \bibfield  {author} {\bibinfo {author} {\bibfnamefont {J.}~\bibnamefont
  {Park}}, \bibinfo {author} {\bibfnamefont {J.-J.}\ \bibnamefont {Zhou}},
  \bibinfo {author} {\bibfnamefont {V.~A.}\ \bibnamefont {Jhalani}}, \bibinfo
  {author} {\bibfnamefont {C.~E.}\ \bibnamefont {Dreyer}}, \ and\ \bibinfo
  {author} {\bibfnamefont {M.}~\bibnamefont {Bernardi}},\ }\href {\doibase
  https://arxiv.org/abs/2003.13782} {\bibfield  {journal} {\bibinfo  {journal}
  {arXiv}\ } (\bibinfo {year} {2020}),\
  https://arxiv.org/abs/2003.13782}\BibitemShut {NoStop}%
\bibitem [{\citenamefont {Gonze}(1997)}]{Gonze1997a}%
  \BibitemOpen
  \bibfield  {author} {\bibinfo {author} {\bibfnamefont {X.}~\bibnamefont
  {Gonze}},\ }\href {https://doi.org/10.1103/physrevb.55.10337} {\bibfield
  {journal} {\bibinfo  {journal} {Phys.~Rev.~B}\ }\textbf {\bibinfo {volume}
  {55}},\ \bibinfo {pages} {10337} (\bibinfo {year} {1997})}\BibitemShut
  {NoStop}%
\bibitem [{\citenamefont {Gonze}\ \emph {et~al.}(1995)\citenamefont {Gonze},
  \citenamefont {Ghosez},\ and\ \citenamefont {Godby}}]{Gonze1995}%
  \BibitemOpen
  \bibfield  {author} {\bibinfo {author} {\bibfnamefont {X.}~\bibnamefont
  {Gonze}}, \bibinfo {author} {\bibfnamefont {P.}~\bibnamefont {Ghosez}}, \
  and\ \bibinfo {author} {\bibfnamefont {R.}~\bibnamefont {Godby}},\ }\href
  {\doibase https://doi.org/10.1103/PhysRevLett.74.4035} {\bibfield  {journal}
  {\bibinfo  {journal} {Phys.~Rev.~Lett.}\ }\textbf {\bibinfo {volume} {74}},\
  \bibinfo {pages} {4035} (\bibinfo {year} {1995})}\BibitemShut {NoStop}%
\bibitem [{\citenamefont {Bassani}\ and\ \citenamefont
  {Parravicini}(1975)}]{Bassani1975}%
  \BibitemOpen
  \bibfield  {author} {\bibinfo {author} {\bibfnamefont {G.~F.}\ \bibnamefont
  {Bassani}}\ and\ \bibinfo {author} {\bibfnamefont {G.~P.}\ \bibnamefont
  {Parravicini}},\ }\href@noop {} {\emph {\bibinfo {title} {Electronic states
  and optical transitions in solids}}},\ \bibinfo {edition} {1st}\ ed.\
  (\bibinfo  {publisher} {Pergamon Press Oxford, New York},\ \bibinfo {year}
  {1975})\BibitemShut {NoStop}%
\bibitem [{Note9()}]{Note9}%
  \BibitemOpen
  \bibinfo {note} {In our implementation, we prefer to perform the
  symmetrization of the periodic part of the wavefunction in ${\protect \bf G}$
  space because the fractional translation can be easily taken into account by
  multiplying by a phase factor. The symmetrization in real space, on the other
  hand, requires a real-space FFT grid compatible with all the fractional
  translations of the space group.}\BibitemShut {Stop}%
\bibitem [{\citenamefont {Hybertsen}\ and\ \citenamefont
  {Louie}(1987)}]{Hybertsen1987}%
  \BibitemOpen
  \bibfield  {author} {\bibinfo {author} {\bibfnamefont {M.~S.}\ \bibnamefont
  {Hybertsen}}\ and\ \bibinfo {author} {\bibfnamefont {S.~G.}\ \bibnamefont
  {Louie}},\ }\href {\doibase 10.1103/PhysRevB.35.5585} {\bibfield  {journal}
  {\bibinfo  {journal} {Phys.~Rev.~B}\ }\textbf {\bibinfo {volume} {35}},\
  \bibinfo {pages} {5585} (\bibinfo {year} {1987})}\BibitemShut {NoStop}%
\bibitem [{\citenamefont {Petretto}\ \emph {et~al.}(2018)\citenamefont
  {Petretto}, \citenamefont {Dwaraknath}, \citenamefont {Miranda},
  \citenamefont {Winston}, \citenamefont {Giantomassi}, \citenamefont {{Van
  Setten}}, \citenamefont {Gonze}, \citenamefont {Persson}, \citenamefont
  {Hautier},\ and\ \citenamefont {Rignanese}}]{Petretto2018}%
  \BibitemOpen
  \bibfield  {author} {\bibinfo {author} {\bibfnamefont {G.}~\bibnamefont
  {Petretto}}, \bibinfo {author} {\bibfnamefont {S.}~\bibnamefont
  {Dwaraknath}}, \bibinfo {author} {\bibfnamefont {H.~P.~C.}\ \bibnamefont
  {Miranda}}, \bibinfo {author} {\bibfnamefont {D.}~\bibnamefont {Winston}},
  \bibinfo {author} {\bibfnamefont {M.}~\bibnamefont {Giantomassi}}, \bibinfo
  {author} {\bibfnamefont {M.~J.}\ \bibnamefont {{Van Setten}}}, \bibinfo
  {author} {\bibfnamefont {X.}~\bibnamefont {Gonze}}, \bibinfo {author}
  {\bibfnamefont {K.~A.}\ \bibnamefont {Persson}}, \bibinfo {author}
  {\bibfnamefont {G.}~\bibnamefont {Hautier}}, \ and\ \bibinfo {author}
  {\bibfnamefont {G.~M.}\ \bibnamefont {Rignanese}},\ }\href {\doibase
  10.1038/sdata.2018.65} {\bibfield  {journal} {\bibinfo  {journal}
  {Sci.~Data.}\ }\textbf {\bibinfo {volume} {5}},\ \bibinfo {pages} {1}
  (\bibinfo {year} {2018})}\BibitemShut {NoStop}%
\bibitem [{\citenamefont {Hamann}(2013)}]{Hamann2013}%
  \BibitemOpen
  \bibfield  {author} {\bibinfo {author} {\bibfnamefont {D.~R.}\ \bibnamefont
  {Hamann}},\ }\href {\doibase 10.1103/physrevb.88.085117} {\bibfield
  {journal} {\bibinfo  {journal} {Phys.~Rev.~B}\ }\textbf {\bibinfo {volume}
  {88}},\ \bibinfo {pages} {085117} (\bibinfo {year} {2013})}\BibitemShut
  {NoStop}%
\bibitem [{\citenamefont {Damle}\ and\ \citenamefont {Lin}(2018)}]{Damle2018}%
  \BibitemOpen
  \bibfield  {author} {\bibinfo {author} {\bibfnamefont {A.}~\bibnamefont
  {Damle}}\ and\ \bibinfo {author} {\bibfnamefont {L.}~\bibnamefont {Lin}},\
  }\href {\doibase 10.1137/17M1129696} {\bibfield  {journal} {\bibinfo
  {journal} {Multiscale Model.~Simul.}\ }\textbf {\bibinfo {volume} {16}},\
  \bibinfo {pages} {1392} (\bibinfo {year} {2018})}\BibitemShut {NoStop}%
\bibitem [{\citenamefont {Pickett}(1989)}]{Pickett1989}%
  \BibitemOpen
  \bibfield  {author} {\bibinfo {author} {\bibfnamefont {W.~E.}\ \bibnamefont
  {Pickett}},\ }\href {\doibase https://doi.org/10.1016/0167-7977(89)90002-6}
  {\bibfield  {journal} {\bibinfo  {journal} {Comput.~Phys.~Rep.}\ }\textbf
  {\bibinfo {volume} {9}},\ \bibinfo {pages} {115} (\bibinfo {year}
  {1989})}\BibitemShut {NoStop}%
\bibitem [{\citenamefont {Calnan}\ and\ \citenamefont
  {Tiwari}(2010)}]{Calnan2010}%
  \BibitemOpen
  \bibfield  {author} {\bibinfo {author} {\bibfnamefont {S.}~\bibnamefont
  {Calnan}}\ and\ \bibinfo {author} {\bibfnamefont {A.~N.}\ \bibnamefont
  {Tiwari}},\ }\href {\doibase 10.1016/j.tsf.2009.09.044} {\bibfield  {journal}
  {\bibinfo  {journal} {Thin Solid Films}\ }\textbf {\bibinfo {volume} {518}},\
  \bibinfo {pages} {1839} (\bibinfo {year} {2010})}\BibitemShut {NoStop}%
\bibitem [{\citenamefont {Lu}\ \emph {et~al.}(2019)\citenamefont {Lu},
  \citenamefont {Zhou},\ and\ \citenamefont {Bernardi}}]{Lu2019}%
  \BibitemOpen
  \bibfield  {author} {\bibinfo {author} {\bibfnamefont {I.-T.}\ \bibnamefont
  {Lu}}, \bibinfo {author} {\bibfnamefont {J.-J.}\ \bibnamefont {Zhou}}, \ and\
  \bibinfo {author} {\bibfnamefont {M.}~\bibnamefont {Bernardi}},\ }\href
  {\doibase 10.1103/PhysRevMaterials.3.033804} {\bibfield  {journal} {\bibinfo
  {journal} {Phys.~Rev.~Mat.}\ }\textbf {\bibinfo {volume} {3}},\ \bibinfo
  {pages} {033804} (\bibinfo {year} {2019})}\BibitemShut {NoStop}%
\end{thebibliography}

%

\end{document}


\title{Phonon-limited electron mobility in Si, GaAs and GaP 
with exact treatment of dynamical quadrupoles \\ Supplemental Material}
\date{\today}
\author{Guillaume Brunin$^1$}
\author{Henrique Pereira Coutada Miranda$^1$}
\author{Matteo Giantomassi$^1$}
\author{Miquel Royo$^2$}
\author{Massimiliano Stengel$^{2,3}$}
\author{Matthieu J. Verstraete$^{4,5}$}
\author{Xavier Gonze$^{1,6}$}
\author{Gian-Marco Rignanese$^1$}
\author{Geoffroy Hautier$^1$}

\affiliation{$^1$UCLouvain, Institute of Condensed Matter and Nanosciences (IMCN), Chemin des \'Etoiles~8, B-1348 Louvain-la-Neuve, Belgium}
\affiliation{$^2$Institut de Ci\`encia de Materials de Barcelona (ICMAB-CSIC), Campus UAB, 08193 Bellaterra, Spain}
\affiliation{$^3$ICREA-Instituci\'o Catalana de Recerca i Estudis Avan\c{c}ats, 08010 Barcelona, Spain}
\affiliation{$^4$NanoMat/Q-Mat/CESAM, Universit{\'e} de Li{\`e}ge (B5), B-4000 Li{\`e}ge, Belgium}
\affiliation{$^5$Catalan Institute of Nanoscience and Nanotechnology (ICN2), Campus UAB, 08193 Bellaterra, Spain}
\affiliation{$^6$Skolkovo Institute of Science and Technology, Moscow, Russia}

\pacs{}

\maketitle

\renewcommand{\figurename}{Supplemental Figure}
\renewcommand{\thefigure}{S\arabic{figure}}
\renewcommand{\theequation}{S\arabic{equation}}
\renewcommand{\tablename}{Supplemental Table}
\renewcommand{\thetable}{S\arabic{table}}

\section{$\qb$-points filtering}

\begin{figure}
    \centering
    \includegraphics[width=.7\textwidth]{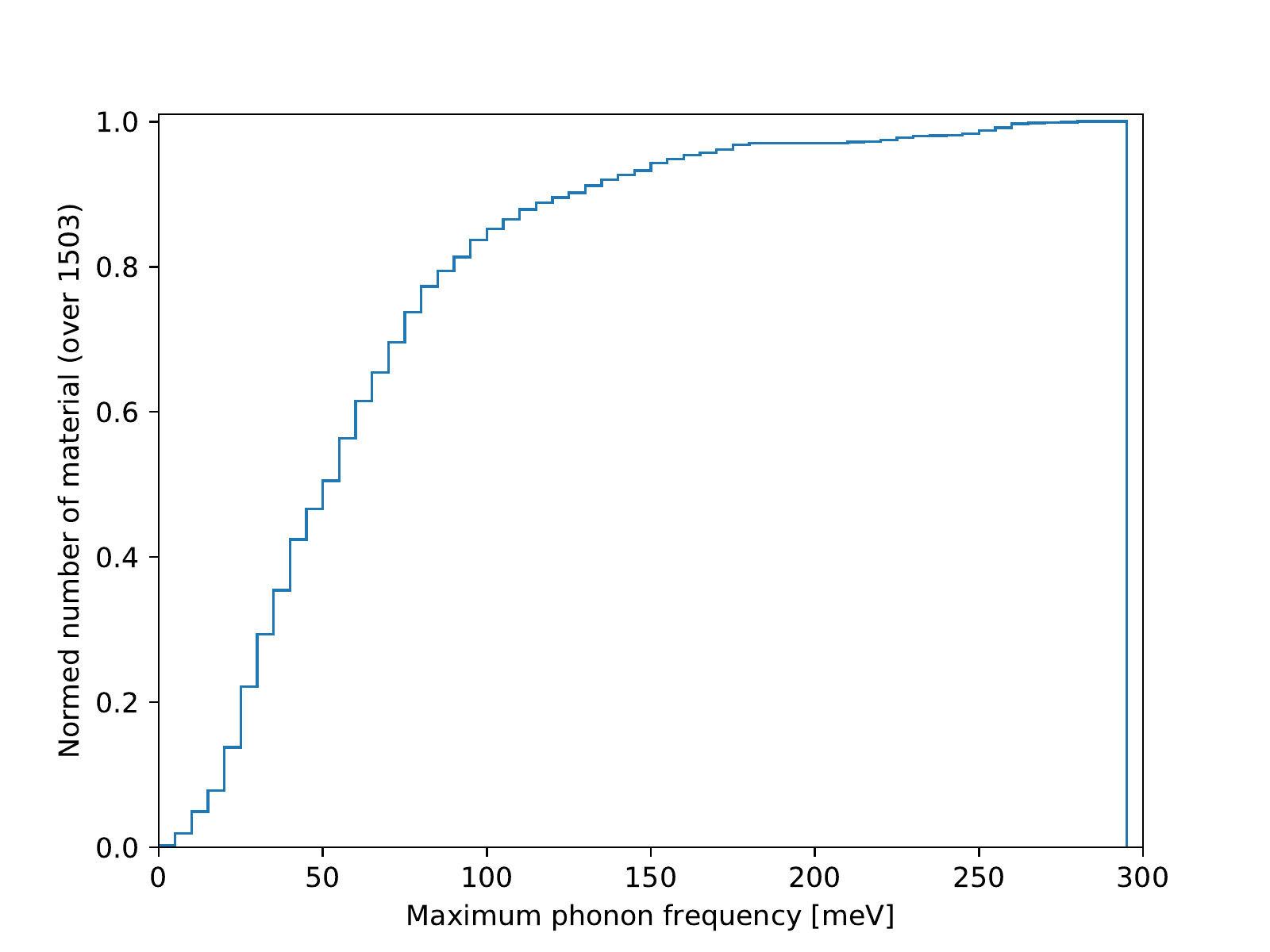}
    
    \includegraphics[width=.7\textwidth]{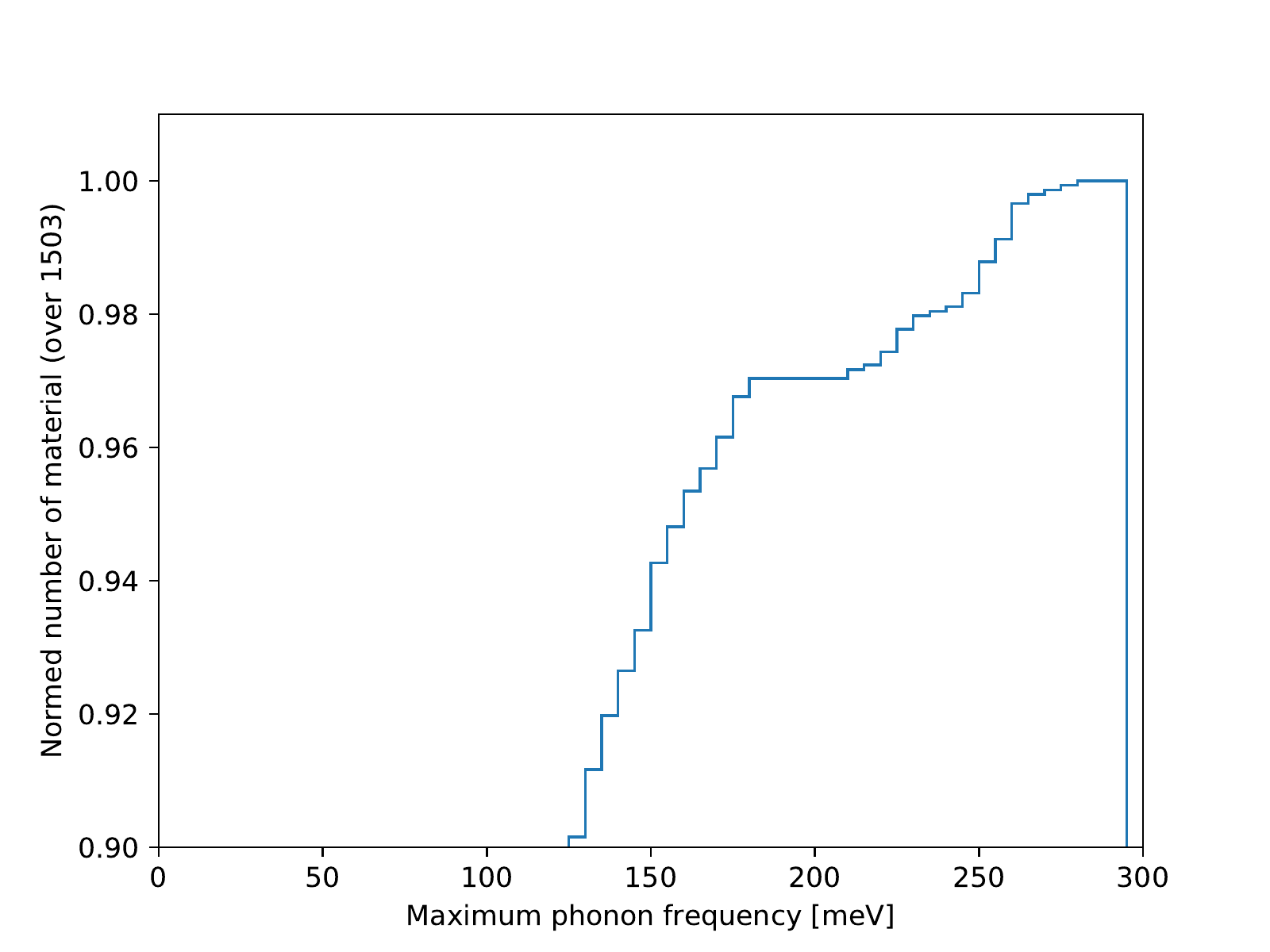}
    \caption{(up) Cumulative histogram of the distribution of the maximum phonon frequency at $\qb = \Gamma$ produced using the DFPT results for the 1503 materials available in the Materials Project database.
    (down) Zoom on the last 10\% of the materials. 90\% of them have phonon frequencies lower than 125~meV.}
    \label{fig:count_omegaLO}
    \end{figure}

The efficiency of the $\qb$-point filtering technique depends on the phase space available for e-ph scattering for the each $\kk$-point.
The ideal case is represented by a semiconductor with a single 
minimum (maximum) at the $\Gamma$ point that is well separated in energy from the other bands.
In this case, indeed, only intra-valley transitions are allowed by energy and momentum conservation
and transitions are restricted to small $\qb$-vectors.
The worst-case scenario is given by
a semiconductor with 
multiple minima (maxima) in the BZ whose energy differences are comparable to the typical phonon frequency of the material, possibly with additional dispersionless states close to the band band edges.
A statistical analysis of the vibrational properties
of the 1503 materials~\cite{Petretto2018} available in the Materials Project database reveals that 
the maximum phonon energy is usually less than 125~meV (see Figure~\ref{fig:count_omegaLO}) which, as expected, corresponds to a small energy window at the electron scale.
On the basis of this qualitative analysis, it is reasonable to conclude that the significant speedup obtained with the $\qb$-point filtering technique
observed for the three semiconductors considered in this work will still apply for a significant fraction of materials provided the electronic dispersion
differs from the worst-case scenario described above.

\section{Validation of our implementation}

\begin{figure}
    \centering
    \includegraphics[clip,trim=0.2cm 0cm 0.2cm 0.2cm,height=.37\textwidth]{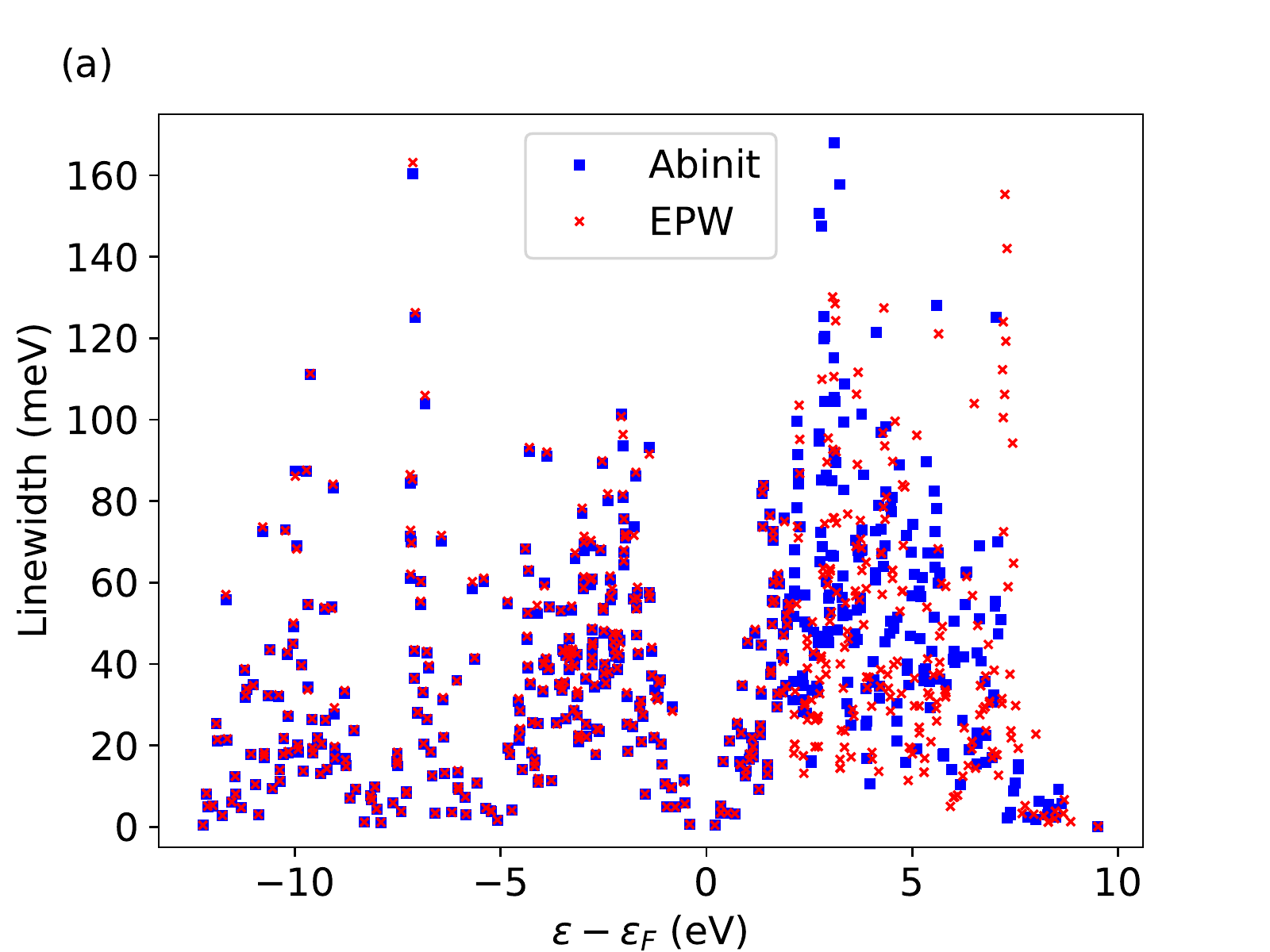}
    \includegraphics[clip,trim=0.2cm 0cm 0.2cm 0.2cm,height=.37\textwidth]{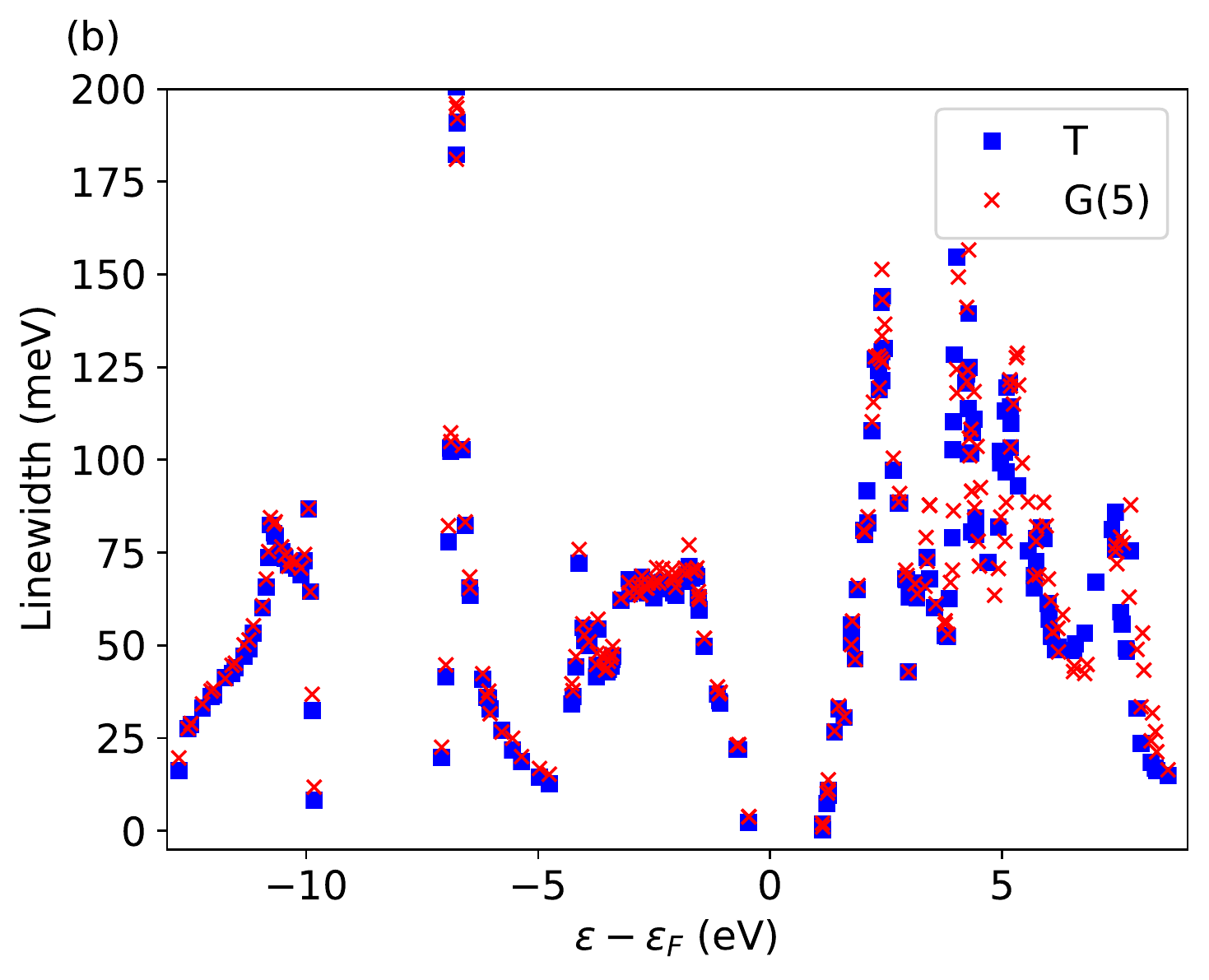}
    \caption{(a) Electron linewidths in Si obtained with \epw (red crosses) and \abinit (blue squares) 
    with a Lorentzian broadening of 5 meV, for \grid{12} $\kb$- and $\qb$-point grids. 
    Only the 4 valence bands and part of the conduction band are included in the frozen window for the Wannierization with \epw.
    (b) Electron linewidths in GaP obtained with \epw with a Gaussian smearing of 5 meV (red
    crosses) for \grid{8} $\kb$-point and \grid{210} $\qb$-point grids and obtained
    with \abinit with the tetrahedron integration method (blue squares) for 
    \grid{8} $\kb$-point and \grid{48} $\qb$-point grids.}
    \label{fig:abinit_vs_epw}
    \end{figure}

We validated our new implementation by comparing in Fig.~\ref{fig:abinit_vs_epw} the linewidths for Si and GaP 
computed with \abinit with the analogous results obtained with the Wannier interpolation implemented in the \epw code~\cite{Ponce2016}.
For this test, the same lattice parameters and pseudopotentials are used in both codes.
More specifically, we employ GGA-PBE
for the XC functional and 
norm-conserving pseudopotentials from the \pseudodojo project~\cite{Vansetten2018, Hamann2013}.
Note that the pseudopotentials used for the validation differ from the ones used in the main article in which LDA is used.
All these calculations are performed without including dynamical quadrupoles in the Fourier interpolation.

\begin{figure}
    \centering
    \includegraphics[width=.7\textwidth]{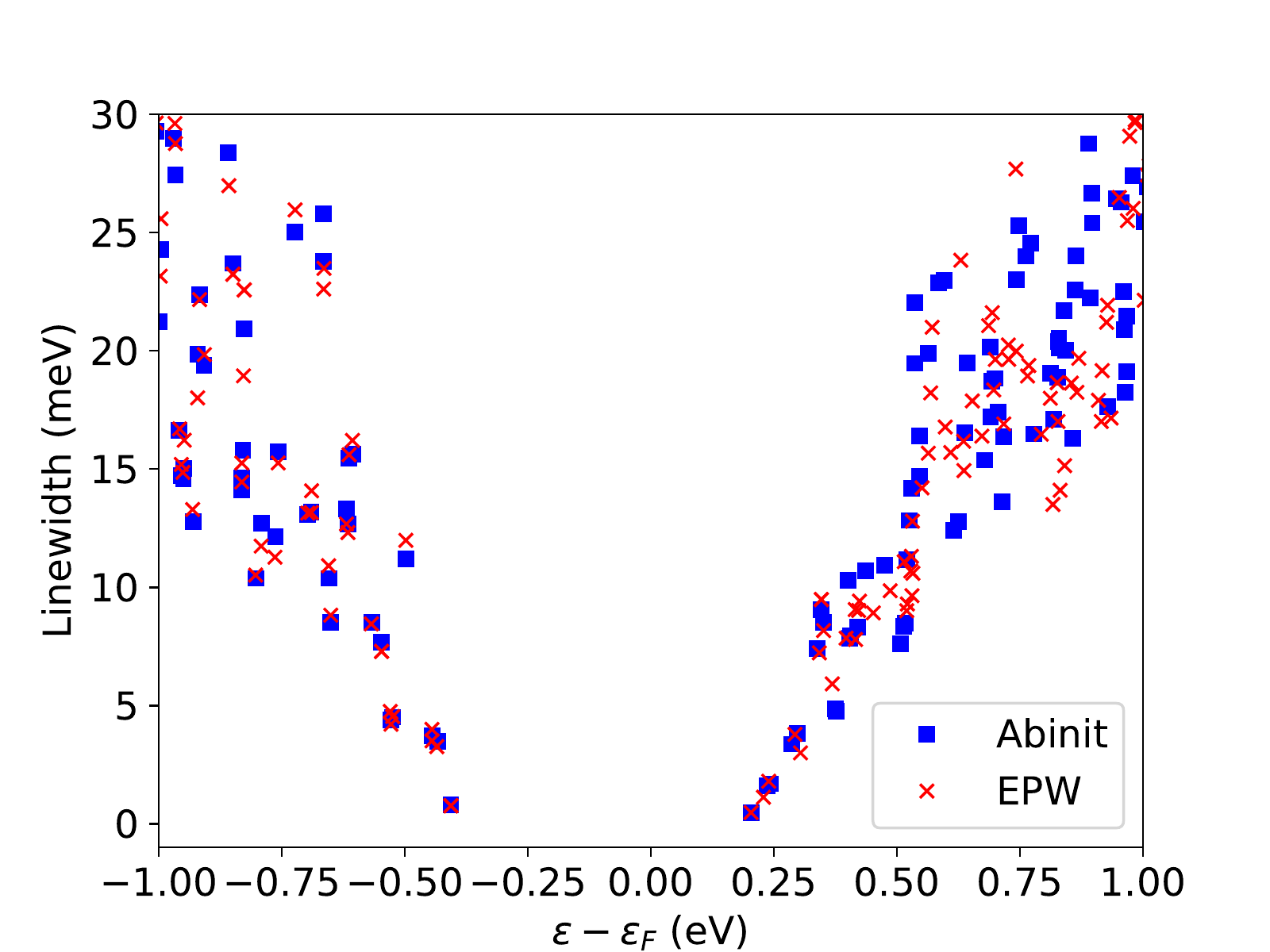}
    \caption{Electronic linewidths in Si obtained with \epw (red crosses) and \abinit (blue squares) with a Lorentzian broadening of 5 meV, 
    for \grid{24} $\kb$- and $\qb$-point grids. Only the 4 valence bands are included in the frozen window 
    for the wannierization with \epw.}
    \label{fig:abinit_vs_epw24}
    \end{figure}

For Si, we use an energy cutoff of 25 Ry for the plane-wave basis set and a lattice parameter of 5.47~\AA.
The linewidths are computed for 4 valence and 4 conduction bands 
starting from coarse $\Gamma$-centered \grid{12} $\kb$-point and \grid{6} $\qb$-point meshes.
Fig.~\ref{fig:abinit_vs_epw}(a) compares the 
linewidths of silicon obtained with \abinit and \epw computed with the same Lorentzian broadening.
Keeping into account the different treatment of the electron states (exact in \abinit, interpolated in \epw) as well as the different approach implemented to interpolate the e-ph scattering potential, 
reasonably good agreement between the two implementations is observed.
For the valence states, indeed, the average difference between \abinit and Wannier-interpolated KS eigenvalues is 0.03 meV while the phonon-induced linewidths differ by 0.41 meV on average.
The agreement is also reasonably good for the conduction states provided we limit the discussion to the conduction states lying inside the frozen energy window employed for the Wannier disentanglement procedure (1.85~eV above the CMB).
At higher energies, the two implementations starts to disagree significantly but this is expected since, by construction, the Wannier interpolation is not supposed to reproduce the KS band structure outside the frozen window.
%
Fig.~\ref{fig:abinit_vs_epw24} reports the same comparison but for a \grid{24} $\kb$-grid and states around the Fermi level.  
The agreement between \abinit and 
Wannier-interpolated values
is good for $\kk$-points belonging to the initial \grid{12} $\kk$-mesh. 
Differences can be noticed for the $\kk$-points that do not belong to the initial coarse $\kk$-mesh although the two sets of data show a similar trend. 
The average difference between \abinit and Wannier-interpolated energies is now
0.7~meV in the VB and 236~meV in the CB, and the linewidths differ on average by 0.83~meV in the VB and 14.21~meV in the CB.

For GaP, we adopt a cutoff energy of 80 Ry, $\Gamma$-centered \grid{8} $\kb$- and \grid{4} $\qb$-point grids, 
and we determine the linewidths for 4 valence and 4 conduction bands.
The lattice parameter of GaP is 5.50~\AA~.
In Fig.~\ref{fig:abinit_vs_epw}b, 
we compare the linewidths in GaP obtained with \epw (Gaussian broadening of 5 meV)
and \abinit (tetrahedron method).
Also in this case, the agreement between the two implementations is reasonably good 
and differences appear for high energy states.

\begin{figure}
    \centering
    \includegraphics[width=.8\textwidth]{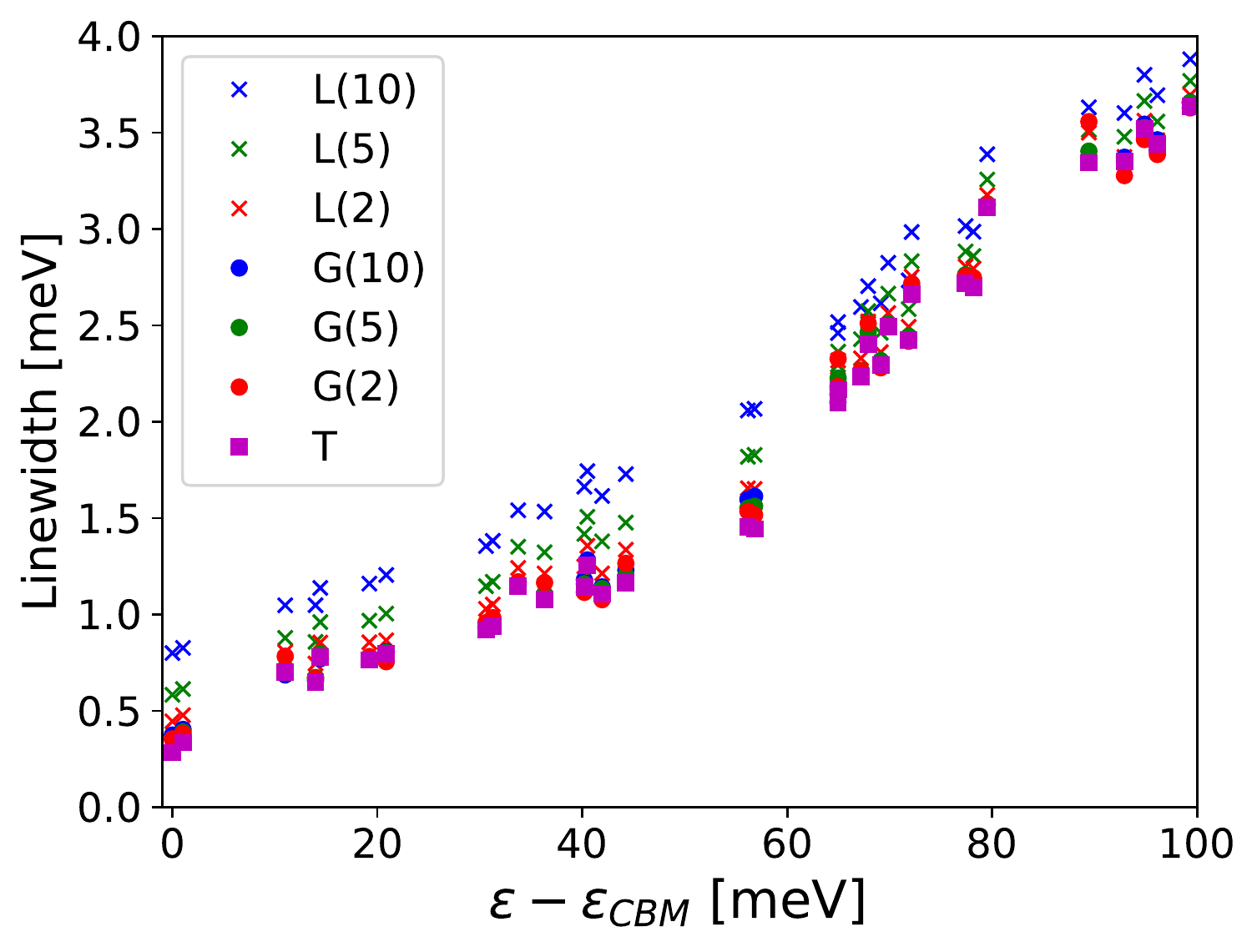}
    \caption{Electron linewidths near the CBM of Si obtained with a \grid{60} $\kb$-point grid at $T=300$\,K. 
    A very dense \grid{180} $\qb$-point grid is used for all curves to achieve well-converged results. 
    Results are obtained with Lorentzian (L), Gaussian (G), 
    and tetrahedron (T) methods. 
    The broadening parameter is given in parentheses in meV.}
    \label{fig:intmethcomp}
    \end{figure}

Similarly to the comparison between the Lorentzian technique and the tetrahedron integration method done in the manuscript, one can compare the results with a Gaussian broadening. This type of broadening is used extensively in the literature, because it converges faster than the Lorentzian method with respect to the broadening parameter, as expressed in Fig.~\ref{fig:intmethcomp}.

\begin{figure}
	\centering
	\includegraphics[clip,trim=0.2cm 0.2cm 0.2cm 0.2cm,width=.6\textwidth]{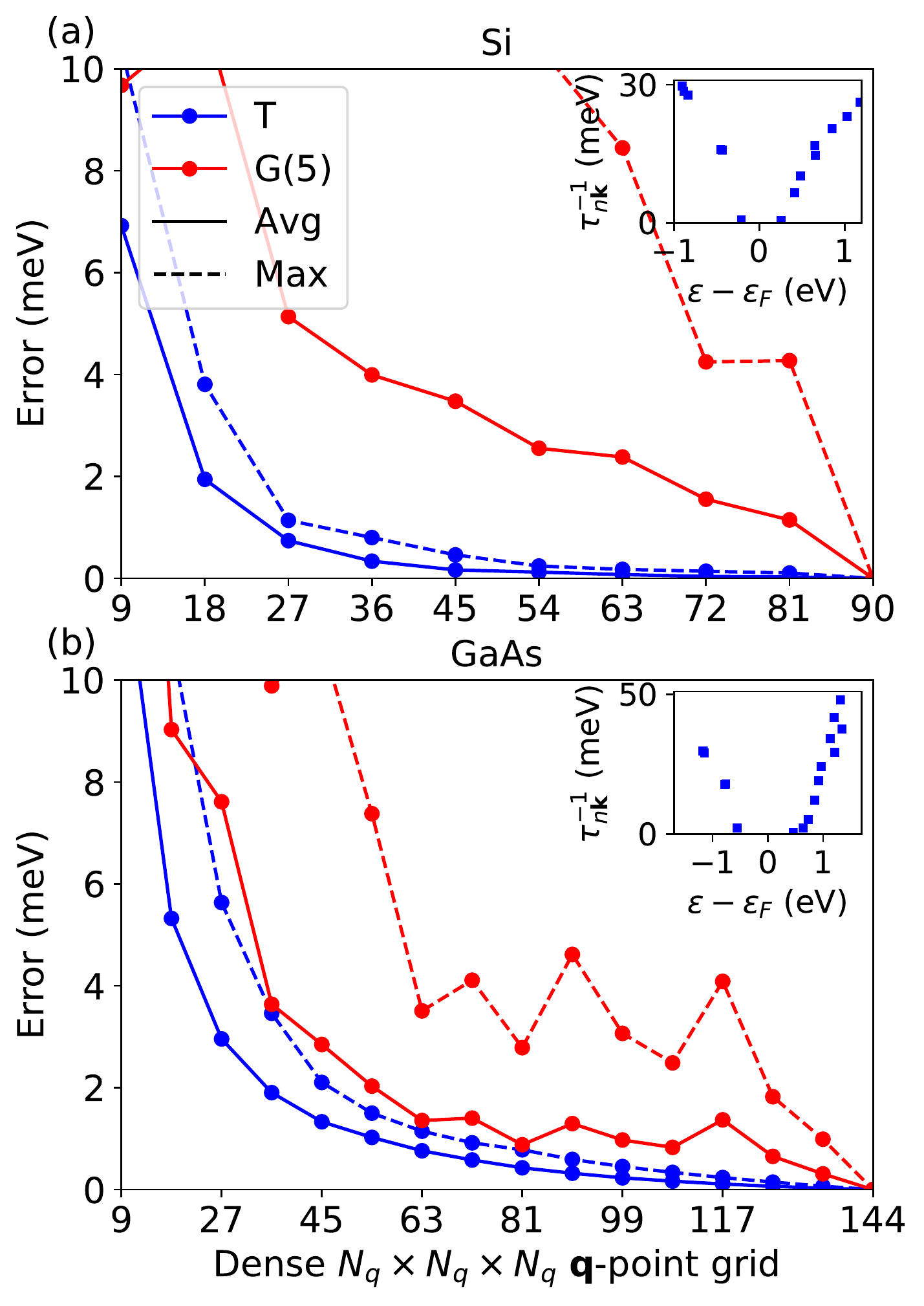}
	\caption{Average (solid line) and maximum (dashed line) error on the linewidths evaluated on a \grid{9} $\kb$-point grid in (a) Si and (b) GaAs as a function of the $\qb$-grid used for the integration. 
	Only states within 1~eV from the CBM and VBM are considered. 
	Results have been obtained with the tetrahedron method (T, blue) and a Gaussian (G, red) broadening of 5~meV.
	The reference linewidths for each curve are the results obtained with the
    densest $\qb$-grid and the corresponding integration technique.
	The insets show the linewidths obtained with the tetrahedron method and the densest $\qb$-grid.}
	\label{fig:errors_SiGaAs}
    \end{figure}

We have also compared the convergence rate with respect to the dense $\qq$-mesh of the Gaussian broadening method with the tetrahedron technique. The results are very similar to the ones of the Lorentzian broadening method, as shown in Fig.~\ref{fig:errors_SiGaAs}.

\section{Values of quadrupoles for Si, GaAs and GaP}

In the materials considered in this work, the quadrupolar tensor assumes the form
\begin{equation}
    Q_{\kappa\alpha}^{\beta\gamma} = (-1)^{\kappa+1} Q_\kappa |\varepsilon_{\beta\gamma\alpha}|,
\end{equation}
with $\varepsilon_{\beta\gamma\alpha}$ the Levi-Civita tensor while
the value of $Q_\kappa$ depends on the atomic type at site  $\kappa$. Working in atomic units (Bohr times the absolute value of the electronic charge),
in Si, the quadrupolar tensor contains a single value $Q_\text{Si} = 13.67$, which is in good agreement with the value reported by Royo \textit{et al.}~\cite{Royo2019}.
In GaP, the quadrupole tensor is given by $Q_\text{Ga} = 12.73$, and $Q_\text{P} = 5.79$.
In GaAs, the quadrupoles are $Q_\text{Ga} = 16.54$ and $Q_\text{As} = 8.57$.

\section{Detailed results for GaP}

\begin{figure}
    \centering
    \includegraphics[clip,trim=0.2cm 0.3cm 0.2cm 0.2cm,width=.7\textwidth]{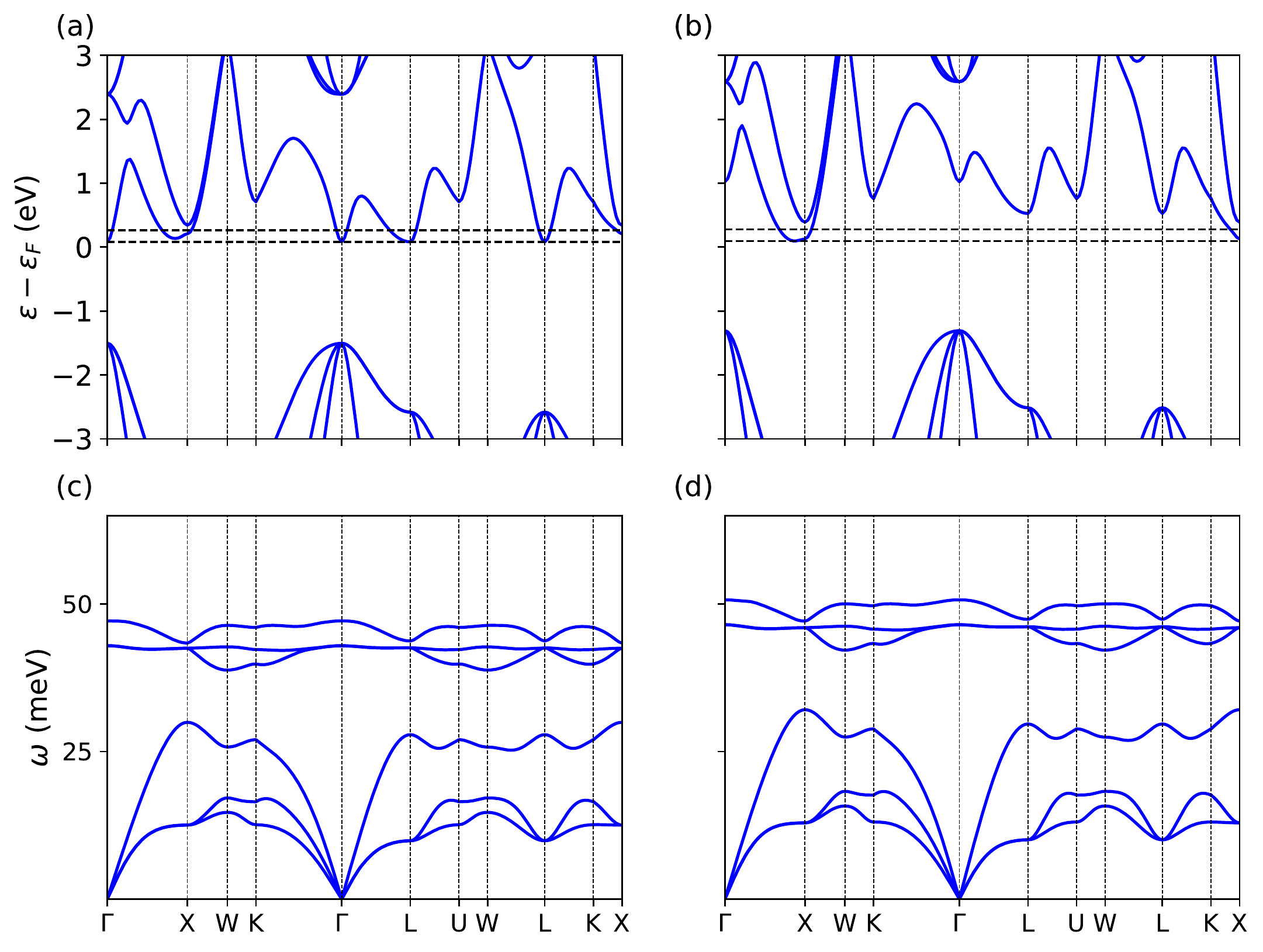}
    \caption{Band structure of GaP using (a) GGA-PBE and (b) LDA pseudopotentials. The Fermi level is located so that $n_e = 10^{18}$ cm$^{-3}$. The energy window used to filter $\kb$-points for transport computations is represented by horizontal dashed lines.
    Phonon dispersions of GaP using (c) GGA-PBE and (d) LDA.
    }
    \label{fig:GaP_GGAvsFHI}
    \end{figure}

The band structures of GaP, obtained with \abinit, are represented in Figs.~\ref{fig:GaP_GGAvsFHI}(a) and (b).
GGA-PBE is used in (a) while LDA FHI pseudopotentials are used in (b).
The band structure of GaP, particularly the conduction band, is quite sensitive to the XC functional. 
The 
band dispersion obtained with GGA-PBE presents additional minima that are slightly higher in energy than 
the CBM 
and this has a significant impact on the mobility.
Note also that the position of the CBM depends of the XC functional: it is located at $\Gamma$ if GGA is used and moves to a $\kk$-point between $\Gamma$ and $X$ in LDA.
The phonon dispersions obtained with both XC functionals are shown in (c) and (d). 
There is a scaling factor between them but the general behavior is similar.
In what follows, we report the results obtained with the LDA pseudopotentials as in the main text, with a cut-off energy of 30 Ha and a lattice constant of 5.32~\AA.

\begin{figure}
	\centering
	\includegraphics[clip,trim=0.2cm 0.2cm 0.2cm 0.2cm,width=.4\textwidth]{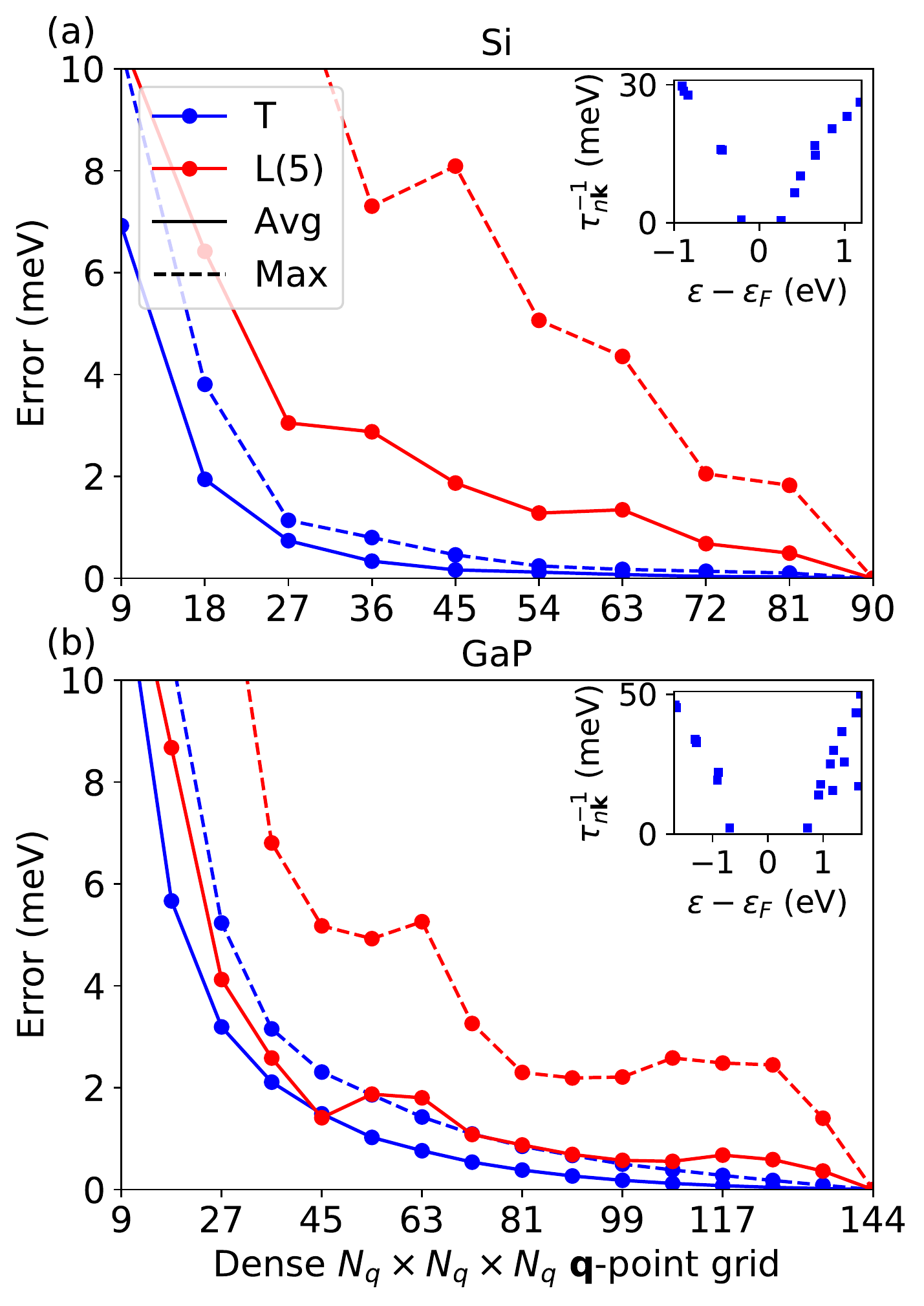}
	\caption{Average (full) and maximum (dashed) error on the linewidths evaluated on a \grid{9} $\kb$-point grid in (a) Si and (b) GaP as a function of the $\qb$-grid used for the integration. 
	Only states within 1~eV from the CBM and VBM are considered. 
	Results have been obtained with the tetrahedron method (T, blue) and a Lorentzian (L, red) broadening of 5 meV.
	The reference linewidths for each curve are the results obtained with the 
    densest $\qb$-grid and the corresponding integration technique.
	The insets show the linewidths obtained with the tetrahedron method and the densest $\qb$-grid.}
	\label{fig:errors_SiGaP}
    \end{figure}

\begin{figure}
    \centering
    \includegraphics[clip,trim=0.2cm 0.2cm 0.2cm 0.2cm,width=.4\textwidth]{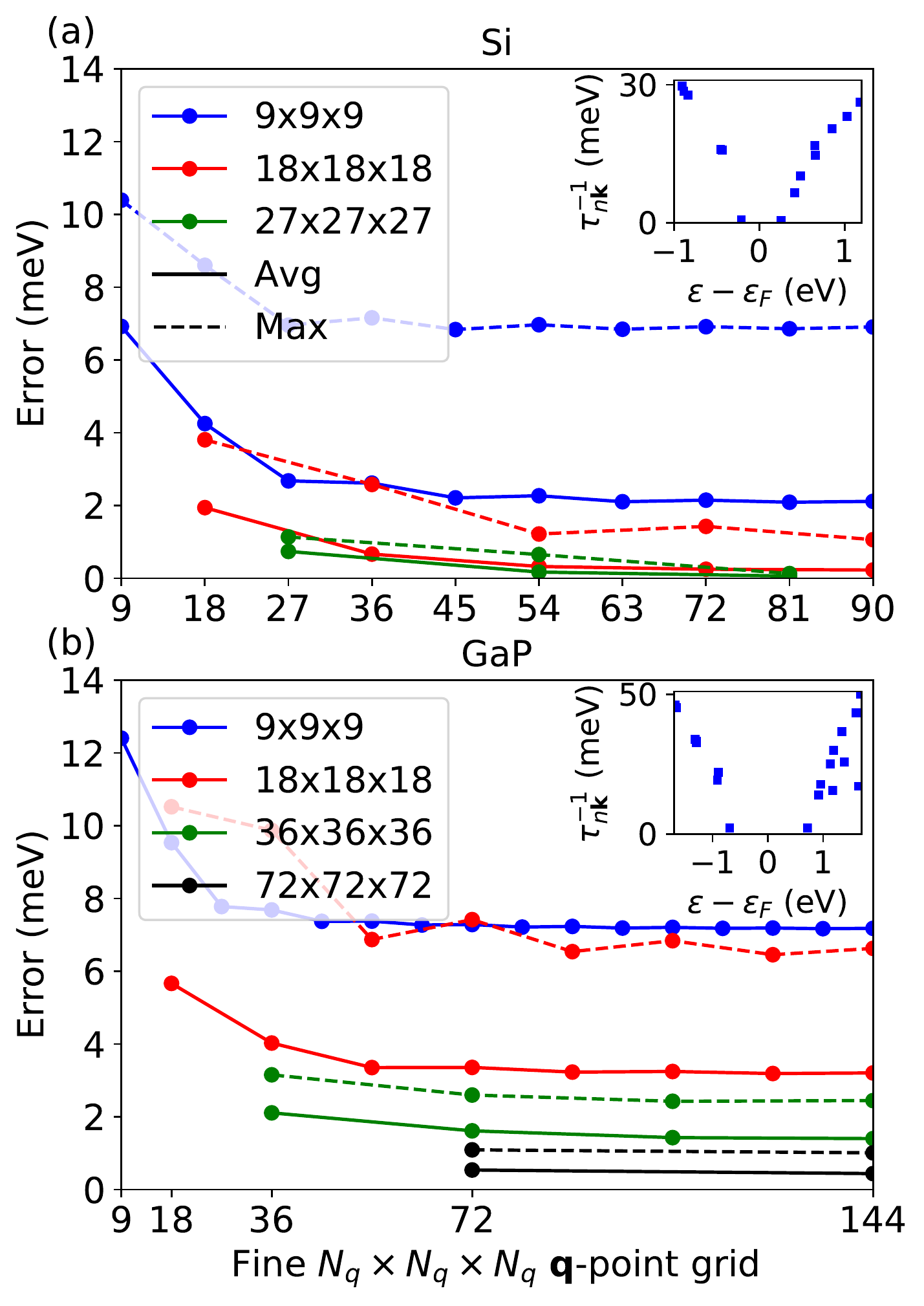}
    \caption{Average (solid line) and maximum (dotted line) error on the linewidths on a \grid{9} $\kb$-point grid in (a) Si and 
    (b) GaP as a function of the fine $\qb$-point grid used for the integration. 
    The matrix elements are obtained on (a) initial coarse \grid{9} (blue), \grid{18} (red) 
    and \grid{27} (green) grids for Si and (b) initial coarse \grid{9} (blue), \grid{18} (red) 
    \grid{36} (green) and \grid{72} (black) grids for GaP. 
    Energies are computed on the dense grids.
    Only states within 1~eV from the CBM and VBM are considered. 
    The reference linewidths are the results obtained with a \grid{90} (\grid{144}) $\qb$-point grid for both matrix elements and 
    the energies for Si (GaP).
    The insets show the considered linewidths obtained with the tetrahedron method for the densest $\qb$-point grid.}
    \label{fig:errors_dg_SiGaP}
    \end{figure}

To compare the convergence rate of the tetrahedron method with the one of a Lorentzian with a 5~meV broadening,
we compute the average and the maximum error on the linewidths in Si and GaP 
as a function of the $\qb$-mesh.
The reference linewidths are those obtained with the densest grids for the corresponding method.
The results are reported in Fig.~\ref{fig:errors_SiGaP}.
In both materials, the tetrahedron method outperforms the Lorentzian method
and convergence is reached with much coarser $\qb$-point grids.
The convergence is slower in GaP than in Si because of the Fr\"ohlich singularity for $\qb \rightarrow \mathbf{0}$ in polar semiconductors.
This singularity is integrable but requires dense $\qb$-meshes for the e-ph matrix elements.
Fig.~\ref{fig:errors_dg_SiGaP} reports the error on the linewidths in Si and GaP,
similarly to Fig.~\ref{fig:errors_SiGaP}, but now obtained with the double-grid technique and tetrahedron integration.
The reference linewidths are those obtained with a single grid corresponding to the densest in Fig.~\ref{fig:errors_SiGaP}.
In the case of Si, using a \grid{9} coarse $\qb$-point grid for the matrix elements and a dense \grid{36} $\qb$-point grid 
for the energies reduces the average error below 1~meV at a very similar computational cost than the computation using \grid{9} $\kb$- and $\qb$-point grids. 
Increasing the density of the coarse grid further reduces the error down to practically zero.
In the case of GaP, the e-ph matrix elements vary more in the BZ 
because of the polar singularity around $\qb = \Gamma$ hence a denser coarse $\qb$-point grid 
is required to reduce the error.
%
%

\section{Effect of the carrier concentration on the mobility}

In this section, we analyze how the results depend on the carrier concentration and show
that, as long as the Fermi level $\ee_F$ is far enough from the CBM, the mobility is not sensitive to the value of $\ee_F$
so that a small doping can be used to make the calculation more stable.
Note that, in principle, the position of $\ee_F$ enters into play both during the computation of the imaginary part of the self-energy as well as in the expression for the mobility via the occupation function.
We first analyze the dependence of the mobility on $\ee_F$ using lifetimes computed with a single fixed $\ee_F$, then we compare results obtained by changing  
$\ee_F$ both in the self-energy and in the equation of the mobility.

\begin{figure}
    \centering
    \includegraphics[width=0.49\textwidth]{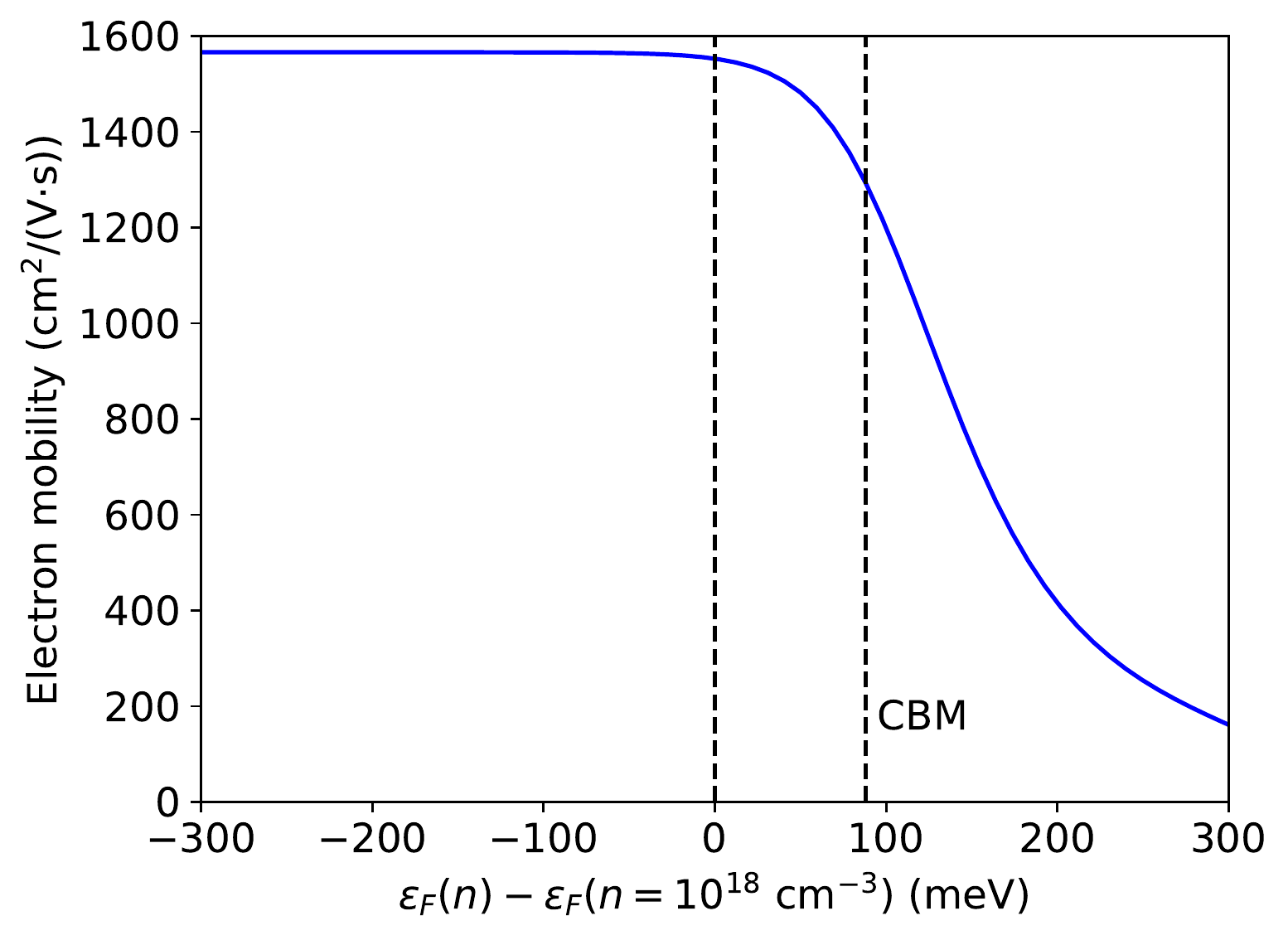}
    \caption{Electron mobility in Si ($T=300$ K) as a function of the Fermi level used in the computation. The lifetimes are computed with a single Fermi level corresponding to an electron concentration of $10^{18}$ cm$^{-3}$ as in the main text. \grid{72} $\kk$-point and \grid{144} $\qq$-point grids are used. The energy of the CBM is reported by a vertical dashed line at 88~meV.}
    \label{fig:Si_mu_ef}
	\end{figure}

\begin{figure}
    \centering
    \includegraphics[width=0.49\textwidth]{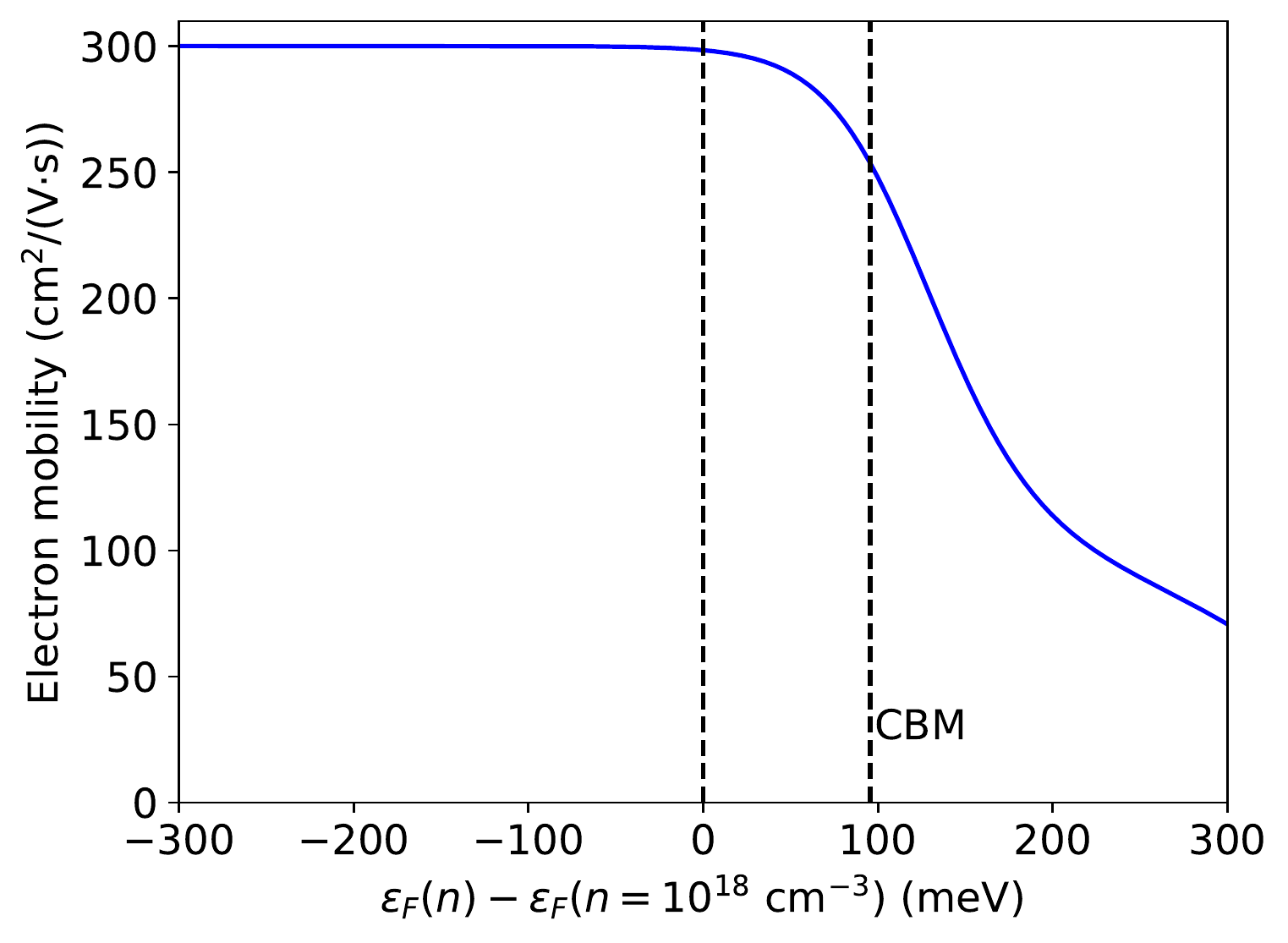}
    \caption{Electron mobility in GaP ($T=300$ K) as a function of the Fermi level used in the computation. The lifetimes are computed with a single Fermi level corresponding to an electron concentration of $10^{18}$ cm$^{-3}$ as in the main text. \grid{78} $\kk$-point and \grid{156} $\qq$-point grids are used. The energy of the CBM is reported by a vertical dashed line at 95~meV.}
    \label{fig:GaP_mu_ef}
	\end{figure}

\begin{figure}
    \centering
    \includegraphics[width=0.49\textwidth]{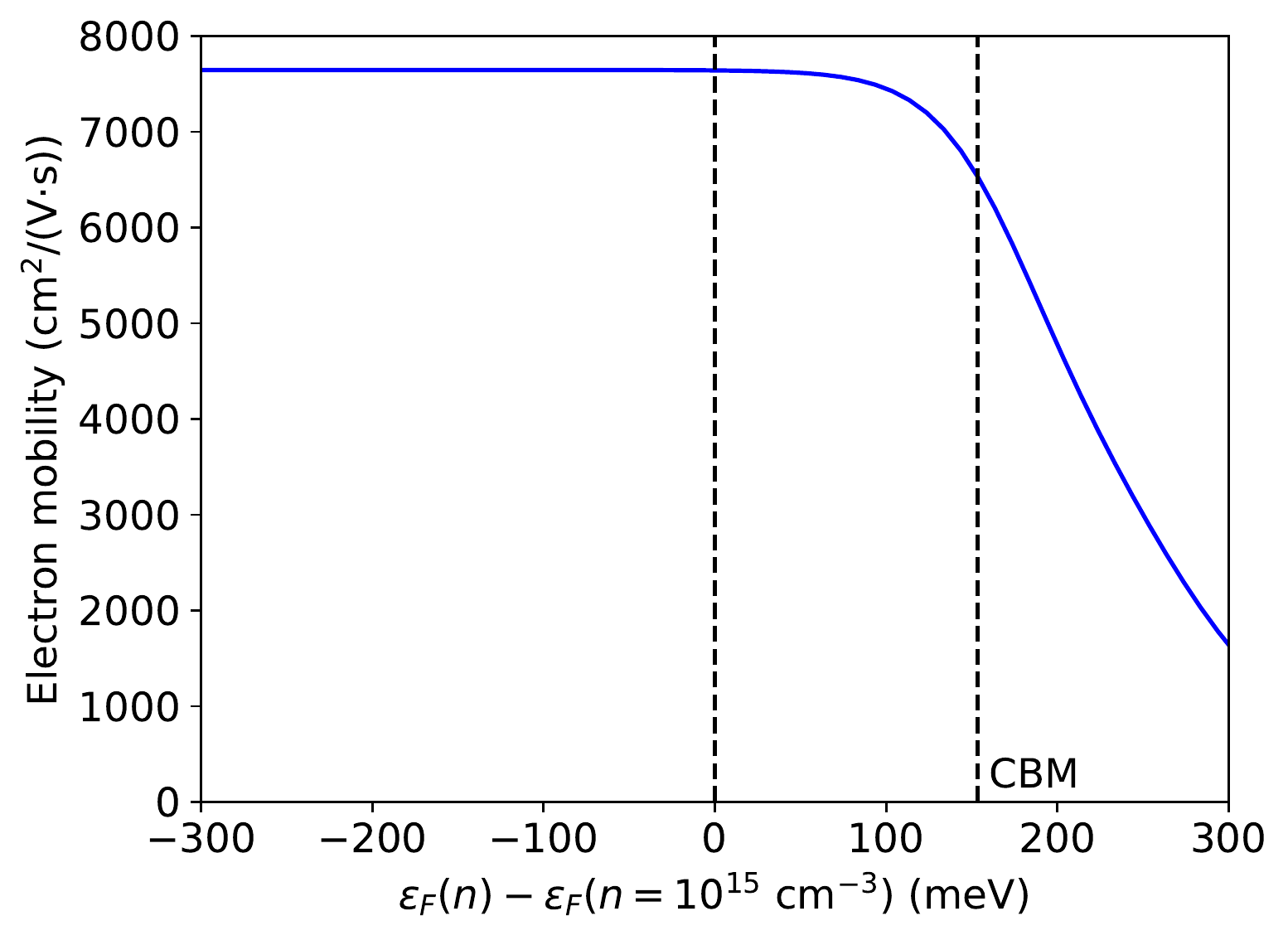}
    \caption{Electron mobility in GaAs ($T=300$ K) as a function of the Fermi level used in the computation. The lifetimes are computed with a single Fermi level corresponding to an electron concentration of $10^{15}$ cm$^{-3}$ as in the main text. \grid{192} $\kk$- and $\qq$-point grids are used. The energy of the CBM is reported by a vertical dashed line at 153~meV.}
    \label{fig:GaAs_mu_ef}
	\end{figure}

Figure~\ref{fig:Si_mu_ef} shows the variation of the mobility with the change of $\ee_F$ for given linewidths
corresponding to an electron concentration of $10^{18}$ cm$^{-3}$. 
Note how the results are insensitive to the value of the $\ee_F$ as long as the Fermi level is below the CBM (up to around 88~meV that corresponds to the value used in our calculations).
In order to quantify the effect of the Fermi level on both the lifetimes and the mobility, we performed full calculations for two different electron concentrations. 
For $10^{18}$ cm$^{-3}$, we obtain 1497~cm$^2$/(V$\cdot$s). For $10^{17}$ cm$^{-3}$, we obtain 1510~cm$^2$/(V$\cdot$s). The difference is smaller than 1\%: we can therefore neglect this effect and consider $10^{18}$ cm$^{-3}$ in all computations for Si.
Figure~\ref{fig:GaP_mu_ef} shows the same test but now for the electron mobility in GaP. We have again computed the lifetimes and mobility for different carrier concentrations to confirm that $\ee_F$ is far enough from the CBM:
an electron concentration of $10^{17}$ cm$^{-3}$ leads to a mobility of 293~cm$^2$/(V$\cdot$s) while a concentration of $10^{18}$ cm$^{-3}$ gives 291~cm$^2$/(V$\cdot$s) (computed with \grid{60} $\kk$- and \grid{120} $\qq$-point grids).
Finally, in Figure~\ref{fig:GaAs_mu_ef}, we report the same test but for the electron mobility in GaAs. Due to the high dispersion, the density of states in the conduction band is very small, and the carrier concentration has to be decreased to $10^{15}$ cm$^{-3}$. We have again computed the lifetimes and mobility for different carrier concentrations to confirm that $\ee_F$ is far enough from the CBM. 
%


%
%

\bibliographystyle{apsrev4-1}